%% file: main.tex
\documentclass[11pt]{article}

\usepackage[margin=1in]{geometry}
\usepackage{amsmath}
\usepackage[numbers,sort&compress]{natbib}
\usepackage{comment}
\usepackage{graphicx}
\usepackage{url}
\usepackage{xcolor}
\providecommand{\bibcommenthead}{}
\usepackage{capt-of}
\usepackage{placeins}
\usepackage{array}
\usepackage{tikz}
\usepackage{pgfplots}
\usetikzlibrary{arrows.meta,calc,decorations.pathmorphing,positioning}
\usepgfplotslibrary{groupplots}
\pgfplotsset{compat=1.17}
\input{figures/Figures.tex}
\definecolor{BLUE}{RGB}{0,90,180}
\usepackage[hidelinks,implicit=false]{hyperref}

\title{\textbf{Thermometry in Near-Degenerate Nitrogen-Vacancy Ensembles
Enabled by Bright--Dark-State Mixing}}
\author{Matt K.~Fu$^{1}$ and John O.~Dabiri$^{1}$\\[0.5em]
\small $^{1}$California Institute of Technology, Pasadena, CA 91125, USA}
\date{}

\begin{document}

\maketitle

\begin{abstract}
\begingroup
\emergencystretch=1em
Nitrogen-vacancy (NV) centers have emerged as a highly sensitive platform for
spatially resolved temperature measurements. When the
$\lvert0\rangle\to\lvert+1\rangle$ and $\lvert0\rangle\to\lvert-1\rangle$
transitions are nearly degenerate and the Rabi frequency is comparable to the
transition splitting, the spin dynamics encountered during pulsed protocols
for temperature measurement can significantly alter the intended accumulation
of temperature-dependent phase. We investigate these triplet
($\lvert0\rangle$, $\lvert\pm1\rangle$) spin dynamics and their effect on
pulsed thermometry protocols under these conditions. We find that the
conventional thermal-echo timing produces negligible temperature-sensitive
signal amplitudes at the common-mode detuning. However, we find that the
nominal spin-echo timing produces resolvable oscillations at the common-mode
detuning, leading to a strong temperature-sensitive carrier signal. This
departure from the expected protocol behavior is qualitatively captured by an
ensemble model of the triplet dynamics, which supports significant dark-state
leakage as a plausible mechanism for the temperature-sensitive response. This
mixing-enabled thermal echo uses a four-phase quadrature readout that resolves
the signed oscillation frequency and is validated using independent sweeps of
the microwave carrier frequency and fluid temperature. We further assess the
ability of a four-refocusing-pulse (LDD4B) low-field dynamical-decoupling
sequence to extend the NV coherence and find that while the echo retains a
sensitivity to common-mode detuning, the effective phase-accumulation interval
corresponds to approximately half the total evolution time. These findings
could support the use of pulsed NV thermometry in biological or mobile
applications where applying magnetic bias fields is impractical and microwave
power may be limited.
\endgroup
\end{abstract}

\section{Introduction}\label{sec:introduction}

Negatively charged nitrogen-vacancy (NV) centers in diamond have been
demonstrated as a versatile quantum sensing platform for a diverse array of
parameters and applications \cite{Degen2017,Schirhagl2014}, particularly
nanoscale magnetometry \cite{Maze2008,Balasubramanian2008,Barry2020} and
thermometry \cite{Kucsko2013,Fujiwara2020InVivo,Hart2026}. The NV ground state
is a spin-1 triplet with long coherence times and sensitivities that persist
in a wide range of conditions, including elevated temperatures
\cite{Balasubramanian2009,Lin2021} and pressures \cite{Wang2021Quadrature}.
The NV spin states can be initialized and read out optically and manipulated
with microwave fields \cite{Degen2017}, making NV centers an accessible
platform for quantum sensing. NV-based thermometry, specifically, has garnered
significant attention \cite{Kucsko2013,Neumann2013,Toyli2013} stemming from
its combination of optical readout, nanoscale spatial resolution, and
operability in ambient and \textit{in vivo} settings
\cite{Kucsko2013,Fujiwara2020InVivo}. While the temperature-dependent
properties of the fluorescence alone have been adapted for a variety of
``all-optical'' temperature measurement
techniques~\cite{Fujiwara2021Thermometry,Plakhotnik2014}, the most common NV
thermometry techniques rely on monitoring the temperature dependence of the
zero-field splitting (ZFS) {frequency $D$} between the $\lvert0\rangle$ and
$\lvert\pm1\rangle$ spin states \cite{Fujiwara2021Thermometry}. {Near} room
temperature, this dependence is expressed as the linearized dependence
$D(T)=D_0+\beta(T-T_0)$, where $D_0\approx 2.87~\mathrm{GHz}$ is the
zero-field splitting frequency at the reference temperature \(T_0\),
$\beta=\mathrm{d}D/\mathrm{d}T\approx -75~\mathrm{kHz}/^{\circ}\mathrm{C}$ is
the temperature coefficient {near} $T_0$ \cite{Acosta2010,Acosta2011Erratum},
and $T$ is the NV temperature. The value of $D$ can often be obtained by
analyzing the optically detected magnetic resonance (ODMR) spectrum of the NV
center and computing the central frequency between the
$\lvert0\rangle\to\lvert+1\rangle$ and $\lvert0\rangle\to\lvert-1\rangle$
resonance frequencies. Similarly, shifts in \(D\) are commonly inferred either
by sweeping the microwave carrier frequency across a fixed range to identify
the resonances or by tracking changes to features or amplitude of the ODMR
line shape relative to a previously characterized
baseline~\cite{Hayashi2018,Fujiwara2020Thermometry}. Both approaches can be
accomplished using either continuous-wave or pulsed protocols
\cite{Fujiwara2021Thermometry,Wojciechowski2018} but differ in their hardware
requirements and sensitivity.

Numerous alternatives to ODMR for spin-based temperature measurement have been
developed with the goal of improving measurement sensitivity, including
thermal-echo and D-Ramsey sequences \cite{Neumann2013,Toyli2013}. Instead of
sweeping the microwave frequency, these thermal-echo protocols sweep the delay
time (free-evolution time) between sequences of microwave pulses with fixed
frequencies. In the conventional thermometry implementation, the
$\lvert0\rangle\leftrightarrow\lvert+1\rangle$ and
$\lvert0\rangle\leftrightarrow\lvert-1\rangle$ transitions are typically split
with a magnetic field bias and are thus spectrally distinct. The two
transitions are then addressed separately with different microwave
frequencies, $f_{+}$ and $f_{-}$, {respectively,} which are the upper and
lower microwave carrier frequencies targeting the two spin state transitions.
These protocols are designed to suppress magnetic-field noise through the use
of refocusing pulses while allowing the NV centers to accumulate {relative}
phase {between $\lvert0\rangle$ and $\lvert\pm1\rangle$ in the microwave
rotating frame} in proportion to the {signed} common-mode {frequency} detuning
given by $\Delta_{\mathrm c}(T)=D(T)-(f_{+} + f_{-})/2$. The accumulated phase
during the pulse sequence manifests as oscillations in the photoluminescence
(PL) of the NV population that vary with the total evolution time. In this
regime, common-mode detuning phase can be accumulated using an eight-pulse
sequence
($\frac{\pi_{-}}{2}-\tau-\pi_{+}\pi_{-}\pi_{+}-\tau-\pi_{-}\pi_{+}\pi_{-}
\frac{\pi_{-}}{2}$), where $\pi$ and $\pi/2$ indicate the rotation angles
induced by the pulse, $\tau$ is a free-evolution period, and the $-$ and $+$
subscripts indicate which resonance each microwave pulse addresses
\cite{Toyli2013,Wang2015Thermometry}. Sequences utilizing additional
repetitions of these $\pi$ pulse trains can further extend the coherence time
and improve sensitivity \cite{Wang2015Thermometry}.

However, splitting the two transitions {(i.e.,
$\lvert0\rangle\leftrightarrow\lvert+1\rangle$ and
$\lvert0\rangle\leftrightarrow\lvert-1\rangle$)} with a bias field is not
tenable in every measurement application. For example, zero- and low-field
magnetic-resonance experiments must preserve intrinsically weak-field
conditions \cite{Cerrillo2021,Vetter2022}. Similarly, there may be
applications (e.g., \textit{in vivo} measurements) that rely on mobile
nanodiamonds, which can translate or rotate, making stable alignment with a
bias field impractical \cite{Sow2025}. Avoiding a dedicated bias source also
simplifies the instrument and removes bias-source drift as a control variable
\cite{Childress2025}. These considerations motivate characterizing the
performance of thermometry methodologies under ambient geomagnetic conditions.
In these conditions, a single microwave pulse often has sufficient bandwidth
to encompass both resonance frequencies and drive both transitions
simultaneously. As such, the two transition frequencies are nearly degenerate,
and the triplet dynamics need to be accounted for when the NV system is used
for sensing. When both transitions are driven equally, it is convenient to
define an alternative basis for the $\lvert\pm1\rangle$ subspace of the
electronic ground-state triplet, comprising the microwave-bright state $\lvert
B\rangle = (\lvert +1\rangle+\lvert -1\rangle)/\sqrt{2}$ and the
microwave-dark state $\lvert D\rangle = (\lvert +1\rangle - \lvert
-1\rangle)/\sqrt{2}$. The microwave field couples the $\lvert0\rangle$ and
$\lvert B\rangle$ states, while the $\lvert D\rangle$ state is not directly
addressed by the microwave field. Instead, a {small but} finite splitting
between the $\lvert\pm1\rangle$ states can produce both differential
bright--dark shifts and bright--dark coupling. Such splitting can arise from
weak magnetic fields, crystal strain, or hyperfine interactions
\cite{Doherty2013}.

The ability to simultaneously address both transitions permits thermal-echo to
be implemented with fewer microwave pulses, typically as the three-pulse
sequence ($\frac{\pi}{2}-\tau-2\pi-\tau-\frac{\pi}{2}$)
\cite{Kucsko2013,Toyli2013,Wang2015Thermometry}. The first outer pulse
prepares a superposition of $\lvert0\rangle$ and $\lvert B\rangle$. The
central pulse exchanges the $\lvert+1\rangle$ and $\lvert-1\rangle$
amplitudes, preserving phase accumulated from the common-mode detuning while
refocusing differential Zeeman phase. The final outer pulse converts the
accumulated phase into a population difference that can be read out optically.
However, simply being able to address both transitions does not guarantee
ideal control. This protocol assumes that the microwave Rabi frequency greatly
exceeds the residual transition splitting so that bright--dark interactions
can be neglected during pulses. As such, many of the low-field thermometry
studies operate in this drive-dominated regime
\cite{Toyli2013,Vetter2022,Sow2025} to avoid these nonidealities.

However, when the splitting and Rabi frequency are comparable, drive and
bright--dark dynamics can occur during finite-duration pulses. This
simultaneous evolution can alter contrast, refocusing performance, and signal
selectivity, particularly in ensembles with a distribution of splittings. The
implications of finite-duration, three-level dynamics under these conditions
have been investigated for coherent control and bias-field-free magnetometry
\cite{LopezGarcia2025,Childress2025}. However, how these dynamics impact the
phase accumulation required for thermometry remains less well characterized.

Here, we explore how finite-duration, three-level dynamics affect the physics
of NV thermometry sequences when the microwave Rabi frequency is comparable to
the difference between the two transition frequencies. Specifically, we
characterize how finite-duration microwave pulses and bright--dark mixing
alter the behavior of thermal-echo and spin-echo-based sequences. We
experimentally find that the temperature-sensitive signal from the
thermal-echo protocol is strongly suppressed in this regime. By contrast, the
spin-echo-like sequence timing produces resolvable oscillations at the
common-mode detuning, as might be expected from a thermal echo, making it
suitable for thermometry measurements. We confirm that the frequency of the
echo waveform from this protocol follows $\Delta_{\mathrm c}(T)$ through
independent sweeps of the microwave carrier frequency and operating
temperature. We extend these findings by evaluating a modified version of the
low-field dynamical-decoupling sequence \cite{Vetter2022} as a four-pulse
extension of this approach and find that, while {some level of} sensitivity is
retained, the resolvable phase accumulates at half the expected rate,
potentially limiting the sensitivity of the approach. Finally, we develop a
finite-pulse three-level ensemble model that qualitatively captures the
protocol-dependent trends and supports bright--dark mixing as a plausible
mechanism underlying this modified behavior.

\section{Methods}\label{sec:methods}

\subsection{Experimental Setup}\label{sec:methods-experimental-setup}

\subsubsection{NV-center spin-control and readout
hardware}\label{sec:methods-spin-control-readout}
Spin initialization, manipulation, and readout were performed using the
synchronized optical, microwave, and photon-counting subsystems diagrammed in
Fig.~\ref{fig:hardware-platform} and based on the system design of Misonou et
al. ~\cite{Misonou2020}. Optical excitation was provided by a
$532~\mathrm{nm}$ diode-pumped laser ($1~\mathrm{W}$, CNI Laser MLL-FN-532-1W)
coupled into a single-mode fiber-optic cable. The excitation-beam power was
attenuated and pulsed with an acousto-optic modulator (AOM with
$200~\mathrm{MHz}$ bandwidth; Gooch \& Housego) before collimation and
delivery into the microscope optical system. A $550~\mathrm{nm}$ cut-on
dichroic mirror (Thorlabs, DMLP550R) reflected the collimated beam into a
$100\times$ microscope objective with $\mathrm{NA}=0.8$ (Olympus,
LMPLFLN100XBD), where it was focused onto the NV center ensemble of interest.
Red NV PL from the stimulated NV centers was collected by the same objective
and then diverted from the excitation-beam pathway by the dichroic mirror. The
transmitted PL passed through an additional band-pass filter stack (Thorlabs,
FELH0650 and FESH0800) to restrict the detected PL to a portion of the
expected fluorescence wavelength range ($650$--$800~\mathrm{nm}$). A 50:50
beamsplitter (Thorlabs, EBS1) in the PL optical path split the fluorescence
signal between a $1.6~\mathrm{MP}$ monochrome Zelux camera (Thorlabs) used for
wide-field alignment, imaging, and calibration and a single-photon detection
module (Thorlabs SPDMH2F) for PL measurement. Photon-detection events from the
photon counter were recorded by a data-acquisition unit (National Instruments,
USB-6343, up to $1~\mathrm{MS/s}$).

A separate radio-frequency (RF) system delivered microwave-frequency pulses
synchronized with the optical excitation and photon-counting events to
coherently manipulate the NV spin states. A vector signal generator (Anritsu
MG3710E) generated a microwave-frequency carrier signal and applied
preprogrammed amplitude and phase modulation via an IQ modulator. The
modulation waveform and timing pulses for gated hardware (i.e., the photon
counter, AOM, and RF switch) were programmed externally according to the
desired pulse sequence and output by an arbitrary waveform generator (AWG,
$160~\mathrm{MS/s}$) integrated into the vector signal generator. The
amplitudes of microwave pulses utilized in these experiments were generally
windowed according to a cosine-squared function to control spectral leakage,
with rectangular windows used where noted.
The resulting waveform (i.e., the microwave pulse
sequence) was gated by a microwave switch (Mini-Circuits, ZYSWA-2-50DR+) and
amplified (Mini-Circuits, ZHL-16W-43+) before being transmitted to the NV
centers through a ring coplanar waveguide antenna \cite{Ma2019}. The antenna
was isolated from the rest of the circuit by a terminated RF circulator
(Orient Microwave, FX00-0329-00) that functioned as an isolator by dissipating
reflected power from the microwave antenna, preventing damage to the rest of
the microwave hardware chain. The onboard AWG maintained hardware
synchronization by supplying the modulation waveform and timing pulses.

\begin{center}
  \includegraphics[width=0.98\textwidth,trim=0bp 180bp 0bp
0bp,clip]{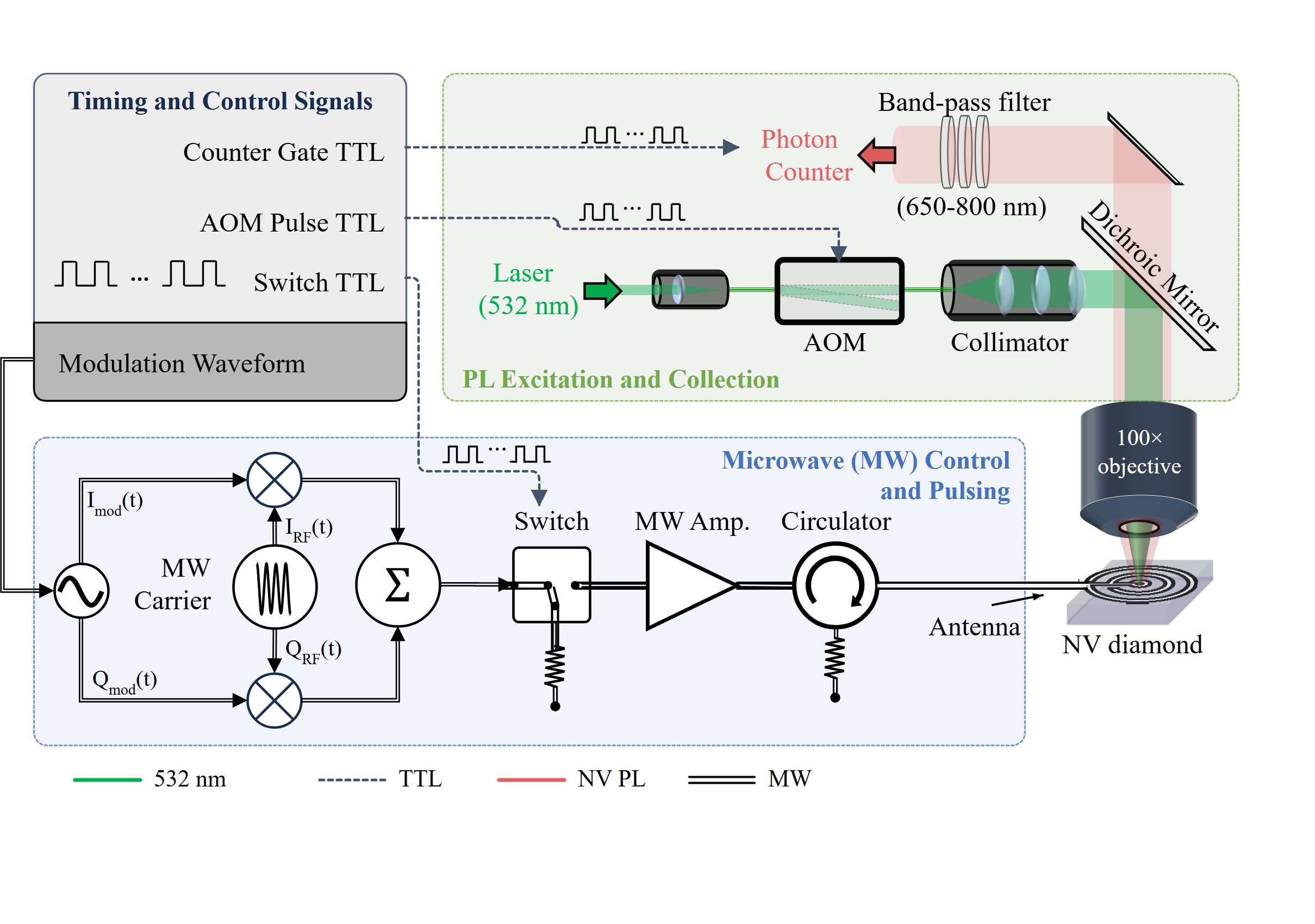}
  \vspace{-0.5em}
  \captionof{figure}{\textbf{Diagram of the NV-center quantum-sensing
platform.} The optical excitation and collection subsystem (green region)
pulses and attenuates a $532~\mathrm{nm}$ laser (green beam) with an
acousto-optic modulator to polarize and read out the NV spin state. The laser
pulses are reflected by a dichroic mirror and focused onto the NV-diamond
sensor by the microscope objective. The spin-dependent NV photoluminescence
(PL; red beam) is collected through the same objective, separated from the
excitation path by the dichroic mirror, spectrally filtered, and detected by a
photon counter. In the microwave subsystem (blue region), the modulated
microwave carrier is gated by a microwave switch, amplified by a +45 dBm
amplifier, passed through an isolator, and delivered to the NV ensemble
through a ring coplanar waveguide. Timing and gating pulses (gray region) are
generated by an arbitrary waveform generator (AWG). These pulses ensure
synchronization between the optical, microwave, and photon-counting
subsystems. The AWG also generates the modulation waveform used to control the
phase and amplitude of the microwave carrier signal.
}\label{fig:hardware-platform}
\end{center}

\subsubsection{NV-diamond sensor and integrated flow
cell}\label{sec:methods-apparatus-flow-cell}
NV measurements were obtained with a
$3~\mathrm{mm}\times3~\mathrm{mm}\times0.25~\mathrm{mm}$ single-crystal
chemical-vapor-deposition diamond substrate (Adamas Nanotechnologies,
DPNVMS-OG33) with the $\langle100\rangle$ crystallographic orientation along
the thickness dimension. The sensing region consisted of a
$0.5~\mu\mathrm{m}$-thick near-surface NV layer on one of the $(100)$ faces,
with a nitrogen concentration below $1~\mathrm{ppm}$. {This orientation
ensures that all four NV axes are tilted at the same angle ($54.7^\circ$) from
the $(100)$ surface normal (the thickness direction).} The NV-containing layer
was exposed directly to the working fluid as a wetted surface, while the
optical excitation and PL collection were performed through the backside of
the diamond, as shown in Fig.~\ref{fig:flow-system}a. NV-PL imaging showed
that the optical interrogation region was approximately $5.3~\mu\mathrm{m}$ in
diameter at half-maximum intensity and $13~\mu\mathrm{m}$ in diameter at
$e^{-2}$ of the maximum intensity, as shown in Fig.~\ref{fig:flow-system}b.

\begin{center}
  \makebox[\linewidth][l]{%
  \hspace{0.51\linewidth}%
  \begin{minipage}[t]{0.49\linewidth}
    \centering
    \vspace{0pt}
    \makebox[\linewidth][c]{\strut(c)\hspace{0.7em}\textbf{Flow
cell}}\\[-0.2em]
    \begin{tikzpicture}[remember picture]
      \node[inner sep=0pt] (figthreea-img)
{\includegraphics[width=\linewidth,trim=0bp 230bp 0bp
220bp,clip]{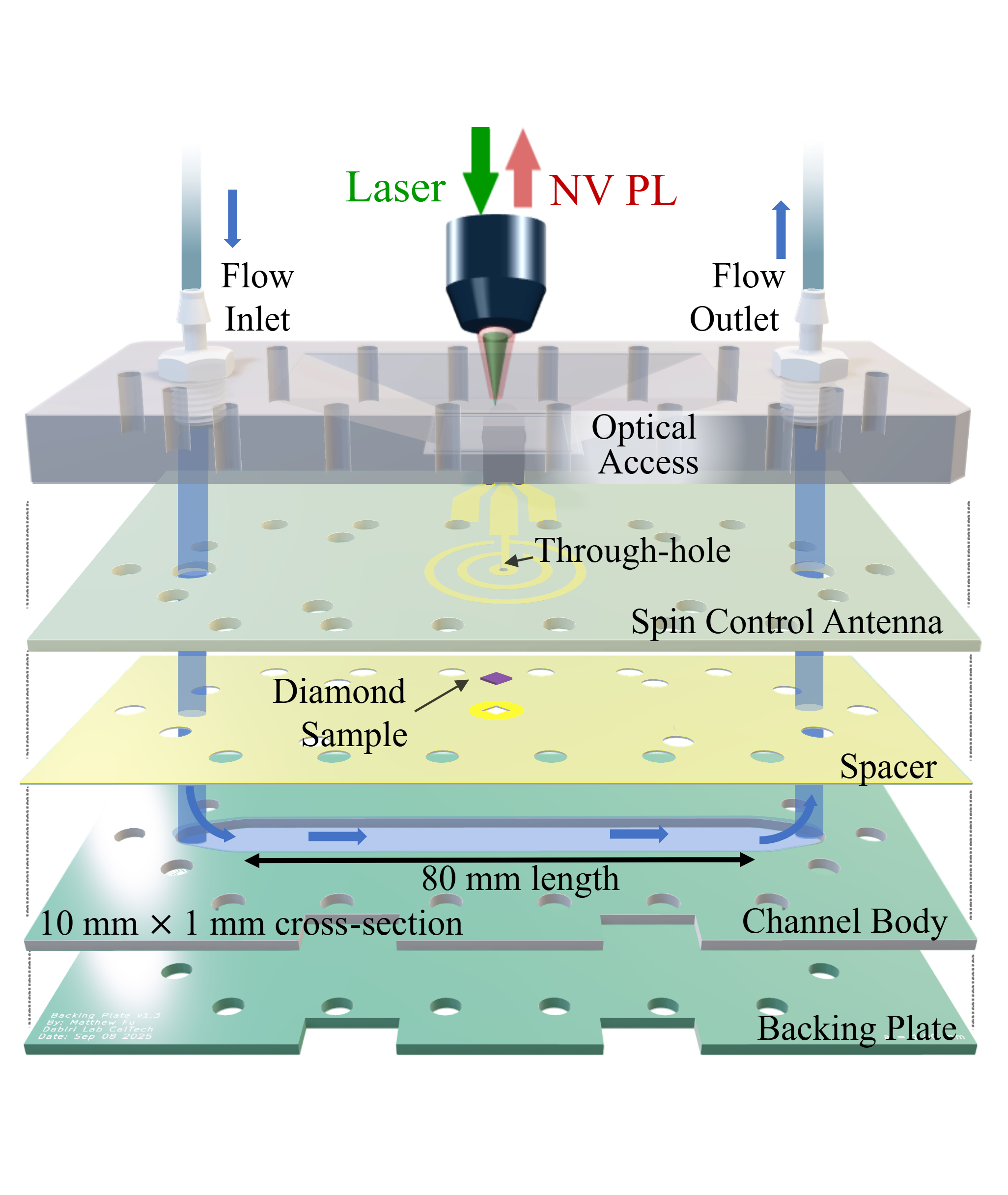}};
      \coordinate (figthreea-diamond-wb) at ($(figthreea-img.south
west)!0.442!(figthreea-img.south east)$);
      \coordinate (figthreea-diamond-wt) at ($(figthreea-img.north
west)!0.442!(figthreea-img.north east)$);
      \coordinate (figthreea-diamond-eb) at ($(figthreea-img.south
west)!0.542!(figthreea-img.south east)$);
      \coordinate (figthreea-diamond-et) at ($(figthreea-img.north
west)!0.542!(figthreea-img.north east)$);
      \coordinate (figthreea-diamond-sw) at
($(figthreea-diamond-wb)!0.387!(figthreea-diamond-wt)$);
      \coordinate (figthreea-diamond-nw) at
($(figthreea-diamond-wb)!0.442!(figthreea-diamond-wt)$);
      \coordinate (figthreea-diamond-se) at
($(figthreea-diamond-eb)!0.387!(figthreea-diamond-et)$);
      \coordinate (figthreea-diamond-ne) at
($(figthreea-diamond-eb)!0.442!(figthreea-diamond-et)$);
      \coordinate (figthreea-callout-upper) at
($(figthreea-diamond-wb)!0.435!(figthreea-diamond-wt)$);
      \coordinate (figthreea-callout-lower) at
($(figthreea-diamond-wb)!0.394!(figthreea-diamond-wt)$);
      \draw[red!80!black,line width=1.2pt,rounded corners=3pt]
(figthreea-diamond-sw) rectangle (figthreea-diamond-ne);
    \end{tikzpicture}
  \end{minipage}%
  \hspace{-\linewidth}%
  \begin{minipage}[t][0.55\linewidth][t]{0.48\linewidth}
    \centering
    \vspace{0pt}
    \makebox[\linewidth][c]{(a)\hspace{0.7em}\textbf{NV geometry}}\\[-0.2em]
    \makebox[\linewidth][l]{%
      \pgfmathsetmacro{\figthreegeoscale}{0.98*\linewidth/(8.35cm)}%
      \begin{tikzpicture}[remember
picture,x=1cm,y=1cm,scale=\figthreegeoscale,transform shape]
        \path[use as bounding box] (0,0) rectangle (8.35,3.05);
        \fill[gray!7] (0.05,0.14) rectangle (8.28,2.95);
        \draw[red,line width=1.2pt,rounded corners=5pt] (0.05,0.14) rectangle
(8.28,2.95);
        \coordinate (figthreeb-callout-upper) at (8.28,2.78);
        \coordinate (figthreeb-callout-lower) at (8.28,0.16);

        \begin{scope}[xshift=0.60cm,yshift=-0.33cm]
        \begin{scope}[scale=1.12,transform shape]
        \path[fill=red!70!black,draw=red!70!black,opacity=0.43,line
width=0.25pt]
          (0.98,2.92) -- (3.10,2.92) -- (2.06,1.01) -- cycle;
        \path[fill=green!55!black,draw=green!55!black,opacity=0.50,line
width=0.25pt]
          (1.55,2.92) -- (2.57,2.92) -- (2.06,1.01) -- cycle;
        \node[font=\footnotesize\bfseries,red!75!black,anchor=east] at
(1.22,2.48) {NV PL};
        \node[font=\footnotesize\bfseries,green!35!black,anchor=center] at
(2.06,2.46) {Laser};

        \fill[gray!38] (0.72,1.30) -- (2.06,1.75) -- (3.40,1.30) --
(2.06,0.87) -- cycle;
        \fill[gray!58] (0.72,1.30) -- (2.06,1.75) -- (2.06,1.40) --
(0.72,0.96) -- cycle;
        \fill[gray!28] (2.06,1.75) -- (3.40,1.30) -- (3.40,0.96) --
(2.06,1.40) -- cycle;
        \fill[purple!60!black,opacity=0.62] (0.72,0.95) -- (2.06,1.39) --
(3.40,0.95) -- (2.06,0.52) -- cycle;
        \draw[purple!70!black,line width=0.45pt] (0.72,0.95) -- (2.06,1.39) --
(3.40,0.95);
        \draw[black!35,line width=0.40pt] (0.72,1.30) -- (2.06,1.75) --
(3.40,1.30) -- (3.40,0.96) -- (2.06,0.52) -- (0.72,0.95) -- cycle;

        \fill[red] (2.06,0.96) ellipse (0.060 and 0.050);
        \draw[black,line width=0.40pt] (2.06,0.96) ellipse (0.070 and 0.058);
        \coordinate (figthreeb-pl-spot-left) at (1.99,0.96);
        \coordinate (figthreeb-pl-spot-right) at (2.13,0.96);
        \draw[black!60,line
width=0.40pt,{Latex[length=0.8mm]}-{Latex[length=0.8mm]}] (0.86,1.46) --
(1.82,1.78);
        \node[font=\scriptsize] at (1.16,1.84) {$3~\mathrm{mm}$};
        \draw[black!60,line
width=0.40pt,{Latex[length=0.8mm]}-{Latex[length=0.8mm]}] (2.30,1.78) --
(3.26,1.46);
        \node[font=\scriptsize] at (2.96,1.84) {$3~\mathrm{mm}$};
        \draw[black!60,line
width=0.40pt,{Latex[length=0.8mm]}-{Latex[length=0.8mm]}] (0.56,0.98) --
(0.56,1.28);
        \node[font=\scriptsize,anchor=east] at (1.06,0.82)
{$0.25~\mathrm{mm}$};

        \draw[purple!70!black,line width=0.85pt] (3.50,0.95) -- (3.80,0.95);
        \draw[purple!70!black,line width=0.85pt] (3.50,1.07) -- (3.80,1.07);
        \draw[purple!70!black,line width=0.70pt,-{Latex[length=1.25mm]}]
(3.65,1.23) -- (3.65,1.07);
        \draw[purple!70!black,line width=0.70pt,-{Latex[length=1.25mm]}]
(3.65,0.79) -- (3.65,0.95);
        \node[font=\scriptsize,purple!70!black,anchor=west,align=left] at
(3.93,1.03) {$0.5~\mu\mathrm{m}$\\($[\mathrm{N}]<1~\mathrm{ppm}$)};

        \node[anchor=west,align=left,font=\small] at (4.18,2.18) {CVD
diamond\\$(100)$ surface};
        \draw[black!60,line width=0.40pt,-{Latex[length=0.9mm]}] (4.34,1.86)
-- (3.33,1.44);
        \end{scope}
        \end{scope}
      \end{tikzpicture}%
    }
    \vspace{0.25em}
    \makebox[\linewidth][c]{(b)\hspace{0.7em}\textbf{NV PL map}}\\[-0.2em]
    \makebox[\linewidth][l]{%
    \begin{tikzpicture}[remember picture]
      \node[inner sep=0pt] (figthreec-img)
{\includegraphics[width=0.966\linewidth,trim=5bp 92bp 59bp
9bp,clip]{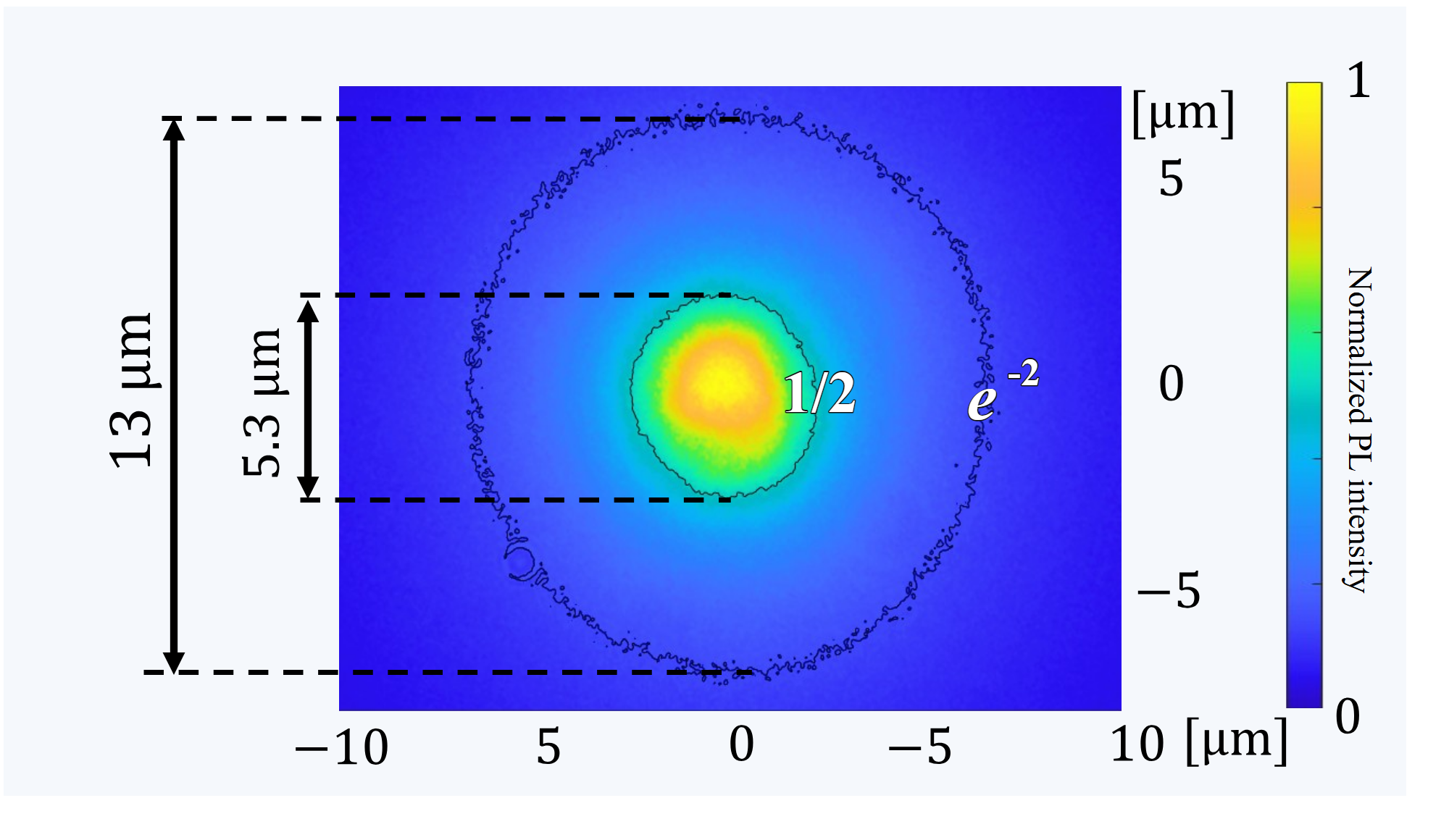}};
      \coordinate (figthreec-frame-sw) at ($(figthreec-img.south
west)+(0,-4pt)$);
      \fill[fill={rgb,255:red,245;green,248;blue,252}] (figthreec-frame-sw)
rectangle (figthreec-img.south east);
      \draw[black,line width=1.2pt,rounded corners=5pt] (figthreec-frame-sw)
rectangle (figthreec-img.north east);
      \node[font=\footnotesize,fill={rgb,255:red,245;green,248;blue,252},
        inner sep=0pt,minimum width=13pt,minimum height=8.5pt]
        at ($(figthreec-img.south west)!0.388719!(figthreec-img.south
east)+(0,4.37pt)$)
        {$-5$};
      \node[font=\footnotesize,fill={rgb,255:red,245;green,248;blue,252},
        inner sep=0pt,minimum width=13pt,minimum height=8.5pt]
        at ($(figthreec-img.south west)!0.653210!(figthreec-img.south
east)+(0,4.37pt)$)
        {$5$};
      \coordinate (figthreec-plot-left-bottom) at ($(figthreec-img.south
west)!0.239!(figthreec-img.south east)$);
      \coordinate (figthreec-plot-left-top) at ($(figthreec-img.north
west)!0.239!(figthreec-img.north east)$);
      \coordinate (figthreec-plot-right-bottom) at ($(figthreec-img.south
west)!0.796!(figthreec-img.south east)$);
      \coordinate (figthreec-plot-right-top) at ($(figthreec-img.north
west)!0.796!(figthreec-img.north east)$);
      \coordinate (figthreec-link-left) at ($(figthreec-img.north
west)+(1.4pt,-1.4pt)$);
      \coordinate (figthreec-link-right) at ($(figthreec-img.north
east)+(-1.4pt,1.1pt)$);
    \end{tikzpicture}%
    }
  \end{minipage}%
  }
  \begin{tikzpicture}[remember picture,overlay]
    \draw[red!80!black,line width=0.65pt,dashed,line cap=round]
(figthreea-callout-upper) -- (figthreeb-callout-upper);
    \draw[red!80!black,line width=0.65pt,dashed,line cap=round]
(figthreea-callout-lower) -- (figthreeb-callout-lower);
    \draw[black,line width=0.65pt,dashed,line cap=round,shorten <=-0.6pt]
(figthreeb-pl-spot-left) -- (figthreec-link-left);
    \draw[black,line width=0.65pt,dashed,line cap=round,shorten <=-0.6pt]
(figthreeb-pl-spot-right) -- (figthreec-link-right);
  \end{tikzpicture}
  \vspace{-2em}
  
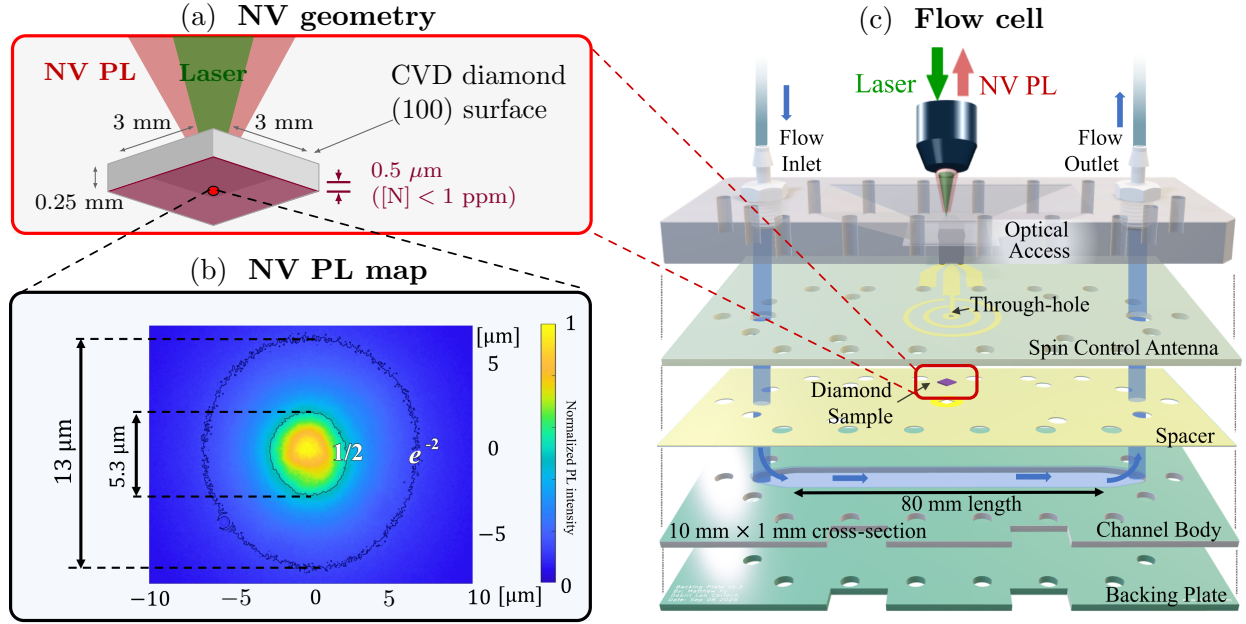
\captionof{figure}{\textbf{Geometry of the flow cell and NV-diamond sensor.}
(a) Diagram of the NV-diamond sensor geometry. The
$3~\mathrm{mm}\times3~\mathrm{mm}\times0.25~\mathrm{mm}$
chemical-vapor-deposition (CVD) diamond has a $(100)$ fluid-facing surface and
contains a $0.5~\mu\mathrm{m}$-thick near-surface layer of nitrogen-vacancy
(NV) centers. The NV layer is optically excited, and its PL is collected
through the surface opposite the NV layer. (b) Normalized PL intensity map at
the diamond surface. The dashed contours mark the intensity boundaries at
half-maximum and at $e^{-2}$ of the maximum, with diameters of approximately
$5.3~\mu\mathrm{m}$ and $13~\mu\mathrm{m}$, respectively, defining the lateral
optical interrogation region. (c) Exploded schematic of the flow cell. Fluid
passes through an 80-mm-long channel with a
$10~\mathrm{mm}\times1~\mathrm{mm}$ cross-section. The NV-diamond sensor is
mounted flush with the channel wall, with the sensing surface exposed to the
passing fluid. The antenna board used for microwave spin control is mounted
behind the diamond and contains a through-hole that provides optical access
through the back side of the diamond for NV excitation and photoluminescence
(PL) collection. Near the center of the ring coplanar waveguide, the microwave
magnetic field is approximately normal to the $(100)$ diamond surface.
Consequently, the four NV orientations experience nominally equal
microwave-field amplitudes, and no NV orientation is preferentially driven.
Schematics in (a) and (c) are not to scale.}\label{fig:flow-system}
\end{center}

The diamond sample was integrated into a tabletop flow cell whose construction
is shown in Fig.~\ref{fig:flow-system}c. The flow system allowed for precise
thermal control of the NV measurement site by providing a constant fluid flow
from a large reservoir of deionized water, whose temperature could be
monitored and controlled. Although deionized water was used as the working
fluid in the experiments presented here, the system is designed to be
compatible with other fluids, such as seawater. Flow within the cell was
driven through a closed loop by a programmable peristaltic pump (Kamoer
DIP1500) at a constant flow rate selected from the
$1$--$300~\mathrm{mL\,min^{-1}}$ range. The flow cell was connected to the
pump by flexible Tygon tubing to reduce pressure spikes from the pump and
further isolated on one side by the reservoir. The fluid temperature was
monitored by PT100 resistance temperature detectors (RTDs) placed in the fluid
reservoir and immediately upstream of the flow channel. The diamond sample was
mounted in the flow cell using a custom printed circuit board (PCB)
incorporating the ring coplanar waveguide design described by Ma et al.
\cite{Ma2019}. This design was chosen because it provides an approximately
uniform microwave intensity and optical access within a
$1.4~\mathrm{mm}$-diameter central hole. The diamond and antenna were
integrated with a $0.3~\mathrm{mm}$-thick FR-4 spacer to form a nearly flush
surface comprising the top wall of the flow channel. This stack was combined
with two additional PCBs and a 3D-printed mounting substrate to create a flow
cell with a $10~\mathrm{mm}\times1~\mathrm{mm}$ cross-section and a length of
approximately $80~\mathrm{mm}$.

\subsubsection{System
Characterization}\label{sec:methods-microwave-pulse-calibration}
Optically detected magnetic resonance (ODMR) measurements were used to confirm
the resonance characteristics of the NV ensemble and identify the nominal
zero-field-splitting frequency of the NV transition, which was found to be
$D\approx2.87~\mathrm{GHz}$ \cite{Degen2017} near ambient temperature and
atmospheric pressure. The ODMR spectrum was acquired using a pulsed protocol,
in which the PL signals of the NV ensemble, $\mathrm{PL}_{1}$ and
$\mathrm{PL}_{2}$, were stimulated and measured before and after a microwave
pulse, respectively. The microwave carrier frequency was stepped through a set
of discrete setpoints within $\pm 8~\mathrm{MHz}$ of $D_0$. In this
implementation, the amplitude of the microwave pulse is windowed with a
cosine-squared envelope to suppress the amplitude of the spectral sidelobes
and accommodate the bandwidth of the system. The resulting PL intensity pairs
were used to compute the PL ratio, $\mathrm{PL}_{2}/\mathrm{PL}_{1}$, for each
frequency.

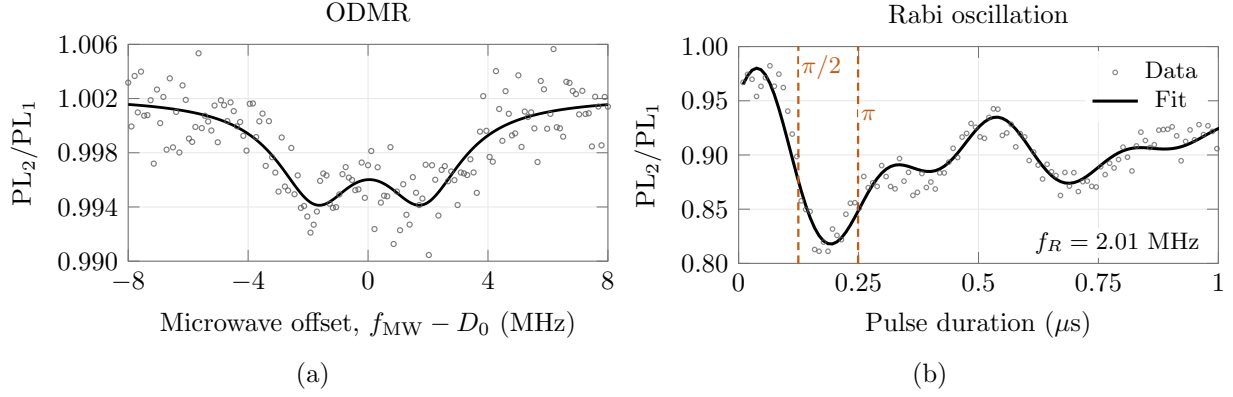
\begin{figure}[!t]
  \centering
  \input{figures/Figure3.tex}
  \caption{\textbf{Microwave resonance and pulse calibration.}
  (a) The pulsed, optically detected magnetic resonance (ODMR) spectrum shows
the PL ratio ($\mathrm{PL}_2/\mathrm{PL}_1$) plotted as a function of the
microwave frequency offset ($f_{\mathrm{MW}}-D_0$). Microwave frequencies are
shown in megahertz relative to the reference zero-field-splitting frequency,
$D_0=2870.5~\mathrm{MHz}$. Open circles show the raw PL ratio. The solid curve
shows the best fit of the raw data to two symmetric Lorentzian functions, with
parameters given in Table~\ref{tab:odmr-two-lobe-fit-parameters}.
  (b) Rabi oscillation in the PL ratio as a function of microwave pulse
duration. The black curve is the damped two-harmonic fit described in the
text. The fit reveals a fundamental frequency at $f_R=2.01~\mathrm{MHz}$.
Orange dashed lines mark the resulting $\pi/2$ and $\pi$ pulse durations of
$125~\mathrm{ns}$ and $250~\mathrm{ns}$, respectively, used to calibrate the
microwave rotations in
Fig.~\ref{fig:quadrature-phasor-extraction}a.}\label{fig:rabi-calibration}
\end{figure}

Because no additional magnetic field was applied beyond the estimated
geomagnetic background
($\lvert\mathbf{B}_{\mathrm{geo}}\rvert\approx46.0~\mu\mathrm{T}$
\cite{WMMHR2025}; see Table~\ref{tab:geomagnetic-nv-splitting} in
Appendix~\ref{sec:geomagnetic-field}), the measured ODMR spectrum around the
zero-field-splitting frequency exhibited two broad, overlapping dips whose
fitted centers were separated by ${3.55}~\mathrm{MHz}{{}\pm0.34~\mathrm{MHz}}$
{(approximate $95\%$ fit confidence interval, CI)} (see
Fig.~\ref{fig:rabi-calibration}a {and Appendix~\ref{sec:odmr-two-lobe-fit}}).
This behavior is consistent with the expected range of transition frequencies
arising from geomagnetic Zeeman splitting along the four NV orientations
within the crystal lattice, together with $^{14}\mathrm{N}$ hyperfine
splitting and additional contributions from transverse strain that require
further characterization. These latter effects were not individually
resolvable because their frequency separations were smaller than the observed
linewidths of the respective transitions. Regardless, the MHz-scale separation
between the two resonance dips that correspond to the $\lvert
0\rangle\to\lvert +1\rangle$ and $\lvert 0\rangle\to\lvert -1\rangle$
transitions indicates that these states are nearly degenerate.

The durations of the microwave pulses required for spin state transitions were
estimated by analyzing the Rabi oscillations of the targeted NV ensemble.
Similar to the ODMR measurement, this process involved stimulating and
measuring the PL before and after a single microwave pulse
(Fig.~\ref{fig:rabi-calibration}b). In this case, however, the microwave
frequency was fixed at the approximate zero-field-splitting frequency
($f_{\mathrm{MW}}=D_0$) while the pulse duration was swept. As before, the
microwave pulse is windowed with a cosine-squared envelope, leading to half
the rotation of a similarly timed rectangular pulse. In a two-state system,
the PL ratio should oscillate with pulse duration at the Rabi frequency. {In
contrast}, the measurements shown in Fig.~\ref{fig:rabi-calibration}b are
considerably more complex and far from ideal. Despite the significant
dephasing behavior, these oscillations contain a second harmonic, which is
consistent with the response of the three-level system when the $\lvert
+1\rangle$ and $\lvert -1\rangle$ transitions are nearly degenerate. However,
this second harmonic is not cleanly resolvable and contains a noticeable phase
shift relative to the fundamental frequency. These deviations are likely due
to the overlapping effects of the multiple splittings in the NV ensemble that
modify the apparent Rabi frequencies.

\begin{samepage}
The resulting oscillations in the normalized contrast
$C=(\mathrm{PL}_1-\mathrm{PL}_2)/(\mathrm{PL}_1+\mathrm{PL}_2)$ were fitted
using a damped, two-harmonic model,
\begin{equation*}
C_{\mathrm{fit}}(t)=c_0+c_1(t-\bar{t})+\exp[-(t-t_{\min})/T_R]\left\{%
-A_1\cos(2\pi f_R t)+A_2\cos(4\pi f_R t-\phi_2)\right\}.
\end{equation*}%
\end{samepage}%
Here, $R_{\mathrm{fit}}(t)=[1-C_{\mathrm{fit}}(t)]/[1+C_{\mathrm{fit}}(t)]$ is
the fitted PL ratio at pulse duration $t$; $c_0$ is the baseline at the mean
sampled duration, $\bar{t}$, $c_1$ is the linear baseline slope, $T_R$ is the
decay time, $f_R$ is the fundamental Rabi frequency, and $A_j$ and $\phi_j$
are the amplitudes and phases of the two harmonics. The decay envelope starts
at the first sampled duration, $t_{\min}$. The fit constrains the
second-harmonic frequency to $2f_R$ but allows $\phi_2$ to be fitted.
The fit revealed a fundamental Rabi frequency of
$f_R={2.008}~\mathrm{MHz}{{}\pm0.045~\mathrm{MHz}}$ {($95\%$ CI)} and
corresponding $\pi$ and $\pi/2$ pulse durations of $250~\mathrm{ns}$ and
$125~\mathrm{ns}$, respectively. Importantly, for the purposes of this study,
this $f_R$ is comparable to the observed ${3.55}~\mathrm{MHz}$ splitting
between the two states.
The presence of a second-harmonic component is consistent with the expected
response of a three-level system in which a linearly polarized microwave field
drives the nearly degenerate $\lvert +1\rangle$ and $\lvert -1\rangle$
transitions, as indicated by the ODMR spectrum. However, the observed phase
shift between the harmonic components is not entirely consistent with the
qutrit behavior described by Vetter et al.~\cite{Vetter2022}. Similarly, the
strong damping observed in the present measurements may result from strong
dark state leakage due to multiple overlapping hyperfine or strain-induced
splittings in the NV ensemble that modify the apparent Rabi frequencies and
transition phases. 

\subsection{Empirical three-level ensemble model}
\label{sec:methods-three-level-ensemble-model}
The experimental measurements were paired with an ensemble model that
numerically evaluates the {response of the NV ensemble} to the different pulse
protocols. The model uses the ordered basis $\{\lvert0\rangle,\lvert
B\rangle,\lvert D\rangle\}$, where the latter two are the microwave-bright and
microwave-dark states defined using the phase convention given above. The
model assumes a linearly polarized microwave field that drives the
$\lvert0\rangle\leftrightarrow\lvert B\rangle$ transition with equal strength
across all four NV orientations. {The model represents the NV ensemble as a
set of branches, each with a specified NV orientation, nuclear-spin
projection, and transverse coupling angle. The branch responses are calculated
separately and averaged to obtain the ensemble response.} The evolution of the
ensemble states is modeled by constructing the three-state, rotating-frame
Hamiltonians for $K$ ensemble branches. Following previous treatments
\cite{Cerrillo2021,Vetter2022,Matsuzaki2016}, the Hamiltonian $H_{k,\phi}(t)$
for ensemble branch $k$ with programmed microwave phase $\phi$ is given by%

\begin{equation}
\frac{H_{k,\phi}(t)}{h}
=
\begin{pmatrix}
0 & \tfrac12\Omega(t)e^{+i\phi} & 0\\
\tfrac12\Omega(t)e^{-i\phi}
  & \Delta_{\mathrm c}(T)+\epsilon_k\cos\chi_k & \epsilon_k\sin\chi_k\\
0 & \epsilon_k\sin\chi_k & \Delta_{\mathrm c}(T)-\epsilon_k\cos\chi_k
\end{pmatrix}.
\label{eq:methods-nv3-hamiltonian}
\end{equation}

Here, $h$ is Planck's constant, $\Omega(t)$ is the instantaneous Rabi
frequency, $f_{\mathrm{MW}}$ is the microwave pulse carrier frequency,
$\Delta_{\mathrm c}(T)=D(T)-f_{\mathrm{MW}}$ is still the common-mode
detuning, $\epsilon_k$ is the branch half-splitting, and $\chi_k$ is the
branch mixing angle. Temperature dependence is assumed to act only through
\(D(T)\) via \(\Delta_{\mathrm c}(T)\). $\epsilon_k\cos\chi_k$ represents the
branch-dependent differential shift of the bright and dark states, while
$\epsilon_k\sin\chi_k$ is the coupling strength between $\lvert B\rangle$ and
$\lvert D\rangle$. The two undriven eigenfrequencies of the bright-dark states
are $\Delta_{\mathrm c}(T)\pm\epsilon_k$ with separation $2\epsilon_k$. %

To simulate the effect of the pulse sequence on an individual branch, the
branch state is evolved over a time increment $\Delta t$ under the
time-evolution operator:

\begin{equation}
U_{k,\phi}(\Delta t;t)
  =\exp\!\left[-i\,2\pi\,
    \frac{H_{k,\phi}(t)}{h}\,\Delta t\right].
\label{eq:methods-nv3-step-propagator}
\end{equation}

The calculations include coherent evolution during finite microwave pulses and
free evolution, but do not include homogeneous dephasing or population
relaxation.

The programmed parameters $\phi$ and $f_{\mathrm{MW}}$, along with the
functional shape of $\Omega(t)$, are known \textit{a priori} from the
experimental sequence; the amplitude of $\Omega(t)$ is calibrated based on the
Rabi data from section~\ref{sec:methods-microwave-pulse-calibration}. The
calculations assume a peak drive $\Omega_{\mathrm{pk}}=4.00~\mathrm{MHz}$.
The remaining branch parameters, $\epsilon_k$ and $\chi_k$, require additional
measurements or assumptions. Specifically, while the splitting contributions
from Zeeman and hyperfine interactions can be estimated from known or modelled
parameters, the contribution from transverse strain is not directly measurable
but can be inferred from the ODMR spectrum.

Estimating the transverse strain proceeds by constructing and fitting a model
for the ODMR spectrum based on the known splitting behavior of the NV ensemble
from the spin Hamiltonian. The model ODMR is constructed from a summation of
24 Lorentzian functions centered on the half-splitting frequencies given by

\begin{equation}
\Delta f_{jm} 
=  
\pm\sqrt{
\left(b_j+m A_\parallel\right)^2
+
E_{\mathrm{iso}}^2
}. \label{eq:methods-nv3-template-half-splittings}
\end{equation}

Here, $b_j=\gamma_e \left|\mathbf B_{\mathrm{geo}}\cdot\hat{\mathbf
n}_j\right|$ is the contribution from the axial Zeeman shift along the four NV
orientations where $j\in\{1,\ldots,4\}$ indexes the four NV orientations,
$\gamma_e$ is the electron gyromagnetic ratio, $\mathbf B_{\mathrm{geo}}$ is
the geomagnetic field, and $\hat{\mathbf n}_j$ is the unit vector along the
axis of the $j$th NV orientation. These components are computed for the
experimental system and listed in Appendix~\ref{sec:geomagnetic-field}. The
three $^{14}\mathrm{N}$ axial hyperfine shifts for each NV orientation are
given by $m A_\parallel$ for $m\in\{-1,0,+1\}$, where
$A_\parallel\approx2.166\;\mathrm{MHz}$ is the axial $^{14}\mathrm{N}$
hyperfine constant. Unlike the magnetic contributions, the final term,
$E_{\mathrm{iso}}$, captures contributions from transverse crystal strain and
cannot be determined \textit{a priori} from the available measurements.
Instead, the value of $E_{\mathrm{iso}}$ is estimated by finding the value of
$E_{\mathrm{iso}}$ that minimizes the residual between a model spectrum of
Lorentzian functions at $\Delta f_{jm}$ and the experimentally measured ODMR
spectrum (see Appendix~\ref{sec:geomagnetic-field}). The resulting fit between
the ODMR spectrum and model finds a corresponding strain splitting frequency
contribution of $E_{\mathrm{iso}}=0.58~\mathrm{MHz}$ ($\pm 0.31~\mathrm{MHz}$
based on {a} $95\%$ {CI}). The splitting contribution from strain is
comparable to the geomagnetic Zeeman splitting and requires inclusion in the
final model. However, in this context, $E_{\mathrm{iso}}$ should be treated as
a physically bounded estimate informed by the ODMR spectrum rather than a
measured quantity.

 While the ODMR spectrum is used to estimate the magnitude of the effective
transverse {strain}, it is insufficient in the present study to define its
direction in the transverse plane. Because the transverse direction is
unresolved, the model prescribes eight equally spaced transverse coupling
angles, \(\alpha_a=\frac{\pi a}{8}\) for
\(a\in\{0,\ldots,7\}\). These angles parameterize the transverse Hamiltonian
coefficients and are a modeling assumption. Thus, for {seed} branch
\(k=(j,m,a)\), the corresponding mixing {angle} can be evaluated as

\[
\chi_k=
{\arccos\!\left(\frac{E_{\mathrm{iso}}\cos\alpha_a}{{\lvert\Delta
f_{jm}\rvert}}\right),}
\qquad {0\leq\chi_k\leq\pi}.
\]

Varying \(\alpha_a\) changes how the fixed half-splitting (${\lvert\Delta
f_{jm}\rvert}$) is partitioned between the differential shift and the
bright--dark coupling. The 96 seed branches (four NV orientations by three
hyperfine splittings by eight angles) are ordered by ${\lvert\Delta
f_{jm}\rvert}$ and assigned {values of} \(\epsilon_k\) {through} deterministic
{sampling of the cumulative distribution derived from the fitted ODMR line
shape}. {This distribution is a normalized Lorentzian profile centered at the
half-splitting from a separate two-Lorentzian ODMR fit and truncated to its
full-width-at-half-maximum interval.} %

The resulting function serves as an empirical distribution, rather than a
precise physical representation, that parameterizes a static branch-to-branch
variation in $\epsilon_k$. Matching the ODMR line shape does not necessarily
guarantee matching the echo dynamics associated with homogeneous or power
broadening, but the static distribution is meant to provide a representation
typical of our experimental diamond. The full model ensemble is then populated
by pairing each \((\epsilon_{k},\chi_k)\) with a mirror branch at
\(\alpha_a+\pi\) having the same \(\epsilon_k\) and equal weight. {Mirror
branches have complementary angles $\chi_k$ and $\pi-\chi_k$, giving opposite
differential shifts and equal bright--dark couplings.} 
{This gives 192 equally weighted branches and 16 transverse coupling angles
for each orientation--hyperfine combination. %
}

Once the parameters for all of the ensemble branches are specified, the NV
spin-state evolution of each branch is calculated by successively applying
Eq.~\eqref{eq:methods-nv3-step-propagator} using the corresponding branch
Hamiltonian and programmed microwave waveform. Each branch is initialized in
\(\lvert0\rangle\), and its final population in that state is evaluated {for
each of} the four final-pulse phases described in
{section}~\ref{sec:results-temperature-sensitivity-calibration}. The
phase-cycled complex branch responses are then averaged over {all 192 ensemble
branches} with equal weighting before their amplitude and phase are evaluated.

{For comparison, a degenerate-limit calculation is also conducted using a
single branch with \(\epsilon_k=0\) to remove both the upper-state splitting
and bright--dark coupling. The states in this degenerate branch are still
subjected to the same finite-pulse propagation and phase-cycled readout as the
192-branch ensemble.} The ensemble model and unit tests were designed by the
authors and implemented in MATLAB with assistance from an AI coding agent
running on OpenAI Codex (GPT-6.1 Sol and GPT-6 Astra).

\section{Results}\label{sec:results}

\subsection{Spin-echo and thermal-echo
performance}\label{sec:results-protocol-calibration}
We first consider how bright--dark mixing affects the behavior of two
canonical pulse sequences, the spin-echo and thermal-echo protocols, and their
ability to refocus and accumulate phase from common-mode detuning,
respectively. This assessment proceeds analogously to the Rabi measurements by
conducting a parametric sweep of the microwave pulse duration but instead
applies three-pulse sequences rather than a single pulse. In the two cases
evaluated here, the durations of the three pulses are not parameterized
independently, but scaled proportionally such that the first:middle:last
pulse-duration ratios were fixed at 1:2:1 and 1:4:1, corresponding to the
canonical timings of a spin-echo protocol and a thermal-echo protocol in the
degenerate limit, respectively. Each set of experimental measurements is
paired with a set of model calculations conducted at matched parameters, i.e.,
the same timing ratios, pulse-duration scales, and microwave offsets.

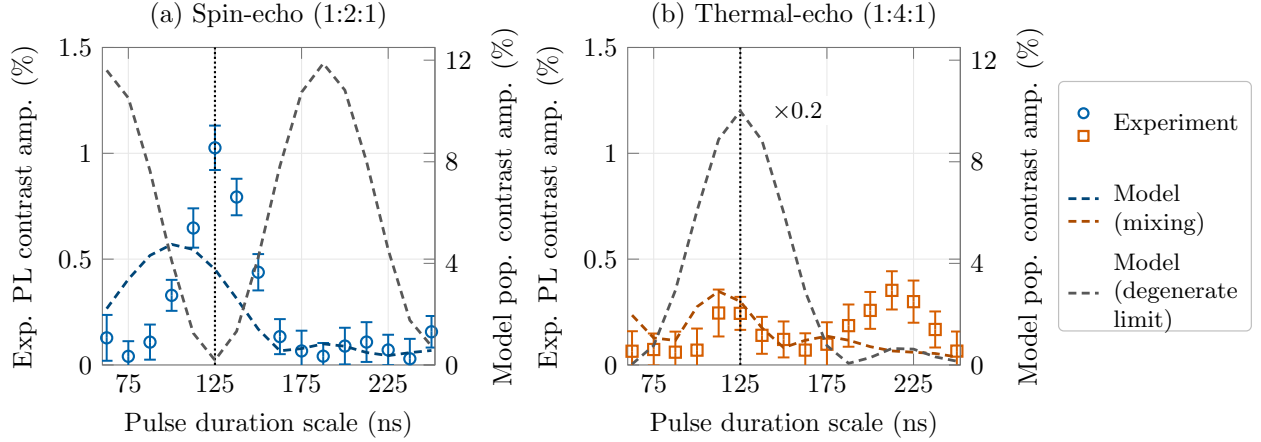
\begin{figure}[!t]
  \centering
  \input{figures/Figure4ab.tex}
  \caption{\textbf{{}Oscillation amplitude{} versus pulse-duration scale for
different protocol timing ratios.}
  (a,b) {}Oscillation amplitude{} versus pulse-duration scale for the
spin-echo (a) and thermal-echo (b) timing ratios. In (a,b), the horizontal
axis gives the duration of the first and last pulses, while the central pulse
is scaled according to the ($1:2:1$) and ($1:4:1$) ratios, respectively.
{}Markers (blue and orange, respectively) show the experimentally measured
oscillation amplitude at the prescribed microwave-offset magnitude and are
referenced to the left axes. Vertical error bars show one standard error of
the fitted experimental amplitude based on a sinusoidal fit at
$0.200~\mathrm{MHz}$ with a fitted offset and linear drift.{} Colored dashed
curves show the corresponding amplitudes predicted by the mixing model, and
dashed gray curves show the predictions of the degenerate-limit model
($\epsilon_k=0$). The scale of the model curves is indicated on the right side
based on $\lvert0\rangle$ population, with the exception of the
degenerate-limit thermal-echo curve in (b){,} which is multiplied by 0.2 for
readability. All model calculations and experiments used a nominal microwave
offset of $-0.200~\mathrm{MHz}$. Dotted black lines indicate the nominal
$125~\mathrm{ns}$ endcap duration inferred from the Rabi calibration.
  }\label{fig:proportional-timing-sweeps}
\end{figure}

This parametric sweep of the pulse duration assesses whether the pulse
durations (i.e., nominal angular rotations) implied by the Rabi oscillations
correspond to the desired protocol behavior and how the operations of these
different protocol families are modified by the more complicated dynamics.
These metrics are evaluated by detuning the microwave carrier frequency for
all measurements by $0.200~\mathrm{MHz}$ below the reference
zero-field-splitting frequency, $D_0$, and measuring the PL before and after
each pulse sequence over a range of pulse-duration scales (up to 250 ns) and
evolution times ($2\tau$) up to $9.75~\mu\mathrm{s}$ in the experiments
($20~\mu\mathrm{s}$ in the model). Coherent oscillations at the detuning
frequency as a function of the evolution time are of specific interest, when
present, as they convey the sensitivity to temperature through changes in
their frequency, $f_{\mathrm{echo}}$. This parametric sweep directly evaluates
whether the canonical timings associated with the spin-echo and thermal-echo
protocols are still capable of refocusing and accumulating phase from
common-mode detuning, respectively, when strong bright--dark mixing is
present. The experimental measurements for the spin-echo and thermal-echo
timing ratios are shown in Fig.~\ref{fig:proportional-timing-sweeps}a,b,
respectively.

For the spin-echo timing (1:2:1; see
Fig.~\ref{fig:proportional-timing-sweeps}a), the experimental measurements
(markers) show {}minimal oscillation amplitude except for a clear peak at 125
ns.{} This timing, as indicated by the dotted line, corresponds to the nominal
Rabi $\frac{\pi}{2}$ time, indicating a nominal
($\frac{\pi}{2}-\tau-\pi-\tau-\frac{\pi}{2}$) pulse sequence. {}{}This peak in
oscillatory amplitude contrasts with the expected behavior of the spin-echo
protocol in the degenerate-limit model (i.e., $\epsilon_k = 0 $), shown in
Fig.~\ref{fig:proportional-timing-sweeps}a by the gray dashed line, where
phase accumulation from common-mode detuning should be refocused at the
calibrated pulse duration and there should be negligible oscillatory
amplitude. However, when the effects of upper-state splitting and coupling are
included in the ensemble model (see dashed blue line in
Fig.~\ref{fig:proportional-timing-sweeps}a), behavior that is qualitatively
consistent with the experimental measurements emerges. Specifically, there is
an increase in the oscillation amplitude at the common-mode detuning frequency
associated with incomplete refocusing; however, the predicted maximum from the
model occurs at a pulse-duration scale approximately \(20\%\) shorter than the
experimental maximum.

The thermal-echo timing (1:4:1; see
Fig.~\ref{fig:proportional-timing-sweeps}b) shows similar deviations from its
canonical behavior. In contrast to the spin-echo timing, which ideally
refocuses the common-mode detuning, the thermal-echo timing is designed to
maximally preserve the common-mode detuning phase accumulation and produce a
large oscillation amplitude at the prescribed microwave-offset magnitude.
{}However, the experimental measurements demonstrate a weak, broad response
near 125 ns, with minimal contrast amplitude. A second weak response occurs
near $212.5~\mathrm{ns}$, where the largest sampled experimental amplitude is
observed.{} In each case, the strength of the common-mode detuning oscillation
is not substantially larger than that of the other spectral content of the
signal, likely making it unsuitable for effective thermometry.
The model calculations in Fig.~\ref{fig:proportional-timing-sweeps}b similarly
{predict} a small oscillation {amplitude}, especially compared to the peak of
the degenerate-limit thermal echo, with its maximum at $112.5~\mathrm{ns}$.
{Even at this maximum, the amplitude of the common-mode-carrying signal
remains exceptionally small, supporting the experimental findings.} This
suggests that finite splitting significantly alters the expected performance
of both protocol families, with the spin-echo timing unexpectedly preserving
phase from the common-mode detuning while the thermal-echo timing fails to do
so effectively under the present conditions. 
These results motivate {further} characterization of the spin-echo (1:2:1)
timing sequence and its functionality as a thermometry protocol. This
temperature-sensitive response will be referred to as a mixing-enabled thermal
echo {(METE)}. %

Finite-pulse detuning effects also occur in effective two-level NV sensing
protocols~\cite{Lang2019}. To assess the contribution of bright--dark mixing,
an additional numerical control calculation was conducted using the model
where the same upper-state splitting and ensemble construction were retained
but the bright--dark mixing terms (i.e., the off-diagonal mixing terms
$\epsilon_k \sin \chi_k$) were instead set to zero and the differential shifts
were set to $\pm\epsilon_k$ for the paired branches. This control removes
mixing during both the pulses and free evolution, so it does not isolate the
contribution from mixing during the pulses. In each case, this control
configuration strongly suppressed oscillations at the common-mode detuning
frequency to the extent that their amplitudes were comparable to or smaller
than the dominant off-carrier spectral components. This result supports the
notion that {bright--dark mixing plays} a role in the {experimentally
observed} temperature-sensitive response in the spin-echo timing.

\subsection{$D(T)$-sensitive
readout}\label{sec:results-temperature-sensitivity-calibration}
As with other temperature-sensitive NV protocols, {the present} {METE}
measures the average frequency difference between the microwave carrier
frequency and the NV ensemble's zero-field-splitting frequency, $D$, which is
temperature dependent. %
Because the microwave carrier frequency is prescribed, measuring the signed
echo frequency provides a precise measure of the zero-field-splitting
frequency and, in effect, of temperature when other parameters that contribute
to shifts in $D$ can be neglected (e.g., strain). While a single echo trace is
sufficient to determine the magnitude of the common-mode detuning, the sign of
the frequency shift is not ascertainable in the absence of additional
information or priors. This ambiguity can be resolved in a number of different
ways, including by acquiring a second echo trace using a different microwave
carrier frequency to constrain the sign of the frequency shift. In practice,
this second-frequency measurement likely requires either acquiring the two
traces asynchronously, modulating the microwave carrier frequency, or
utilizing a second microwave source.

Alternatively, the sign of the frequency shift can be determined by measuring
the quadrature component of the echo contrast alongside the in-phase component
to construct the complex phasor and determine its signed frequency. This
approach has been utilized previously in NV-based protocols to acquire both
signal quadratures \cite{Wang2021Quadrature}. Here, we adapt the {METE}
protocol described in the previous section to obtain quadrature readout and
resolve the sign of the common-mode detuning. This method uses the same
three-pulse sequence timings described in the previous section (see, for
example, the upper-left sequence in
Fig.~\ref{fig:quadrature-phasor-extraction}a), but varies the phase of the
final $\frac{\pi}{2}$ pulse relative to the other two pulses in the sequence.
The in-phase component of the echo waveform is measured using pulses whose
rotations are all about the same axis, as in the sequence designated by the
$+x$ subscript. The quadrature component of the echo contrast can be measured
by applying a 90{-}degree phase offset to the final pulse relative to the
first pulse in the sequence, which changes the rotation axis of the final
pulse to the $y$-axis (see the upper-right sequence in
Fig.~\ref{fig:quadrature-phasor-extraction}a). While this approach can be
implemented in principle with only two readouts (i.e., 0{-} and 90{-}degree
phase offsets), we extend this idea to a four-phase quadrature cycle in which
the {METE} protocol is repeated four times for each free-evolution time
($\tau$), with the final $\frac{\pi}{2}$ pulse offset by $0$, $180$, $90$, and
$270$ degrees relative to the first $\frac{\pi}{2}$ pulse in the sequence,
corresponding to the four sequences in
Fig.~\ref{fig:quadrature-phasor-extraction}a. Here, the additional $180${-}
and $270${-}degree readout phases yield oscillations comparable to those
produced by $0$ and $90$ degrees, but with opposite signs, thereby improving
noise rejection and signal-to-noise ratio.

\begin{figure}[!t]
    \centering
    \makebox[\linewidth][c]{%
      \begin{minipage}[t]{\linewidth}
        \centering
        \vspace{0pt}
        \input{figures/Figure5a.tex}\\[0.4em]
        {\qtaPlotPanelFont (a) Four-phase quadrature cycle}
      \end{minipage}%
    }
    \par\vspace{0.55em}
    \begin{minipage}{0.99\linewidth}
      \centering
      \input{figures/Figure5bcd.tex}
    \end{minipage}
    \caption{\textbf{Mixing-enabled thermal echo protocol and signed-frequency
estimation from quadrature measurements of echo contrast,
$Z=C_I+\mathrm{i}C_Q$.} (a) Four-phase quadrature cycle for the mixing-enabled
thermal echo protocol. At each evolution time, the sequence is acquired with
four projection phases denoted by the $+x$, $-x$, $+y$, and $-y$ subscripts,
with corresponding PL ratios $R_{+x}$, $R_{-x}$, $R_{+y}$, and $R_{-y}$,
respectively. Each ratio is defined as $R=\mathrm{PL}_2/\mathrm{PL}_1$, where
$\mathrm{PL}_1$ and $\mathrm{PL}_2$ are the reference and signal readouts,
respectively. The ratios for opposite readout phases are differenced to
determine the in-phase and quadrature contrasts,
$C_I=\tfrac{1}{2}(R_{+x}-R_{-x})$ and $C_Q=\tfrac{1}{2}(R_{+y}-R_{-y})$,
respectively. The complete four-phase quadrature cycle is repeated for each
total evolution time, $2\tau$, and averaged over $N$ repetitions. (b,c)
In-phase ($C_I$, dark and solid) and quadrature ($C_Q$, light and dashed)
components of the echo-contrast traces. Each trace is vertically offset to
have zero mean. (b) PL contrast acquired at a negative microwave offset,
$f_{\mathrm{MW}}-D_0=-0.37~\mathrm{MHz}$. In this regime, the quadrature
component lags the in-phase component, giving
$f_{\mathrm{echo}}=+{0.355}~\mathrm{MHz}{{}\pm0.006~\mathrm{MHz}}$. (c)
Corresponding measurement acquired at a positive microwave offset,
$f_{\mathrm{MW}}-D_0=+0.35~\mathrm{MHz}$. Here, the quadrature component leads
the in-phase component, giving
$f_{\mathrm{echo}}=-{0.373}~\mathrm{MHz}{{}\pm0.005~\mathrm{MHz}}$ {($95\%$
CIs)}. (d) Normalized squared Fourier magnitude,
$\lvert\mathrm{FFT}(Z)\rvert^2$, for the negative-offset (blue) and
positive-offset (red) measurements, with peaks at $+0.35~\mathrm{MHz}$ and
$-0.37~\mathrm{MHz}$, respectively. The signs of the two peaks show that the
complex contrast, $Z$, resolves the sign of the oscillation frequency, which
is opposite the microwave offset. All displayed curves use a three-point
moving mean for clarity, but reported fitted frequencies are obtained from the
raw measurements.}\label{fig:quadrature-phasor-extraction}
\end{figure}

The resulting four PL ratios given by $R_{+x}$, $R_{-x}$, $R_{+y}$, and
$R_{-y}$, each defined as $R=\mathrm{PL}_2/\mathrm{PL}_1$ for the
corresponding sequence, are combined to form the in-phase
[$C_I=\tfrac{1}{2}(R_{+x}-R_{-x})$] and quadrature
[$C_Q=\tfrac{1}{2}(R_{+y}-R_{-y})$] components of the complex-valued echo
phasor, $Z=C_I+iC_Q$. This complex signal resolves the instantaneous phase of
the oscillations and thus their signed frequency. The sign of the oscillation
is determined by whether the quadrature component leads or lags its in-phase
counterpart. Thus, acquiring both the in-phase and quadrature components of
the echo waveform resolves the sign of the echo-frequency oscillations and,
correspondingly, the sign of the common-mode detuning. 

Examples of the complex echo waveforms in this instance are shown in
Fig.~\ref{fig:quadrature-phasor-extraction}b,c. The respective
microwave-carrier offsets from $D_0=2.8705\;\mathrm{GHz}$ are
$-0.37~\mathrm{MHz}$ and $+0.35~\mathrm{MHz}$. The two panels show how both
components of the waveform oscillate at the same frequency but $90$ degrees
out of phase relative to one another. To estimate the signed echo frequency,
the complex waveform ($Z$) is fit to a decaying phasor model of the form

\begin{equation}
Z_{\mathrm{fit}}(t)
= d_{\mathrm{c}}
+ A \exp\left[-(t/T_{\mathrm{echo}})
  + i\left(2\pi f_{\mathrm{echo}}t+\phi\right)\right],
\qquad t\equiv 2\tau,
\label{eq:decaying-phasor}
\end{equation}

where $d_{\mathrm{c}}$ is a complex-valued offset constant, $A$ is the maximum
oscillation amplitude, $T_{\mathrm{echo}}$ is the envelope decay time,
$f_{\mathrm{echo}}$ is the echo oscillation frequency, and $\phi$ is an
arbitrary phase offset. The fitted echo frequencies for the two cases are
$f_{\mathrm{echo}}=+{0.355}~\mathrm{MHz}{{}\pm0.006~\mathrm{MHz}}$ and
$f_{\mathrm{echo}}=-{0.373}~\mathrm{MHz}{{}\pm0.005~\mathrm{MHz}}$,
respectively {($95\%$ CIs)}, with $T_{\mathrm{echo}}\approx 8~\mu\mathrm{s}$.
In this instance, $T_{\mathrm{echo}}$ describes the decay related to the
{METE} pathway and is distinct from the $T_2$ that describes the decay of the
refocused echo component, which is insensitive to the common-mode detuning.
For the diamond used here, $T_2$ is nearly an order of magnitude longer than
$T_{\mathrm{echo}}$.

In the case of negative microwave offset, where $f_{\mathrm{MW}} < D$, the
quadrature component lags the in-phase component, indicating that the echo
oscillates with a positive frequency
(Fig.~\ref{fig:quadrature-phasor-extraction}b). Conversely, in the case of
positive microwave offset, the quadrature component instead leads the in-phase
component, indicating a negative echo frequency. In addition to resolving the
sign of the oscillation frequency, these measurements confirm that the sign of
the echo frequency is opposite to the sign of the microwave offset. %

These frequency estimates are consistent with the power spectral densities
(PSDs) of the complex waveforms, which show the distribution of power as a
function of both positive and negative frequency components.
In contrast to the PSDs of real-valued functions, which are symmetric about
$f=0$, the PSDs of complex-valued functions need not have this symmetry and
can resolve the sign of the echo frequency.
Figure~\ref{fig:quadrature-phasor-extraction}d shows the PSDs of both complex
echo waveforms, computed from the squared magnitude of their respective
Fourier transforms, $\lvert\mathrm{FFT}(Z)\rvert^2$, and normalized by their
maximum values. The sinusoidal nature of each complex waveform manifests as a
PSD peak at $+0.35~\mathrm{MHz}$ or $-0.37~\mathrm{MHz}$, consistent with the
corresponding fitted echo frequency. Importantly, the signed oscillation
frequency is resolved by the Fourier transform with values and signs
consistent with those obtained by fitting the complex waveform to the decaying
phasor equation. Altogether, these two measurements demonstrate that the
chosen quadrature approach can produce a complex quadrature signal that is
parameterized by the decaying phasor given in Eq.~\eqref{eq:decaying-phasor}
and resolves the signed frequency of oscillation.

Two additional parametric sweeps were conducted to establish that the measured
echo frequency accurately represents the desired quantity
$f_{\mathrm{echo}}=D(T)-f_{\mathrm{MW}}$. In the first test, the sampling
protocol was held constant while the microwave frequency was swept over a
microwave-offset range from nominally $-1.8~\mathrm{MHz}$ to
$+1.8~\mathrm{MHz}$ to verify a one-to-one correlation between the measured
quantity $f_{\mathrm{echo}}$ and the negative of the prescribed microwave
offset, $D_0-f_{\mathrm{MW}}$. In this test, the acquisition settings were
slightly shortened to 100, 200, and 100 ns to {accommodate} a correspondingly
higher microwave power. The results of this sweep are shown in
Fig.~\ref{fig:echo-frequency-temperature-calibration}a. Across the full tested
range, $f_{\mathrm{echo}}$ varied linearly with microwave offset, consistent
with an expected slope of $-1$. A linear fit to the data using
$f_{\mathrm{echo}}=D_0-f_{\mathrm{MW}}+\Delta f$, where $\Delta
f=-15~\mathrm{kHz}$ is a fitted offset parameter corresponding to the average
offset of $D$ from $D_0$ during the test, returns $R^2$ near unity
($R^2=0.9995$). Over the microwave-offset range, the fitted
oscillation-envelope amplitude increases symmetrically (within the fit
uncertainty) with the magnitude of the microwave offset from $1.2\%$ up to
$1.6\%$ at the largest frequency differences. This trend is qualitatively
reproduced by the mixing model, which predicts a symmetric increase in
oscillation amplitude about zero microwave offset. The model prediction is
shown as the solid orange line in
Fig.~\ref{fig:echo-frequency-temperature-calibration}a after rescaling by a
single empirical factor ($A_{\mathrm{mult}}=0.279$) to relate the model
population to the experimental contrast. 

In the second test, the microwave carrier frequency of the pulses was instead
held fixed while the temperature of the surrounding fluid was varied from
$20.4~^{\circ}\mathrm{C}$ to $38.1~^{\circ}\mathrm{C}$.
Figure~\ref{fig:echo-frequency-temperature-calibration}b shows the estimated
frequency shift of $D(T)$ relative to $D_0$, $\Delta f = f_{\mathrm{echo}} -
D_0 + f_{\mathrm{MW}}$, obtained by fitting the complex quadrature signal to
Eq.~\eqref{eq:decaying-phasor}, as a function of temperature. The measured
slope of $\Delta f$ versus temperature is
$-71.41\pm1.24~\mathrm{kHz}/^{\circ}\mathrm{C}$ (95\% CI), compared with the
literature value of $-75.0~\mathrm{kHz}/^{\circ}\mathrm{C}$
\cite{Acosta2010,Acosta2011Erratum}. Taken together, these two tests support a
calibrated relation between the measured echo frequency and the fluid
temperature over the measured range.

\begin{center}
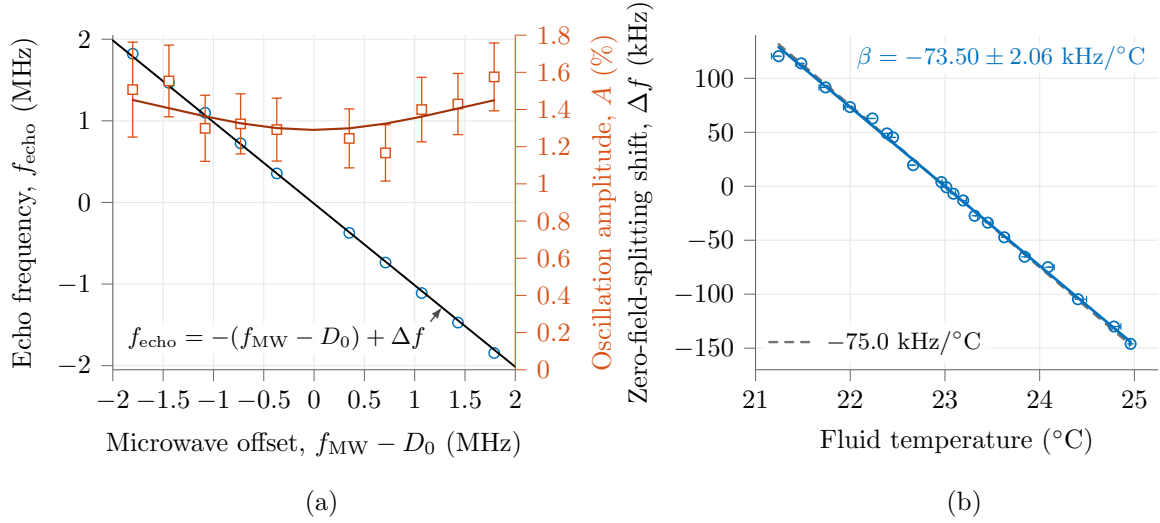

  \begin{minipage}{\linewidth}
    \centering
    \begin{minipage}[t]{0.485\linewidth}
      \centering
      \vspace{0pt}
      \begin{minipage}[c][6.51cm][c]{\linewidth}
        \centering
        \input{figures/Figure6a.tex}
      \end{minipage}\\[0.25em]
      {\qtaPlotPanelFont (a)}
    \end{minipage}\hfill
    \begin{minipage}[t]{0.485\linewidth}
      \centering
      \vspace{0pt}
      \begin{minipage}[c][6.51cm][c]{\linewidth}
        \centering
        \input{figures/Figure6b.tex}
      \end{minipage}\\[0.25em]
      {\qtaPlotPanelFont (b)}
    \end{minipage}
    
    \captionof{figure}{\textbf{Echo frequency as a function of microwave
offset and fluid temperature.} (a) Echo frequency (blue circles),
$f_{\mathrm{echo}}$, versus the frequency offset of the microwave carrier from
the nominal zero-field splitting. $f_{\mathrm{echo}}$ is determined by fitting
$Z$ to the complex phasor model in Eq.~\eqref{eq:decaying-phasor}. The linear
fit (black line) is constrained to the expected slope of $-1$,
$f_{\mathrm{echo}}=-(f_{\mathrm{MW}}-D_0)+\Delta f$, yielding the fitted
frequency offset $\Delta f=-14.78~\mathrm{kHz}$ ($R^2=0.9995$, $n=10$). The
corresponding oscillation amplitudes (orange squares), $A$, from the same
accepted traces are shown with error bars indicating the $95\%$ {CIs} based on
the local covariance of the fit. The amplitude prediction from the ensemble
model (solid dark-orange curve) is shown after rescaling by a single
multiplicative factor ($A_{\mathrm{mult}}=0.279$).  (b) Zero-field-splitting
shift versus fluid temperature (blue circles). These measurements used
rectangular microwave pulses and yielded a temperature coefficient consistent
with that obtained using cosine-squared pulses. A fit to $\Delta f=\beta T+b$
(blue line) yields $\beta=-71.41\pm1.24~\mathrm{kHz}/^{\circ}\mathrm{C}$
($n=25$; 95\% CI). Horizontal bars span the recorded RTD extrema; vertical
bars show local 95\% frequency-fit CIs. A reference line (dashed gray) based
on the literature with slope $-75.0~\mathrm{kHz}/^{\circ}\mathrm{C}$
\cite{Acosta2010,Acosta2011Erratum} is
included.}\label{fig:echo-frequency-temperature-calibration}
  \end{minipage}
\end{center}

\subsection{Multi-pulse sequence}\label{sec:results-multi-pulse}
To further assess the viability of the {METE} approach, it is important to
evaluate whether the empirically determined pulse timings can be used in a
sequence of refocusing pulses to extend coherence (and {potentially improve}
sensitivity). {Vetter et al.}~\cite{Vetter2022} recently derived and
demonstrated a family of low-field dynamical decoupling (LDD) sequences that
are designed to cancel progressively higher orders of pulse errors through the
use of additional pulses. The simplest protocols that provide cooperative
first-order error cancellation form the four-pulse LDD4 family. This family is
similar to the sequences used by {Genov et al.}~\cite{Genov2017} and includes
two specific, four-pulse implementations: LDD4A and LDD4B. Here, we adapt and
implement the palindromic LDD4B phase pattern as a multi-pulse version of the
{{METE}} approach and compare it with the performance of the {{METE}} sequence
demonstrated above. In this case, the modified LDD4B is given by the nominal
pulse sequence:
$({\left(\frac{\pi}{2}\right)_{\!x}-\tau-{}}\pi_x{{}-2\tau-{}}\pi_{-x}{{}%
-2\tau-{}}\pi_{-x}{{}-2\tau-{}}\pi_{x}{{}-\tau-\left(\frac{\pi}{2}\right)_%
{\!x}})$, where the first and last refocusing {pulse centers} are separated by
a delay time $\tau$ from the endcap {pulse centers} and consecutive refocusing
{pulse centers} are separated by $2\tau$ (see the LDD4B sequence diagram in
Fig.~\ref{fig:results-hahn-ldd4b}a). As with the {{METE}} sequence, the
durations of the $\frac{\pi}{2}$ endcap (i.e., first and last) pulses and the
refocusing $\pi$ pulses have been modified with the $\pi$ pulse timings found
in section~{\ref{sec:results-protocol-calibration}} that maximize the
temperature-sensitive echo oscillation. Similarly, the LDD4B sequence is
implemented with four-phase cycling to construct the complex echo phasor and
resolve the signed echo frequency.

Representative complex-phasor traces, including both in-phase and quadrature
components, obtained with the {{METE}} and modified LDD4B protocols at the
same {microwave offset} ($f_{\mathrm{MW}}-D_0=-0.243~\mathrm{MHz}$) are shown
in Fig.~\ref{fig:results-hahn-ldd4b}b{,}c, respectively. Both traces are shown
as functions of their {total} evolution times, corresponding to $2\tau$ and
$8\tau$ for the {{METE}} and LDD4B sequences, respectively. The total
evolution time in the case of the latter was extended to {approximately}
$40~\mu\mathrm{s}$ (compared to {approximately} $25~\mu\mathrm{s}$) to
accommodate the expected extension of the echo-envelope time
($T_{\mathrm{echo}}$) from the multi-pulse sequence. While the oscillations in
the LDD4B variant visually appear to extend to longer times compared with the
{{METE}}, there is a noticeable difference in the dominant tone of each of the
signals. As expected, the oscillation in the {{METE}} case yields a frequency
consistent with the {microwave-offset magnitude} ($f_{\mathrm{echo}}\approx
0.23~\mathrm{MHz}$). However, the modified LDD4B exhibits a signal with an
oscillation frequency that is approximately half the {microwave-offset
magnitude}. This frequency reduction is inconsistent with the {ideal
thermometry response of the LDD4B sequence used by Vetter et
al.}~\cite{Vetter2022} and suggests that {phase from the common-mode detuning}
is not properly accumulated over the duration of the protocol. This behavior,
however, is consistent with the corresponding predictions of the {ensemble}
model for the LDD4B variant {of} the {METE}, in which multiple refocusing
pathways arising from the strong bright--dark mixing can interrupt the
coherent (and constant) accumulation of {phase from the common-mode detuning}
over the entire evolution period. 

To quantify the new relationship between the observed LDD4B echo frequency and
the applied {microwave offset}, a microwave-frequency sweep analogous to that
shown in Fig.~\ref{fig:echo-frequency-temperature-calibration}a was performed
over $-0.44~\mathrm{MHz} \leq f_{\mathrm{MW}}-D_0 \leq 0.36~\mathrm{MHz}$,
using paired {{METE}} and LDD4B measurements. As before, the complex echo
traces for each microwave carrier were fitted to the decaying phasor (see
Eq.~\eqref{eq:decaying-phasor}) as a function of the total evolution time. The
echo frequencies extracted from both protocols are shown in
Fig.~\ref{fig:results-hahn-ldd4b}d. Consistent with
Fig.~\ref{fig:echo-frequency-temperature-calibration}a, the frequencies
extracted from the {{METE}} measurements exhibit the expected linear
dependence on the applied {microwave offset}, with an imposed slope of $-1$.
By contrast, the LDD4B data show a clear reduction in the slope of the echo
frequency versus {microwave offset} to approximately $-1/2$. This shallower
slope suggests that the dominant pathway is effectively accumulating phase
over half of the total evolution time, indicating that the phase associated
with the {common-mode detuning} accumulates over an effective interval of
$4\tau$ rather than the $8\tau$ expected {for that ideal thermometry
sequence}. However, despite the discrepancy in the {microwave-offset
dependence}, the LDD4B measurements still preserve a sensitivity to {phase
from the common-mode detuning}, albeit with functionally reduced sensitivity
compared to the ideal case.

Based on the Figure~\ref{fig:results-hahn-ldd4b} echo fits and acquisition
parameters, we estimate the photon-shot-noise-limited temperature
sensitivities of $254$ and $229~\mathrm{mK}/\sqrt{\mathrm{Hz}}$ for METE and
ME-LDD4B, respectively (see
Appendix~\ref{sec:appendix-photon-shot-noise-sensitivity}). This estimate
includes the reduced temperature coefficient of the ME-LDD4B as indicated by
its measured microwave-offset response. The longer echo-decay time is largely
offset by the reduced frequency response, leaving the projected sensitivities
comparable. Thus, these results do not establish a clear sensitivity advantage
from adding refocusing pulses under the present acquisition conditions.

\begin{figure}[!htbp]
  \centering
  \begin{minipage}{0.90\linewidth}
    {\qtaPlotPanelFont (a) {ME-}LDD4B pulse sequence\par}
    \centering
    \input{figures/Figure7a.tex}\par
  \end{minipage}
  \vspace{0.35em}

  \begin{minipage}{0.90\linewidth}
    \centering
    \input{figures/Figure7bcd.tex}
  \end{minipage}
  \caption{\textbf{{Mixing-enabled thermal} echo and LDD4B quadrature signals
and {microwave-offset} responses.} {ME denotes mixing-enabled.}
  (a) Diagram of the ME-LDD4B protocol {adapted from} {Vetter et
al.}~\cite{Vetter2022}, which utilizes
  four cosine-squared $\pi$ refocusing pulses
  for a total evolution time ${t}=8\tau${, measured between the centers of the
preparation and readout pulses}. {All pulse-rotation labels are nominal.}
  As in the quadrature-readout diagram in
  Fig.~\ref{fig:quadrature-phasor-extraction}, the phase of
  the final readout $\pi/2$ pulse is advanced in increments of $\pi/2$ to
  construct $Z$; only the $+x$ projection is illustrated here.
  (b,c) Complex waveforms, $Z=C_I+\mathrm{i}C_Q$, measured using
{mixing-enabled thermal} echo (b) and
  LDD4B (c) at a {fixed microwave offset}
  $f_{\mathrm{MW}}-D_0\approx-0.24~\mathrm{MHz}$. Solid and dashed curves
  correspond to $C_I$ and $C_Q$, respectively. {For display, a constant
baseline was subtracted from each component, followed by a three-point moving
average.}
  {The annotated $T_{\mathrm{echo}}$ values are fitted envelope decay times.}
  (d) Fitted $f_{\mathrm{echo}}$ from Eq.~\eqref{eq:decaying-phasor} for the
$Z$ traces at nine carrier frequencies per protocol. Error bars are $95\%$
{CIs}. Markers indicate measurements using {mixing-enabled thermal} echo (blue
circles) and LDD4B (red squares), respectively. Lines correspond to the
least-squares fits given by $f_{\mathrm{echo}}={m}(f_{\mathrm{MW}}-D_0)+b$
with slopes fixed to ${m}=-1$ for the {mixing-enabled thermal} echo (blue) and
${m}=-0.5$ for LDD4B (red). Fitted offsets in each case are
${b_{\mathrm{ME}}}=-0.0214~\mathrm{MHz}$ ($R^2=0.99909$) and
${b_{\mathrm{LDD4B}}}=-0.0106~\mathrm{MHz}$ ($R^2=0.99943$), respectively. The
dotted vertical line denotes the {microwave offset} used in (b,c).}
  \label{fig:results-hahn-ldd4b}
\end{figure}

\section{Discussion}\label{sec:discussion}

The performance of canonical spin-echo and thermal-echo protocols changes
significantly when the $\lvert+1\rangle$ and $\lvert-1\rangle$ states are
nearly degenerate and the microwave Rabi frequency is comparable to their
splitting. Under these conditions, the conventional 1:4:1 thermal-echo timing
does not experimentally produce a clearly resolvable oscillation at the
common-mode detuning $\Delta_{\mathrm c}(T)$. The mixed-ensemble calculation
likewise predicts only a small {{oscillation amplitude} at the common-mode
detuning {frequency}}, largely because the bright--dark mixing fails to
preserve phase accumulation along the desired pathway. By contrast, the 1:2:1
spin-echo timing produces a distinct oscillatory signal at the common-mode
detuning. Further, the experimental {}oscillation amplitude{} peaks near the
nominal pulse duration inferred from the Rabi period
(Fig.~\ref{fig:proportional-timing-sweeps}a). This result is initially
surprising as these specific pulse timings in the absence of finite mixing
(i.e., in the degenerate limit) should ideally refocus common-mode detuning
and thus exhibit no oscillatory amplitude. However, in the presence of strong
bright--dark mixing, the model predicts that the common-mode phase is no
longer completely refocused, leading to a finite phase accumulation and the
emergence of a detuning carrier. While the model qualitatively predicts the
failure of the pulses to refocus the common-mode detuning, it fails to
accurately locate the correct pulse duration scale that maximizes the
amplitude of the {common-mode} carrier signal peak, predicting a pulse
duration that is approximately $20\%$ shorter. This discrepancy may reflect an
overestimate of the NV ensemble Rabi frequency using the two-harmonic fitting
which may oversimplify the actual dynamics of the triplet system with strong
leakage into the dark state, leading to an incorrect scaling of the Rabi
frequency. Decreasing the Rabi frequency in the model in proportion to the
pulse duration error reduces the error in the predicted optimal pulse
duration, but this adjustment would need to be reconciled with a more complete
modelling of the underlying dynamics that contribute to Rabi measurements.
These dynamics may not be ascertainable with the current measurement system.
Additional parametric analysis of the ensemble model examined the sensitivity
of the METE carrier amplitude and optimal pulse duration to drive strength,
transverse strain, and mirror-pair weighting (see
Appendix~\ref{sec:appendix-mete-robustness}) and supports the findings that a
METE signal persists throughout the tested parameter variations and that the
predicted optimal pulse duration depends on the assumed drive strength. This
dependence is consistent with uncertainty in the drive calibration as a
possible contributor to the timing discrepancy. Extending the model to other
diamond samples remains an area of ongoing work and requires sample-specific
calibration. Despite this discrepancy, the mixing model importantly predicts
the same sensitivity to phase accumulated from $D(T)$ observed in the
experimental measurements, yielding a 1:1 correlation between the common-mode
detuning and the signed frequency of the complex echo trace. This behavior is
experimentally confirmed through controlled sweeps of the microwave carrier
frequency and fluid temperature at the NV ensemble (see
Fig.~\ref{fig:echo-frequency-temperature-calibration}). 

{Results from the} mixing model {suggest that} this new behavior {arises from}
competing pathways with distinct phase sensitivities. As with the conventional
spin echo, the {METE} signal retains a refocused contribution that is
insensitive to shifts in $D(T)$. Bright--dark mixing opens an additional
pathway that preserves phase accumulated from the common-mode detuning while
refocusing the branch-dependent splitting, thereby producing the observed
carrier. Other mixed pathways that are sensitive to both quantities and would
otherwise confound this $D(T)$-dependent phase accumulation are hypothesized
to dephase rapidly across the inhomogeneously broadened ensemble compared to
these previous two pathways, leaving the more coherent contributions
resolvable. The resulting signal comprises a short-lived {oscillation
sensitive to the common-mode detuning} superimposed on a longer-lived
background spin-echo decay. Though the existence of the original refocusing
pathway competes for NV population, and thus signal, at the expense of the
temperature-sensitive pathway, the temperature-sensitive oscillation can be
resolved against the longer-lived background signal. Interestingly, this
composition resembles the signal shape reported by {Wang et
al.}~\cite{Wang2015Thermometry} for thermal-echo measurements under nominally
degenerate conditions. Whereas that study attributed the additional frequency
content to possible coupling with residual paramagnetic defects, the present
results identify finite bright--dark mixing as a plausible mechanism that
could also produce a temperature-sensitive signal.

The complexity behind this {METE} presents challenges when trying to extend
coherence through using multiple refocusing pulses. For the modified LDD4B
protocol, the measured echo-frequency dependence indicates an effective
common-mode phase-accumulation time of approximately $4\tau$ over the total
evolution time of $8\tau$ (see Fig.~\ref{fig:results-hahn-ldd4b}). The mixing
model reproduces this same echo-frequency dependence stemming from the
competing pathways exhibiting different phase sensitivities. The model further
predicts that a $2\tau$ component becomes increasingly prominent as the
magnitude of the common-mode detuning increases, although the present data are
insufficient to resolve this contribution conclusively. The inability to
accumulate phase from $D(T)$ across the full $8\tau$ interval supports the
notion that mixing redistributes populations among competing pathways. {As
such, this behavior limits the potential sensitivity benefit of the measured
coherence-time extension.}

Applying METE to measurements using mobile nanodiamonds, particularly in
biological applications, remains a longer-term prospect. Interpreting the
signal as a local temperature requires assessing nonthermal frequency shifts,
including those caused by changes in strain and electric fields. The
electric-field contribution can depend on the NV centers' proximity to the
surface and on surface functionalization (see
Plakhotnik~\cite{Plakhotnik2026}).

\section{Conclusion}\label{sec:conclusion}
We examined the behavior of spin-echo and thermal-echo protocols for
near-degenerate NV ensembles when the microwave Rabi frequency was comparable
to the upper-manifold splitting frequency. In this regime, the model suggests
that leakage from the bright state into the dark state is a leading-order
effect and produces substantial departures from idealized protocol behavior.
In instances where this regime is unavoidable, such as where microwave power
may be limited, the {METE} may present an attractive alternative for NV
thermometry.

The mixing model presented here functions as an effective qualitative
diagnostic for interpreting the experimental measurements and parameterizing
the {role of bright--dark coupling and shifts in protocol performance}. The
three most salient predictions from the model are the collapse of the
{canonical 1:4:1} thermal-echo protocol signal, the emergence of a strong
$D(T)$-sensitive pathway using the spin-echo timing, and the reduction in the
effective phase accumulation time when multiple refocusing pulses are used.
All three predictions are supported by the experimental measurements, with
some notable quantitative discrepancies, particularly in the optimal pulse
durations. The pulse-duration discrepancy may reflect uncertainty in the
Rabi-frequency calibration or other limitations of the ensemble model. %

While the present results indicate that naively applying the modified pulse
timings {to} an LDD4B sequence scheme exhibits an {effective} phase
accumulation over a reduced {interval} (i.e., $4\tau$ rather than $8\tau$), it
is possible that varying the intensity or phase of the refocusing pulses using
cooperative optimal control~\cite{Konzelmann2018} may permit phase
accumulation over the full evolution interval.

\section*{Data and code availability}\label{sec:data-code-availability}

\looseness=1\relax
The data that support the findings of this study and the custom code used to
process and analyze them will be made publicly available through CaltechDATA
(\href{https://doi.org/10.22002/mvvy2-a1f31}{DOI~10.22002/mvvy2-a1f31})
upon publication. %

\section*{Acknowledgements}\label{sec:acknowledgements}

\looseness=1\relax
The authors would like to thank Dr. Sean Mendoza and Dr. Eisuke Abe for their
helpful discussions.

\section*{Funding}\label{sec:funding}

This work was supported by the Gordon and Betty Moore Foundation.

\section*{Author contributions}\label{sec:author-contributions}
M.K.F.: Conceptualization, methodology, investigation, formal analysis,
visualization, writing, and project administration. J.O.D.: Conceptualization,
methodology, formal analysis, writing, and project administration.

\section*{Competing interests}\label{sec:competing-interests}

The authors declare no competing interests.

\section*{Use of large language models}\label{sec:llm-use}

 Generative AI coding agents (OpenAI Codex, GPT-5.6 Sol, GPT-6.1 Sol and GPT-6
Astra) assisted with drafting, proofreading, and editing the manuscript text.
The AI coding agents were instrumental in ensuring that algorithms, settings,
and parameters in the experimental and model codes were accurately described
in the Results and Methods. No generative image models were used in drafting
this manuscript or figures; however, the coding agents were employed to encode
user-generated diagrams (e.g.{,} PowerPoint schematics) and MATLAB figures as
native \LaTeX-compilable versions using `pgfplots' and `TikZ'. The authors
{retain} responsibility for the correctness of this content. 
 
 The core data acquisition and processing pipeline codes were written by the
authors. AI coding agents were later integrated into the experimental platform
to automate writing of experimental configuration files and data plotting
scripts. An AI coding agent assisted in creating the ensemble model used for
comparison with the experimental results. The authors prescribed the desired
algorithmic methodology and mathematical foundation, which were converted into
a MATLAB pipeline by the coding agent. The purpose of this model is to provide
mathematical and physical insight into the observed experimental results.

\clearpage
\appendix

\section{ODMR fitting}
\label{sec:odmr-two-lobe-fit}
\def\odmrPhysicalBaseline{1.00160414}
\def\odmrPhysicalSlopePerMHz{0.0000009880}
\def\odmrPhysicalDepth{0.02276332}
\def\odmrPhysicalMeanFrequencyMHz{2870.50000}
\def\odmrPhysicalMinFrequencyMHz{2862.50000}
\def\odmrPhysicalMaxFrequencyMHz{2878.50000}
\def\odmrPhysicalCenterGHz{2.87057791}
\def\odmrPhysicalCenterMHz{2870.57791}
\def\odmrPhysicalCenterDetuningMHz{0.078}
\def\odmrEffectiveStrainMHz{0.58125}
\def\odmrEffectiveStrainCiLowMHz{0.27211}
\def\odmrEffectiveStrainCiHighMHz{0.89039}
\def\odmrPhysicalHwhmMHz{0.76001}
\def\odmrPhysicalFwhmMHz{1.52002}
\def\odmrPhysicalMinBranchMHz{2867.19173}
\def\odmrPhysicalMaxBranchMHz{2873.96410}
\def\odmrPhysicalRmsPercent{0.17764}
\def\odmrPhysicalFitRSquared{0.698402}
\def\odmrPhysicalAicc{-1900.03}
\def\odmrPhysicalBranchCount{24}
\def\odmrPhysicalN{151}
\def\odmrPhysicalK{6}
\def\odmrPhysicalReportedMinFrequencyMHz{2862.5}
\def\odmrPhysicalReportedMaxFrequencyMHz{2878.5}
\def\odmrPhysicalReportedMeanFrequencyMHz{2870.5}
\def\odmrPhysicalReportedBaseline{1.0016}
\def\odmrPhysicalReportedSlopePerMHz{0.000001}
\def\odmrPhysicalReportedDepth{0.0228}
\def\odmrPhysicalReportedCenterMHz{2870.58}
\def\odmrEffectiveStrainReportedMHz{0.58}
\def\odmrEffectiveStrainReportedCiLowMHz{0.27}
\def\odmrEffectiveStrainReportedCiHighMHz{0.89}
\def\odmrPhysicalReportedHwhmMHz{0.76}
\def\odmrPhysicalReportedMinBranchMHz{2867.19}
\def\odmrPhysicalReportedMaxBranchMHz{2873.96}
\def\odmrPhysicalReportedRmsPercent{0.178}
\def\odmrPhysicalReportedFitRSquared{0.698}
\def\odmrPhysicalReportedAicc{-1900.03}
\def\odmrTwoLobeBaseline{1.00216345}
\def\odmrTwoLobeSlopePerMHz{0.0000000000}
\def\odmrTwoLobeLowerAmplitude{0.00685128}
\def\odmrTwoLobeUpperAmplitude{0.00685128}
\def\odmrTwoLobeMeanFrequencyMHz{2870.50000}
\def\odmrTwoLobeCenterMHz{2870.53981}
\def\odmrTwoLobeHalfSplittingMHz{1.77500}
\def\odmrTwoLobeHalfSplittingCiLowMHz{1.60351}
\def\odmrTwoLobeHalfSplittingCiHighMHz{1.94649}
\def\odmrTwoLobeLowerCenterMHz{2868.76481}
\def\odmrTwoLobeUpperCenterMHz{2872.31481}
\def\odmrTwoLobeLowerDetuningMHz{-1.73519}
\def\odmrTwoLobeUpperDetuningMHz{1.81481}
\def\odmrTwoLobeLowerHwhmMHz{1.60040}
\def\odmrTwoLobeUpperHwhmMHz{1.60040}
\def\odmrTwoLobeRmsPercent{0.17809}
\def\odmrTwoLobeFitRSquared{0.696887}
\def\odmrTwoLobeAicc{-1901.44}
\def\odmrTwoLobeN{151}
\def\odmrTwoLobeK{5}
\def\odmrTwoLobeCommonAmplitude{0.00685128}
\def\odmrTwoLobeCommonHwhmMHz{1.60040}
\def\odmrTwoLobeReportedMinFrequencyMHz{2862.5}
\def\odmrTwoLobeReportedMaxFrequencyMHz{2878.5}
\def\odmrTwoLobeReportedMeanFrequencyMHz{2870.5}
\def\odmrTwoLobeReportedBaseline{1.0022}
\def\odmrTwoLobeReportedSlopePerMHz{0.000000}
\def\odmrTwoLobeReportedLowerAmplitude{0.0069}
\def\odmrTwoLobeReportedUpperAmplitude{0.0069}
\def\odmrTwoLobeReportedCenterMHz{2870.54}
\def\odmrTwoLobeReportedHalfSplittingMHz{1.78}
\def\odmrTwoLobeReportedHalfSplittingCiLowMHz{1.60}
\def\odmrTwoLobeReportedHalfSplittingCiHighMHz{1.95}
\def\odmrTwoLobeReportedLowerCenterMHz{2868.76}
\def\odmrTwoLobeReportedUpperCenterMHz{2872.31}
\def\odmrTwoLobeReportedLowerHwhmMHz{1.60}
\def\odmrTwoLobeReportedUpperHwhmMHz{1.60}
\def\odmrTwoLobeReportedRmsPercent{0.178}
\def\odmrTwoLobeReportedFitRSquared{0.697}
\def\odmrTwoLobeReportedAicc{-1901.44}
\def\odmrTwoLobeReportedCommonAmplitude{0.0069}
\def\odmrTwoLobeReportedCommonHwhmMHz{1.60}
\def\odmrMovingMeanWindow{5}
\def\odmrCenterGHz{2.87054}
\def\odmrCenterMHz{2870.54}
\def\odmrCenterCiLowMHz{2870.34}
\def\odmrCenterCiHighMHz{2870.74}
\def\odmrLowerResonanceMHz{2868.76}
\def\odmrUpperResonanceMHz{2872.31}
\def\odmrDoubletSeparationMHz{3.55}
\def\odmrFitRSquared{0.697}

\begin{table}[htbp]
\centering
\setlength{\belowcaptionskip}{3pt}
\footnotesize
\caption{{\textbf{Phenomenological two-lobe ODMR fit.} The phenomenological
two-lobe fit prescribes the curve-fit shown in
Fig.~\ref{fig:rabi-calibration}a. Its fitted amplitude and linewidth describe
the broad ODMR features without assigning individual NV orientations or
hyperfine transitions. Thus, the empirically fitted half-splitting
$E_{2\mathrm{L}}$ characterizes the aggregate spectral structure rather than
specific sources such as transverse strain.}}
\label{tab:odmr-two-lobe-fit-parameters}
\begin{tabular}{@{}>{\raggedright\arraybackslash}p{0.21\linewidth}>%
{\raggedright\arraybackslash}p{0.19\linewidth}>{\raggedright\arraybackslash}%
p{0.53\linewidth}@{}}
\hline
\noalign{\vskip 3pt}
{Quantity} & Symbol & Value \\
\noalign{\vskip 3pt}
\hline
\noalign{\vskip 3pt}
\multicolumn{3}{@{}>{\raggedright\arraybackslash}p{0.96\linewidth}@{}}{%
\begin{minipage}{\linewidth}
\textbf{Two-lobe fit equation.}
\begin{equation}
\begin{aligned}
{R_{\mathrm{fit}}}({f_{\mathrm{MW}}})
  &=
{c_0}{{}-a\left[L_\Gamma(f_{\mathrm{MW}}-f_-)+L_\Gamma(f_{\mathrm{MW}}-f_%
+)\right]},\\
L_{{\Gamma}}(x)&=\left[1+(x/{\Gamma})^2\right]^{-1},
\qquad f_\pm=D_{2\mathrm{L}}\pm E_{2\mathrm{L}}.
\end{aligned}
\label{eq:appendix-odmr-two-lobe-fit}
\end{equation}
{Here $\bar f_{\mathrm{MW}}$ is the mean scan frequency, and $c_0$ is a
constant baseline. The two lobes share depth $a$ and half-width at
half-maximum $\Gamma$; their centers $f_\pm$ have midpoint $D_{2\mathrm{L}}$
and separation $2E_{2\mathrm{L}}$.}
\end{minipage}} \\
\noalign{\vskip 3pt}
\hline
\noalign{\vskip 3pt}
{Data and baseline} & {$({f_{\mathrm{MW}}},{\bar f_{\mathrm{MW}}},{c_0})$} &
{${f_{\mathrm{MW}}}=\odmrTwoLobeReportedMinFrequencyMHz$-%
-$\odmrTwoLobeReportedMaxFrequencyMHz~\mathrm{MHz}$; ${\bar
f_{\mathrm{MW}}}=\odmrTwoLobeReportedMeanFrequencyMHz~\mathrm{MHz}$;
${c_0}=\odmrTwoLobeReportedBaseline$} \\
{Lobe parameters} & {$({a,\Gamma})$} &
{${a=\odmrTwoLobeReportedCommonAmplitude}$;
${\Gamma=\odmrTwoLobeReportedCommonHwhmMHz}~\mathrm{MHz}$} \\
{Center and splitting} & {$(D_{2\mathrm{L}},E_{2\mathrm{L}})$} &
{$D_{2\mathrm{L}}={\odmrTwoLobeReportedCenterMHz}~\mathrm{MHz}$;
$E_{2\mathrm{L}}={\odmrTwoLobeReportedHalfSplittingMHz}~\mathrm{MHz}$ (95\%
{CI}
${\odmrTwoLobeReportedHalfSplittingCiLowMHz}$--$%
{\odmrTwoLobeReportedHalfSplittingCiHighMHz}~\mathrm{MHz}$)} \\
{Fit quality} & {RMS residual, $R^2$} & {${\odmrTwoLobeReportedRmsPercent}\%$,
${\odmrTwoLobeReportedFitRSquared}$; $n=\odmrTwoLobeN$, ${\odmrTwoLobeK}$
{fitted parameters}} \\
\noalign{\vskip 3pt}
\hline
\end{tabular}
\end{table}
\FloatBarrier
\section{Geomagnetic field conditions}
\label{sec:geomagnetic-field}

The geomagnetic projections in Table~\ref{tab:geomagnetic-nv-splitting}
provide the fixed axial Zeeman contributions to the constrained ODMR fit in
Table~\ref{tab:odmr-physical-fit-parameters}. 

\begin{table}[htbp]
\centering
\setlength{\belowcaptionskip}{3pt}
\caption{{\textbf{Modeled geomagnetic splitting.} Nominal values conditional
on the WMMHR field and cardinal-aligned mounting geometry.}}
\label{tab:geomagnetic-nv-splitting}
\begin{tabular}{lc}
\hline
\noalign{\vskip 3pt}
NV orientation & $2\gamma_e\lvert B_{\parallel}\rvert$ (MHz) \\
\noalign{\vskip 3pt}
\hline
\noalign{\vskip 3pt}
North & {2.34} \\
South & {0.21} \\
East & {1.49} \\
West & {1.06} \\
\noalign{\vskip 3pt}
\hline
\end{tabular}
\end{table}
\FloatBarrier

The magnetic field vector at the measurement site was estimated with the World
Magnetic Model High Resolution 2025 (WMMHR-2025) \cite{WMMHR2025} at
$34.13667^{\circ}\mathrm{N}$, $118.12361^{\circ}\mathrm{W}$ and an elevation
of $263~\mathrm{m}$ above mean sea level. In north--east--down coordinates,
the modeled field was $(B_N,B_E,B_D)=(23.34,4.66,{39.36})~\mu\mathrm{T}$,
corresponding to a total magnitude
$\lvert\mathbf{B}_{\mathrm{geo}}\rvert={46.00}~\mu\mathrm{T}$. {The WMMHR
total-intensity uncertainty is $0.13~\mu\mathrm{T}$ (one standard deviation);
this model uncertainty does not constrain apparatus-scale magnetic
perturbations at the sensor.} %

The nominal mounting geometry used in the model assumed that the $(100)$
diamond surface was horizontal and its in-plane diagonals were
cardinal-aligned. Table~\ref{tab:geomagnetic-nv-splitting} gives the nominal
Zeeman splitting magnitude in MHz for each NV orientation ($\Delta
f_B=2\gamma_e\lvert B_{\parallel}\rvert$ where
$\gamma_e=28.0~\mathrm{GHz\,T^{-1}}$) {for this cardinal-aligned geometry}.
{The fit fixes the nominal geometry; it does not estimate the diamond azimuth
or local-field perturbations.}

The four NV orientations and three $^{14}\mathrm{N}$ spin projections define
12 orientation--hyperfine branches, each with two transitions, giving 24
equally weighted Lorentzian components. Fitting this constrained spectrum
estimates the effective transverse strain coupling parameter
$E_{\mathrm{iso}}$. This parameter, in conjunction with the coupling angles,
is used {to} determine the branch mixing angles $\chi_k$ in the ensemble model
of section~\ref{sec:methods-three-level-ensemble-model}.

\begingroup
\begin{table}[!t]
\centering
\setlength{\belowcaptionskip}{3pt}
\footnotesize
\caption{{\textbf{24-component constrained ODMR fit for strain estimation{.}}
The 24-component equal-weight model was fitted to
$R=\mathrm{PL}_2/\mathrm{PL}_1$ in Fig.~\ref{fig:rabi-calibration}a; values
are rounded, and $E_{\mathrm{iso}}$ is model-conditioned.}}
\label{tab:odmr-physical-fit-parameters}
\begin{tabular}{@{}>{\raggedright\arraybackslash}p{0.21\linewidth}>%
{\raggedright\arraybackslash}p{0.19\linewidth}>{\raggedright\arraybackslash}%
p{0.53\linewidth}@{}}
\hline
\noalign{\vskip 3pt}
{Quantity} & Symbol & Value \\
\noalign{\vskip 3pt}
\hline
\noalign{\vskip 3pt}
\multicolumn{3}{@{}>{\raggedright\arraybackslash}p{0.96\linewidth}@{}}{%
\begin{minipage}{\linewidth}
\textbf{ODMR fit equation.}
\begin{equation}
\begin{aligned}
{R_{\mathrm{fit}}}({f_{\mathrm{MW}}})
  &= c_0+c_1({f_{\mathrm{MW}}}-{\bar f_{\mathrm{MW}}})
     -\frac{d}{{24}}\sum_{j=1}^{4}\sum_{{m}=-1}^{1}\sum_{\sigma=\pm1}
       L_\Gamma\!\left({f_{\mathrm{MW}}}-{D_{\mathrm{fit}}-\Delta
f_{jm}^{\sigma}}\right),\\
L_\Gamma(x)
  &= \left[1+(x/\Gamma)^2\right]^{-1},\\
{\Delta f_{jm}^{\sigma}}
  &= \sigma\left\{E_{\mathrm{iso}}^2
     +\left[{b_j}
     +{m}A_\parallel\right]^2\right\}^{1/2}{,}\\
{b_j}
  &{=
\gamma_e\left\lvert\mathbf{B}_{\mathrm{geo}}\!\cdot\!\hat{\mathbf{n}}%
_j\right\rvert.}
\end{aligned}
\label{eq:appendix-odmr-physical-fit}
\end{equation}
{The superscript $\sigma=\pm1$ labels the two signs of $\Delta f_{jm}$ in
Eq.~\eqref{eq:methods-nv3-template-half-splittings}; $D_{\mathrm{fit}}$ is the
fitted estimate of $D(T)$, and $\Gamma$ is the Lorentzian component half-width
at half-maximum.}
\end{minipage}} \\
\noalign{\vskip 3pt}
\hline
\noalign{\vskip 3pt}
{Data and baseline} & {$({f_{\mathrm{MW}}},{\bar f_{\mathrm{MW}}},$\newline
$c_0,c_1,d)$} &
{${f_{\mathrm{MW}}}=\odmrPhysicalReportedMinFrequencyMHz$-%
-$\odmrPhysicalReportedMaxFrequencyMHz~\mathrm{MHz}$; ${\bar
f_{\mathrm{MW}}}=\odmrPhysicalReportedMeanFrequencyMHz~\mathrm{MHz}$;
$c_0=\odmrPhysicalReportedBaseline$;
$c_1=\odmrPhysicalReportedSlopePerMHz~\mathrm{MHz}^{-1}$;
$d=\odmrPhysicalReportedDepth$} \\
{Fixed inputs} & {$(\mathbf B_{\mathrm{geo}},\hat{\mathbf
n}_j,\gamma_e,A_\parallel)$} & {$\mathbf
B_{\mathrm{geo}}=(23.34,4.66,39.36)~\mu\mathrm{T}$; $\hat{\mathbf
n}_{1,2}=(\pm\sqrt{2/3},0,1/\sqrt3)$, $\hat{\mathbf
n}_{3,4}=(0,\pm\sqrt{2/3},1/\sqrt3)$ (north, south, east, west);
$\gamma_e=0.028~\mathrm{MHz}\,\mu\mathrm{T}^{-1}$;
$A_\parallel=2.166~\mathrm{MHz}$} \\
{Fitted spectrum} & {$(D_{\mathrm{fit}},E_{\mathrm{iso}},\Gamma)$} &
{$D_{\mathrm{fit}}=\odmrPhysicalReportedCenterMHz~\mathrm{MHz}$;
$E_{\mathrm{iso}}=\odmrEffectiveStrainReportedMHz~\mathrm{MHz}$ (conditional
95\% {CI}
$\odmrEffectiveStrainReportedCiLowMHz$-%
-$\odmrEffectiveStrainReportedCiHighMHz~\mathrm{MHz}$);
$\Gamma=\odmrPhysicalReportedHwhmMHz~\mathrm{MHz}$} \\
{Fit quality} & {RMS residual, $R^2$} & {$\odmrPhysicalReportedRmsPercent\%$,
$\odmrPhysicalReportedFitRSquared$; $n=\odmrPhysicalN$, $\odmrPhysicalK$
{fitted parameters}} \\
\noalign{\vskip 3pt}
\hline
\end{tabular}
\end{table}
\endgroup
\FloatBarrier

The 192-branch ensemble resamples empirical branch half-splittings
$\epsilon_k$ sampled from a truncated Lorentzian distribution nominally
matched to the two-lobe fit in {Eq.~(}\ref{eq:appendix-odmr-two-lobe-fit}{)}.
$E_{\mathrm{iso}}$ enters via the mixing angle of each branch rather than
through the distribution of $\epsilon_k$. These inputs define the model
ensemble composition used for modeling the spin-echo, thermal-echo, and
modified LDD4B calculations.

\FloatBarrier
\clearpage

\section{Photon-shot-noise-limited temperature sensitivity}
\label{sec:appendix-photon-shot-noise-sensitivity}

We project photon-shot-noise sensitivity for calibrated local phase
readout~\cite{Toyli2013,Wang2015Thermometry,Barry2020}:
$\eta_T=\sigma_T\sqrt{t_{\mathrm{acq}}}$, where $\sigma_T$ is the temperature
standard deviation and $t_{\mathrm{acq}}=4Nt_{\mathrm{phase}}$ is the
acquisition time at one sensing interval
for $N$ repetitions per phase.
Independent Poisson counts give $\operatorname{var}(R)\simeq\bar R(1+\bar
R)/(N\overline{\mathrm{PL}}_1)$.
Opposite-phase differencing halves this variance in the balanced,
small-contrast limit, giving
\begin{equation}
    \eta_{T,4}(t)
    \simeq\sqrt{\frac{2\bar R(1+\bar R)}{\overline{\mathrm{PL}}_1}}\,
    \frac{\sqrt{t_{\mathrm{phase}}}\exp(t/T_{\mathrm{echo}})}
    {2\pi\lvert\beta\rvert tA}.
    \label{eq:appendix-four-phase-shot-noise-sensitivity}
\end{equation}
Here $\bar R=\overline{\mathrm{PL}}_2/\overline{\mathrm{PL}}_1$ is the
signal/reference
photon-yield ratio; $A$ and $T_{\mathrm{echo}}$ follow
Eq.~\eqref{eq:decaying-phasor}.
The phase duration $t_{\mathrm{phase}}$ includes optical and control time;
$t=2\tau$ for METE and $8\tau$ for ME-LDD4B.
We use $\beta=df_{\mathrm{echo}}/dT$ from the METE calibration
(Fig.~\ref{fig:echo-frequency-temperature-calibration}b) and assume half this
response
for ME-LDD4B, following its microwave-offset response
(Fig.~\ref{fig:results-hahn-ldd4b}d).
For fixed $t_{\mathrm{phase}}$, the optimum $t=T_{\mathrm{echo}}$ lies within
the sampled intervals (Table~\ref{tab:appendix-photon-shot-noise-parameters}).

\begin{table}[!htbp]
    \centering
    \small
    \setlength{\tabcolsep}{4pt}
    \renewcommand{\arraystretch}{1.1}
    \caption{\textbf{Inputs to
Eq.~\eqref{eq:appendix-four-phase-shot-noise-sensitivity} from
Fig.~\ref{fig:results-hahn-ldd4b}b,c.}}
    \label{tab:appendix-photon-shot-noise-parameters}
    \begin{tabular}{@{}lcccc@{}}
        \hline
        Description & Symbol & Unit & METE & ME-LDD4B \\
        \hline
        Reference photon yield & $\overline{\mathrm{PL}}_1$ & photons/readout
& 11.81 & 16.50 \\
        Signal/reference ratio & $\bar R$ & -- & 0.9283 & 0.9298 \\
        Oscillation amplitude & $A$ & \% & 2.03 & 2.00 \\
        Envelope decay time & $T_{\mathrm{echo}}$ & $\mu\mathrm{s}$ & 6.49 &
14.54 \\
        Temperature coefficient & $\beta$ & $\mathrm{kHz\,K^{-1}}$ &
$-71.41\pm1.24$ & $-35.70\pm0.62$ \\
        Recorded phase duration & $t_{\mathrm{phase}}$ & $\mu\mathrm{s}$ & 100
& 140 \\
        Sampled sensing interval & $t$ & $\mu\mathrm{s}$ & 1--24.75 & 2--39.8
\\
        \hline\hline
        Projected sensitivity & $\eta_{T,4}$ & $\mathrm{mK\,Hz^{-1/2}}$ & 254
& 229 \\
        \hline
    \end{tabular}
\end{table}

The projected sensitivities are $254$ and $229~\mathrm{mK\,Hz^{-1/2}}$ for
METE and ME-LDD4B, respectively.

For pulsed ODMR, the spectral-slope relation~\cite{Dreau2011,Barry2020}
with the same ratio noise and one reference/signal pair per cycle gives
\begin{equation}
    \eta_{T,\mathrm{ODMR}}(f_{\mathrm{MW}})
    \simeq\sqrt{\frac{R_{\mathrm{fit}}(1+R_{\mathrm{fit}})}{\overline%
{\mathrm{PL}}_1}}\,
    \frac{\sqrt{t_{\mathrm{cyc}}}}
    {\lvert\beta_D\rvert\,\lvert\partial R_{\mathrm{fit}}/\partial
f_{\mathrm{MW}}\rvert}.
    \label{eq:appendix-odmr-shot-noise-sensitivity}
\end{equation}
Here $R_{\mathrm{fit}}$ follows Eq.~\eqref{eq:appendix-odmr-two-lobe-fit},
and $t_{\mathrm{cyc}}$ includes initialization, microwave control, both
readouts and padding.
We assume both lobes shift equally with temperature, with fixed shape and
$\beta_D=dD_{2\mathrm{L}}/dT$.
Table~\ref{tab:appendix-odmr-sensitivity-parameters} gives the optimum
fitted-flank
projection; a full-spectrum measurement must include time at every frequency.

\begin{table}[!htbp]
    \centering
    \small
    \setlength{\tabcolsep}{4pt}
    \renewcommand{\arraystretch}{1.1}
    \caption{\textbf{Inputs for pulsed ODMR in
Fig.~\ref{fig:rabi-calibration}a.}
    The ratio and slope
    are evaluated at the frequency setting corresponding to the lowest
sensitivity (e.g., steepest slope).
    The assumed $\beta_D$ uses the METE calibration in
    Fig.~\ref{fig:echo-frequency-temperature-calibration}b (95\% interval).}
    \label{tab:appendix-odmr-sensitivity-parameters}
    \begin{tabular}{@{}lccc@{}}
        \hline
        Description & Symbol & Unit & Pulsed ODMR \\
        \hline
        Reference photon yield & $\overline{\mathrm{PL}}_1$ & photons/readout
& 25.00 \\
        Fitted signal/reference ratio & $R_{\mathrm{fit}}$ & -- & 0.9961 \\
        Slope magnitude & $\lvert\partial R_{\mathrm{fit}}/\partial
f_{\mathrm{MW}}\rvert$ & $\mathrm{MHz}^{-1}$ & $3.092\times10^{-3}$ \\
        Assumed temperature coefficient & $\beta_D$ & $\mathrm{kHz\,K^{-1}}$ &
$-71.41\pm1.24$ \\
        Reference/signal cycle & $t_{\mathrm{cyc}}$ & $\mu\mathrm{s}$ &
169.375 \\
        \hline\hline
        Projected sensitivity & $\eta_{T,\mathrm{ODMR}}$ &
$\mathrm{K\,Hz^{-1/2}}$ & 16.6 \\
        \hline
    \end{tabular}
\end{table}

\FloatBarrier
The ODMR acquisition was not optimized for thermometry.
Its shallow lobes and low readout duty limit the projection; optimizing
microwave
excitation and readout requires balancing contrast, linewidth, photon yield and
cycle duration~\cite{Dreau2011,Barry2020}.

\FloatBarrier
\clearpage
\section{Sensitivity of METE amplitude and pulse timing}
\label{sec:appendix-mete-robustness}

We conducted parametric sweeps using the empirical ensemble model described in
Sec.~\ref{sec:methods-three-level-ensemble-model} for drive strengths
$\Omega_{\mathrm{pk}}=\{1,2,3,4,5,6\}~\mathrm{MHz}$, transverse strain values
$E_{\mathrm{iso}}=\{0,0.272,0.581,0.891,1.2\}~\mathrm{MHz}$, and mirror-pair
population fractions $p_\chi=\{0,0.25,0.5,0.75,1\}$. In these calculations,
varying $E_{\mathrm{iso}}$ changes the mixing angles while holding the
empirical splittings fixed. The three interior $E_{\mathrm{iso}}$ values span
its conditional $95\%$ fit interval. Here, $p_\chi$ is the fraction assigned
to the larger-$\chi$ member of each nominal mirror pair, with the total
population of each pair held fixed. 

These sweeps were conducted using the 1:2:1 sequence at $\Delta_{\mathrm
c}=0.2~\mathrm{MHz}$ neglecting any additional sources of dephasing. The
$-\Delta_{\mathrm c}$ carrier amplitude $A(t_p)$ was extracted from the
complex signal over $0$--$20~\mu\mathrm{s}$ using fixed phasors at
$\pm\Delta_{\mathrm c}$. Figure~\ref{fig:appendix-mete-robustness} shows the
effect of varying each parameter separately.

\begin{figure}[!htbp]
    \centering
    \input{figures/Figure8.tex}
    \caption{\textbf{METE amplitude and peak timing under parameter
variations.}
    Panels (a), (c), and (d) vary one parameter, with the others held at their
nominal values:
    $\Omega_{\mathrm{pk}}=4~\mathrm{MHz}$,
    $E_{\mathrm{iso}}=0.581~\mathrm{MHz}$, and $p_\chi=0.5$.
    Amplitudes are normalized by
    $A_0=A_{\mathrm{nom}}(125~\mathrm{ns})$.
    Dashed vertical lines mark the experimental endcap duration.
    Within each panel, darker curves indicate larger values of the varied
parameter.
    (b) Endcap duration at the largest sampled amplitude of each drive-strength
    curve in (a), plotted against $1/\Omega_{\mathrm{pk}}$ with matching
colors.}
    \label{fig:appendix-mete-robustness}
\end{figure}

The results from the parametric sweep reveal that a METE signal survives
across the tested parameter ranges, although the optimal pulse duration can
change substantially, particularly with variations in drive strength (see
Figure~\ref{fig:appendix-mete-robustness}a,b). Decreasing the drive strength
not only increases the optimal pulse duration, but can also reduce the peak
amplitude. These results are consistent with uncertainty in the modelled drive
strength contributing to the discrepancy between the nominal model optimum of
100 ns and the experimental optimum of 125 ns.
Figure~\ref{fig:appendix-mete-robustness}c,d reveals that the METE signal is
relatively insensitive to variations in $p_\chi$ and $E_{\mathrm{iso}}$, with
neither substantially affecting the location of the peak pulse duration.
Increasing $E_{\mathrm{iso}}$ appears to modestly reduce the peak amplitude,
but the effect is small compared to the amplitude reduction observed when the
drive strength is decreased.

Figure~\ref{fig:appendix-power-endcap}a,b shows additional experimental
measurements using rectangular 1:2:1 pulses analogous to the parametric sweep
in Figure~\ref{fig:appendix-mete-robustness}a,b. The sampled peak in the
noise-corrected carrier amplitude shifts from $106.25~\mathrm{ns}$ at
$f_R=2.84~\mathrm{MHz}$ to $162.5~\mathrm{ns}$ at $f_R=1.67~\mathrm{MHz}$. At
$f_R=1.11~\mathrm{MHz}$, the largest sampled amplitude occurs at
$231.25~\mathrm{ns}$, although several pulse durations give comparable
amplitudes and the optimum is not uniquely resolved. Similar to the model
prediction, as the microwave power and Rabi drive strength decrease, the
location of the peak amplitude response shifts to longer pulse durations and
the amplitude of the METE response weakens. 

\begin{figure}[!htbp]
    \centering
    \input{figures/Figure9.tex}
    \caption{\textbf{Experimental METE amplitude and peak timing.}
    (a) Darker curves indicate higher fitted Rabi frequency. 
    Each point gives the noise-corrected positive-frequency carrier amplitude
    extracted from the complex contrast over $2\tau=1$--$8~\mu\mathrm{s}$.
    The noise floor is estimated as the sum of the real and imaginary
    coefficient variances from the fit residuals. This contribution is
subtracted
    in power before conversion to amplitude; negative corrected powers are set
to zero.
    Error bars show the original one-standard-error amplitude intervals after
    the same correction, conditional on the carrier frequency used in the fit,
$f_c$, and the estimated noise floor.
    Lines connect measured points. All curves use the same contrast scale.
    (b) The endcap duration at the largest sampled amplitude in (a), plotted
    against the fitted fundamental Rabi period $T_R=1/f_R$.
    Colors match (a); peak locations are taken from the $6.25~\mathrm{ns}$
sampling grid.}
    \label{fig:appendix-power-endcap}
\end{figure}

\end{document}

%% file: figures/Figures.tex
\providecommand{\qtaPlotFont}{%
  \normalfont\fontfamily{cmr}\fontsize{10}{12}\selectfont}
\providecommand{\qtaPlotSmallFont}{%
  \normalfont\fontfamily{cmr}\fontsize{9}{10.8}\selectfont}
\providecommand{\qtaPlotPanelFont}{%
  \normalfont\fontfamily{cmr}\fontsize{10}{12}\selectfont}

\pgfplotsset{
  qta plot/.style={
    tick label style={font=\qtaPlotFont},
    label style={font=\qtaPlotFont},
    title style={font=\qtaPlotFont},
    legend style={font=\qtaPlotSmallFont}
  },
  qta full-width export/.style={
    font=\qtaPlotSmallFont,
    tick label style={font=\qtaPlotSmallFont},
    label style={font=\qtaPlotSmallFont},
    title style={font=\qtaPlotSmallFont},
    legend style={font=\qtaPlotSmallFont}
  },
  every axis/.append style={qta plot}
}

\tikzset{
  qta plot annotation/.style={font=\qtaPlotSmallFont},
  qta plot panel label/.style={font=\qtaPlotPanelFont}
}

%% file: figures/Figure3.tex
\begingroup

\def\odmrXMin{-8}
\def\odmrXMax{8}
\def\odmrYMin{0.990}
\def\odmrYMax{1.006}
\def\odmrXTicks{-8,-4,0,4,8}
\def\odmrYTicks{0.990,0.994,0.998,1.002,1.006}
\def\odmrLegendColumns{2}
\def\odmrFitColor{black}

\input{figures/Figure3ab.tex}
\endgroup

%% file: figures/Figure3ab.tex
\providecommand{\odmrReferenceGHz}{2.8705}
\providecommand{\odmrXMin}{-15}
\providecommand{\odmrXMax}{15}
\providecommand{\odmrYMin}{1.072}
\providecommand{\odmrYMax}{1.094}
\providecommand{\odmrXTicks}{-12,-6,0,6,12}
\providecommand{\odmrYTicks}{1.072,1.076,1.080,1.084,1.088,1.092}
\providecommand{\odmrLegendColumns}{2}
\providecommand{\odmrFitColor}{black}

\input{figures/data/Figure3a_physical_fit.tex}
\input{figures/data/Figure3a_two_lobe_fit.tex}
\input{figures/data/Figure3b.tex}
\definecolor{rabiorange}{RGB}{197,90,17}
\definecolor{cosblue}{RGB}{0,100,171}
\noindent
\begin{minipage}[t]{0.49\linewidth}
  \centering
\makebox[\linewidth][c]{%
\begin{tikzpicture}
  \begin{axis}[
    width=0.98\linewidth,
    height=4.45cm,
    xmin=\odmrXMin,
    xmax=\odmrXMax,
    ymin=\odmrYMin,
    ymax=\odmrYMax,
    xtick={\odmrXTicks},
    ytick={\odmrYTicks},
    xlabel={Microwave offset, $f_{\mathrm{MW}}-D_0$ (MHz)},
    ylabel={$\mathrm{PL}_2/\mathrm{PL}_1$},
    title={ODMR},
    title style={font=\qtaPlotFont,yshift=-1pt},
    axis line style={black!55},
    tick style={black!55},
    xticklabel style={
      font=\qtaPlotFont,
      /pgf/number format/fixed,
      /pgf/number format/precision=0
    },
    yticklabel style={
      font=\qtaPlotFont,
      /pgf/number format/fixed,
      /pgf/number format/fixed zerofill,
      /pgf/number format/precision=3
    },
    label style={font=\qtaPlotFont},
    grid=major,
    major grid style={black!9},
    scaled ticks=false,
    legend columns=\odmrLegendColumns,
    legend style={
      font=\qtaPlotSmallFont,
      draw=none,
      fill=white,
      fill opacity=0.85,
      text opacity=1,
      at={(0.02,0.015)},
      anchor=south west,
      column sep=5pt,
      row sep=-1pt,
      inner sep=1pt,
      overlay
    }
  ]
    \addplot[
      only marks,
      mark=o,
      mark size=0.95pt,
      draw=black!65,
      fill=white,
      line width=0.45pt,
      opacity=0.78
    ] table[
      x expr={1000*(\thisrow{frequency_GHz}-\odmrReferenceGHz)},
      y=pl_ratio
    ]
      {figures/data/Figure3a.dat};

    \addplot[\odmrFitColor,solid,line width=1.05pt,
      domain=\odmrXMin:\odmrXMax,samples=401]
      {\odmrTwoLobeBaseline
       +\odmrTwoLobeSlopePerMHz
        *(x+1000*\odmrReferenceGHz-\odmrTwoLobeMeanFrequencyMHz)
       -\odmrTwoLobeLowerAmplitude
        /(1+((x-\odmrTwoLobeLowerDetuningMHz)
             /\odmrTwoLobeLowerHwhmMHz)^2)
       -\odmrTwoLobeUpperAmplitude
        /(1+((x-\odmrTwoLobeUpperDetuningMHz)
             /\odmrTwoLobeUpperHwhmMHz)^2)};

  \end{axis}
\end{tikzpicture}}
  \par\vspace{0.10em}
  {\qtaPlotPanelFont (a)\par}
\end{minipage}\hfill%
\begin{minipage}[t]{0.49\linewidth}
  \centering
\makebox[\linewidth][c]{%
\begin{tikzpicture}

  \begin{axis}[
    width=0.98\linewidth,
    height=4.45cm,
    xmin=0,
    xmax=1.0,
    ymin=0.80,
    ymax=1.00,
    xtick={0,0.25,0.5,0.75,1.0},
    ytick={0.80,0.85,0.90,0.95,1.00},
    yticklabels={0.80,0.85,0.90,0.95,1.00},
    xlabel={Pulse duration ($\mu\mathrm{s}$)},
    ylabel={$\mathrm{PL}_2/\mathrm{PL}_1$},
    title={Rabi oscillation},
    title style={font=\qtaPlotFont,yshift=-1pt},
    axis line style={black!55},
    tick style={black!55},
    tick label style={font=\qtaPlotFont},
    label style={font=\qtaPlotFont},
    grid=major,
    major grid style={black!9},
    scaled ticks=false,
    legend columns=1,
    legend style={
      font=\qtaPlotSmallFont,
      draw=none,
      fill=white,
      fill opacity=0.90,
      text opacity=1,
      at={(0.98,0.98)},
      anchor=north east,
      row sep=-1pt,
      inner sep=2pt
    }
  ]
    \addplot[
      only marks,
      mark=o,
      mark size=0.85pt,
      draw=black!65,
      fill=white,
      line width=0.45pt,
      opacity=0.72
    ] table[x=time_us,y=pl_ratio]
      {figures/data/Figure3b.dat};
    \addlegendentry{Data}

    \addplot[black,line width=1.15pt]
      table[x=time_us,y=fit_ratio]
      {figures/data/Figure3b.dat};
    \addlegendentry{Fit}

    \addplot[rabiorange,densely dashed,line width=0.85pt,forget plot]
      coordinates {(\rabiPiOverTwoUs,0.80) (\rabiPiOverTwoUs,1.00)};
    \addplot[rabiorange,densely dashed,line width=0.85pt,forget plot]
      coordinates {(\rabiPiUs,0.80) (\rabiPiUs,1.00)};
    \node[qta plot annotation,text=rabiorange,anchor=north west,
      fill=white,fill opacity=0.88,text opacity=1,inner sep=1pt]
      at (axis cs:\rabiPiOverTwoUs,0.995) {$\pi/2$};
    \node[qta plot annotation,text=rabiorange,anchor=north west,
      fill=white,fill opacity=0.88,text opacity=1,inner sep=1pt]
      at (axis cs:\rabiPiUs,0.945) {$\pi$};
    \node[qta plot annotation,anchor=south east,
      fill=white,fill opacity=0.88,text opacity=1,inner sep=1.5pt]
      at (axis cs:0.97,0.805) {$f_R=\rabiFitMHz~\mathrm{MHz}$};
  \end{axis}
\end{tikzpicture}}
  \par\vspace{0.10em}
  {\qtaPlotPanelFont (b)\par}
\end{minipage}

%% file: figures/Figure4ab.tex
\begingroup
\definecolor{timingblue}{RGB}{0,100,171}
\definecolor{timingorange}{RGB}{214,94,0}
\definecolor{timingcontrol}{RGB}{88,88,88}
\pgfplotsset{
  timing sweep base/.style={
    scale only axis, width=4.4cm, height=4.2cm,
    xmin=60, xmax=252, scaled ticks=false,
    axis line style={black!55}, tick style={black!55},
    tick label style={font=\qtaPlotFont},
    label style={font=\qtaPlotFont},
    tick align=inside
  },
  timing sweep experiment/.style={
    timing sweep base, ymin=0, ymax=1.5,
    xtick={75,125,175,225}, ytick={0,0.5,1.0,1.5},
    grid=major, major grid style={black!9}
  },
  timing sweep model/.style={
    timing sweep base, ymin=0, ymax=12.5,
    xtick=\empty, ytick={0,4,8,12},
    axis x line=none, axis y line*=right
  }
}
{}{}
\begin{tikzpicture}
  \path[fill=none,draw=black!25,line width=0.45pt,rounded corners=2pt]
    (12.65,0.45) rectangle (15.20,3.75);
  \draw[timingblue,line width=0.85pt] (12.98,3.33) circle[radius=0.065];
  \draw[timingorange,line width=0.85pt] (12.915,2.965) rectangle
(13.045,3.095);
  \node[anchor=west,font=\qtaPlotSmallFont,text width=1.65cm,align=left]
    at (13.24,3.18) {Experiment};
  \draw[timingblue!70!black,densely dashed,line width=1.05pt]
(12.78,2.20)--(13.18,2.20);
  \draw[timingorange!80!black,densely dashed,line width=1.05pt]
(12.78,1.90)--(13.18,1.90);
  \node[anchor=west,font=\qtaPlotSmallFont,text width=1.65cm,align=left]
    at (13.24,2.05) {{Model (mixing)}};
  \draw[timingcontrol,densely dashed,line width=1.05pt]
(12.78,0.95)--(13.18,0.95);
  \node[anchor=west,font=\qtaPlotSmallFont,text width=1.65cm,align=left]
    at (13.24,0.95) {{Model}\\{(degenerate} limit{)}};

  \node[font=\qtaPlotFont] at (2.20,4.63) {(a) Spin-echo (1:2:1)};
  \node[font=\qtaPlotFont] at (9.15,4.63) {(b) Thermal-echo (1:4:1)};

  \begin{axis}[
    timing sweep experiment, at={(0,0)}, anchor=south west,
    ylabel={{}Exp. PL contrast amp. (\%){}}
  ]
{}    \addplot+[
      only marks, mark=o, mark size=2.1pt, color=timingblue,
      mark options={draw=timingblue,fill=white}, line width=0.8pt,
      error bars/.cd, y dir=both, y explicit,
      error bar style={draw=timingblue,line width=0.75pt},
      error mark=-,
      error mark options={draw=timingblue,rotate=90,mark size=2.2pt,line
width=0.75pt}
    ] table[
      x=endcap_ns, y expr=50*\thisrow{amplitude},
      y error expr=50*\thisrow{amplitude_se},
      restrict expr to domain={\thisrow{ratio}}{2:2}, col sep=comma
    ] {figures/data/Figure4ab_experiment.csv};
    \draw[black,densely dotted,line width=0.8pt]
      (axis cs:125,0)--(axis cs:125,1.5);
{}  \end{axis}
  \begin{axis}[
    timing sweep model, at={(0,0)}, anchor=south west,
    ylabel={Model pop. contrast amp. (\%)},
  ]
    \addplot[timingblue!70!black,densely dashed,line width=1.05pt,no marks]
      table[x=duration_ns,y expr=100*\thisrow{authoritative_hahn},col
sep=comma]
      {figures/data/Figure4ab_model.csv};
    \addplot[timingcontrol,densely dashed,line width=1.05pt,no marks]
      table[x=duration_ns,y expr=100*\thisrow{textbook_hahn},col sep=comma]
      {figures/data/Figure4ab_model.csv};
  \end{axis}

  \begin{axis}[
    timing sweep experiment, at={(6.95cm,0)}, anchor=south west,
    ylabel={{}Exp. PL contrast amp. (\%){}},
  ]
{}    \addplot+[
      only marks, mark=square, mark size=2.1pt, color=timingorange,
      mark options={draw=timingorange,fill=white}, line width=0.8pt,
      error bars/.cd, y dir=both, y explicit,
      error bar style={draw=timingorange,line width=0.75pt},
      error mark=-,
      error mark options={draw=timingorange,rotate=90,mark size=2.2pt,line
width=0.75pt}
    ] table[
      x=endcap_ns, y expr=50*\thisrow{amplitude},
      y error expr=50*\thisrow{amplitude_se},
      restrict expr to domain={\thisrow{ratio}}{4:4}, col sep=comma
    ] {figures/data/Figure4ab_experiment.csv};
    \draw[black,densely dotted,line width=0.8pt]
      (axis cs:125,0)--(axis cs:125,1.5);
{}  \end{axis}
  \begin{axis}[
    timing sweep model, at={(6.95cm,0)}, anchor=south west,
    ylabel={Model pop. contrast amp. (\%)}
  ]
    \addplot[timingorange!80!black,densely dashed,line width=1.05pt,no marks]
      table[x=duration_ns,y expr=100*\thisrow{authoritative_thermal},col
sep=comma]
      {figures/data/Figure4ab_model.csv};
    \addplot[timingcontrol,densely dashed,line width=1.05pt,no marks]
      table[x=duration_ns,y expr=20*\thisrow{textbook_thermal},col sep=comma]
      {figures/data/Figure4ab_model.csv};
    \node[anchor=west,font=\qtaPlotSmallFont,fill=white,inner sep=2pt]
      at (axis cs:140,10.0) {$\times 0.2$};
  \end{axis}
  \node[font=\qtaPlotFont] at (2.20,-0.78) {Pulse duration scale (ns)};
  \node[font=\qtaPlotFont] at (9.15,-0.78) {Pulse duration scale (ns)};
\end{tikzpicture}
\endgroup

%% file: figures/Figure5a.tex
\begin{tikzpicture}[
  x=0.515cm,
  y=1.05cm,
  line cap=round,
  line join=round,
  font=\normalfont\footnotesize
]
  \definecolor{protocolgreen}{RGB}{181,215,167}
  \definecolor{traceblue}{RGB}{77,173,219}
  \definecolor{tracedarkblue}{RGB}{0,79,158}
  \definecolor{traceorange}{RGB}{219,117,44}
  \definecolor{tracedarkorange}{RGB}{170,78,16}
  \definecolor{pulseblue}{RGB}{139,202,232}
  \definecolor{pulseorange}{RGB}{244,187,130}
  \definecolor{pulsebluefaint}{RGB}{246,251,254}
  \definecolor{pulseorangefaint}{RGB}{255,250,244}
  \definecolor{softblue}{RGB}{238,248,252}
  \definecolor{softorange}{RGB}{254,243,232}
  \def\protocolfont{\normalfont\footnotesize}
  \def\pulsefont{\normalfont\footnotesize}
  \def\headerfont{\normalfont\footnotesize}
  \def\endcapfont{\normalfont\scriptsize}
  \def\contrastfont{\normalfont\footnotesize}
  \def\phasefont{\normalfont\footnotesize}
  \path[use as bounding box] (-0.35,0.13) rectangle (31.45,5.10);

  \newcommand{\sequenceheaders}[1]{%
    \node[font=\headerfont\bfseries,text=black!65] at (#1+3.58,4.82)
      {initialize};
    \node[font=\headerfont\bfseries,text=black!65] at (#1+7.40,4.82)
      {refocus};
    \node[font=\headerfont\bfseries,text=black!65] at (#1+10.30,4.82)
      {projection};
    \node[font=\headerfont\bfseries,text=black!65] at (#1+13.30,4.82)
      {readout};
  }
  \sequenceheaders{0}
  \sequenceheaders{15.00}

  \newcommand{\cossquaredpulse}[6]{%
    \path[fill=#5,draw=#6,line width=0.45pt]
      plot[domain=-1:1,samples=41,smooth]
        ({#1+#2*\x},{#4+#3*(cos(90*\x))^2})
      -- ({#1+#2},#4)
      -- ({#1-#2},#4)
      -- cycle;
  }

  \newcommand{\phaseecho}[8]{%
    \begin{scope}[shift={(#2,0)}]
      \def\rowbottom{-0.78}%
      \def\rowtop{1.02}%
      \ifdim #8pt<0pt
        \def\analyzerlabeloffset{0.62}%
      \else
        \def\analyzerlabeloffset{-0.22}%
      \fi
      \fill[#7] (0.80,{#1+\rowbottom}) rectangle (14.60,{#1+\rowtop});
      \draw[#6!28,line width=0.40pt,rounded corners=2pt]
        (0.80,{#1+\rowbottom}) rectangle (14.60,{#1+\rowtop});

      \node[font=\contrastfont\bfseries,text=#6,anchor=west,inner sep=0pt]
        at (1.00,#1+0.33) {$#4$};

      \draw[black!70,line width=0.55pt] (0.90,#1) -- (14.45,#1);

      \fill[protocolgreen] (2.30,#1) rectangle (4.85,#1+0.82);
      \fill[red!88] (4.07,#1) rectangle (4.85,#1+0.55);
      \draw[green!40!black,line width=0.30pt]
        (2.30,#1) rectangle (4.85,#1+0.82);
      \node[font=\pulsefont] at (3.19,#1+0.57) {Laser};
      \node[font=\endcapfont,text=black] at (4.46,#1-0.22)
{PL\textsubscript{1}};

      \cossquaredpulse{5.50}{0.55}{0.58}{#1}
        {pulseblue}{tracedarkblue}
      \cossquaredpulse{7.90}{1.00}{0.58}{#1}
        {pulseblue}{tracedarkblue}
      \node[font=\endcapfont,text=tracedarkblue]
        at (5.50,#1-0.22)
        {$\left(\frac{\pi}{2}\right)$\rlap{\ensuremath{{}_{\!x}}}};
      \node[font=\pulsefont,text=tracedarkblue]
        at (7.90,#1-0.22) {$\pi_x$};

      \draw[black!45,line
width=0.38pt,{Latex[length=1.25mm]}-{Latex[length=1.25mm]}]
        (6.12,#1+0.29) -- (7.90,#1+0.29);
      \draw[black!45,line
width=0.38pt,{Latex[length=1.25mm]}-{Latex[length=1.25mm]}]
        (7.90,#1+0.29) -- (9.68,#1+0.29);
      \node[font=\pulsefont,fill=#7,inner sep=0.6pt]
        at (7.01,#1+0.29) {$\tau$};
      \node[font=\pulsefont,fill=#7,inner sep=0.6pt]
        at (8.79,#1+0.29) {$\tau$};

      \cossquaredpulse{10.30}{0.55}{#8}{#1}{#5}{#6}
      \node[font=\endcapfont\bfseries,text=#6]
        at (10.30,{#1+\analyzerlabeloffset})
        {$\left(\frac{\pi}{2}\right)_{\!#3}$};

      \fill[protocolgreen] (11.65,#1) rectangle (14.20,#1+0.82);
      \shade[left color=black!42,right color=red!88]
        (11.65,#1) rectangle (12.43,#1+0.55);
      \draw[green!40!black,line width=0.30pt]
        (11.65,#1) rectangle (14.20,#1+0.82);
      \node[font=\pulsefont] at (13.32,#1+0.57) {Laser};
      \node[font=\endcapfont,text=black] at (12.04,#1-0.22)
{PL\textsubscript{2}};
    \end{scope}
  }

  \phaseecho{3.50}{0}{+x}{R_{+x}}
    {pulseblue}{tracedarkblue}{softblue}{0.58}
  \phaseecho{3.50}{15.00}{+y}{R_{+y}}
    {pulseorange}{tracedarkorange}{softorange}{0.58}
  \phaseecho{1.55}{0}{-x}{R_{-x}}
    {pulseblue}{tracedarkblue}{softblue}{-0.58}
  \phaseecho{1.55}{15.00}{-y}{R_{-y}}
    {pulseorange}{tracedarkorange}{softorange}{-0.58}

  \draw[line width=1.15pt]
    (0.22,4.62) .. controls (0.00,4.62) and (-0.06,4.42) .. (-0.06,4.07)
    -- (-0.06,1.20)
    .. controls (-0.06,0.85) and (0.00,0.65) .. (0.22,0.65);
  \draw[line width=1.15pt]
    (29.82,4.62) .. controls (30.04,4.62) and (30.10,4.42) .. (30.10,4.07)
    -- (30.10,1.20)
    .. controls (30.10,0.85) and (30.04,0.65) .. (29.82,0.65);
  \node[font=\protocolfont,anchor=west] at (30.29,4.62) {$N$};

  \draw[line width=0.65pt,-{Latex[length=2.1mm,width=1.5mm]}]
    (0.92,0.48) -- (14.45,0.48);
  \node[fill=white,inner xsep=4pt,font=\pulsefont] at (7.69,0.48) {Time};
  \draw[line width=0.65pt,-{Latex[length=2.1mm,width=1.5mm]}]
    (15.92,0.48) -- (29.45,0.48);
  \node[fill=white,inner xsep=4pt,font=\pulsefont] at (22.69,0.48) {Time};
\end{tikzpicture}

%% file: figures/Figure5bcd.tex
\pgfmathsetlengthmacro{\scPhasorCanvasHeight}{0.3108108108*\linewidth}
\pgfmathsetlengthmacro{\scPhasorAxisBottom}{0.315*\scPhasorCanvasHeight}
\pgfmathsetlengthmacro{\scPhasorAxisHeight}{0.480*\scPhasorCanvasHeight}
\pgfmathsetlengthmacro{\scPhasorAxisWidth}{0.235*\linewidth}
\pgfmathsetlengthmacro{\scPhasorLeftOne}{0.100*\linewidth}
\pgfmathsetlengthmacro{\scPhasorLeftTwo}{0.405*\linewidth}
\pgfmathsetlengthmacro{\scPhasorLeftThree}{0.745*\linewidth}
\pgfmathsetlengthmacro{\scPhasorPanelCenterOne}{0.2175*\linewidth}
\pgfmathsetlengthmacro{\scPhasorPanelCenterTwo}{0.5225*\linewidth}
\pgfmathsetlengthmacro{\scPhasorPanelCenterThree}{0.8625*\linewidth}
\pgfmathsetlengthmacro{\scPhasorPanelLabelY}{0.0475*\scPhasorCanvasHeight}
\begin{tikzpicture}
  \path[use as bounding box] (0,0) rectangle (\linewidth,\scPhasorCanvasHeight);
  \begin{pgfinterruptboundingbox}
  \definecolor{scphasordarkblue}{RGB}{0,79,158}
  \definecolor{scphasorlightblue}{RGB}{77,173,219}
  \definecolor{scphasordarkred}{RGB}{176,18,28}
  \definecolor{scphasorlightred}{RGB}{230,99,110}

  \begin{axis}[
    at={(\scPhasorLeftOne,\scPhasorAxisBottom)},
    anchor=south west,
    scale only axis,
    width=\scPhasorAxisWidth,
    height=\scPhasorAxisHeight,
    xmin=0.8,
    xmax=10,
    ymin=-15,
    ymax=15,
    xtick={0,2,4,6,8,10},
    ytick={-15,-10,-5,0,5,10,15},
    xlabel={Total evolution time, $2\tau$ ($\mu\mathrm{s}$)},
    ylabel={Contrast ($\times 10^{-3}$)},
    title={$f_{\mathrm{echo}}=+0.36\,\mathrm{MHz}$},
    tick label style={font=\qtaPlotFont},
    label style={font=\qtaPlotFont},
    title style={font=\qtaPlotFont},
    axis lines*=left,
    axis line style={black!60},
    tick style={black!60},
    tick align=outside,
    scaled ticks=false,
    every axis plot/.append style={line cap=round},
    legend style={
      draw=none,
      fill=none,
      font=\qtaPlotSmallFont,
      at={(0.98,0.04)},
      anchor=south east,
      inner sep=0pt,
      row sep=0pt
    }
  ]
    \addplot[scphasordarkblue,line width=1.10pt]
      table[x=time_us,y=contrast_i_milli]
      {figures/data/Figure5b.dat};
    \addlegendentry{$C_I$}
    \addplot[scphasorlightblue,line width=1.10pt,dashed]
      table[x=time_us,y=contrast_q_milli]
      {figures/data/Figure5b.dat};
    \addlegendentry{$C_Q$}
    \addplot[black!58,dotted,line width=0.48pt,forget plot]
      coordinates {(0.8,0) (10,0)};
  \end{axis}

  \begin{axis}[
    at={(\scPhasorLeftTwo,\scPhasorAxisBottom)},
    anchor=south west,
    scale only axis,
    width=\scPhasorAxisWidth,
    height=\scPhasorAxisHeight,
    xmin=0.8,
    xmax=10,
    ymin=-15,
    ymax=15,
    xtick={0,2,4,6,8,10},
    ytick={-15,-10,-5,0,5,10,15},
    xlabel={Total evolution time, $2\tau$ ($\mu\mathrm{s}$)},
    title={$f_{\mathrm{echo}}=-0.37\,\mathrm{MHz}$},
    tick label style={font=\qtaPlotFont},
    label style={font=\qtaPlotFont},
    title style={font=\qtaPlotFont},
    axis lines*=left,
    axis line style={black!60},
    tick style={black!60},
    tick align=outside,
    scaled ticks=false,
    every axis plot/.append style={line cap=round},
    legend style={
      draw=none,
      fill=none,
      font=\qtaPlotSmallFont,
      at={(0.98,0.04)},
      anchor=south east,
      inner sep=0pt,
      row sep=0pt
    }
  ]
    \addplot[scphasordarkred,line width=1.10pt]
      table[x=time_us,y=contrast_i_milli]
      {figures/data/Figure5c.dat};
    \addlegendentry{$C_I$}
    \addplot[scphasorlightred,line width=1.10pt,dashed]
      table[x=time_us,y=contrast_q_milli]
      {figures/data/Figure5c.dat};
    \addlegendentry{$C_Q$}
    \addplot[black!58,dotted,line width=0.48pt,forget plot]
      coordinates {(0.8,0) (10,0)};
  \end{axis}

  \begin{axis}[
    at={(\scPhasorLeftThree,\scPhasorAxisBottom)},
    anchor=south west,
    scale only axis,
    width=\scPhasorAxisWidth,
    height=\scPhasorAxisHeight,
    xmin=-0.8,
    xmax=0.8,
    ymin=0,
    ymax=1.02,
    xtick={-0.8,-0.4,0,0.4,0.8},
    ytick={0,0.25,0.5,0.75,1},
    xlabel={Frequency (MHz)},
    ylabel={Normalized $|\mathrm{FFT}(Z)|^2$},
    tick label style={font=\qtaPlotFont},
    label style={font=\qtaPlotFont},
    axis lines*=left,
    axis line style={black!60},
    tick style={black!60},
    tick align=outside,
    scaled ticks=false,
    every axis plot/.append style={line cap=round},
    legend style={
      draw=none,
      fill=none,
      font=\qtaPlotSmallFont,
      at={(0.5,1.04)},
      anchor=south,
      inner sep=0pt,
      row sep=0pt
    }
  ]
    \addplot[scphasordarkblue,line width=1.10pt]
      table[x=frequency_mhz,y=normalized_power]
      {figures/data/Figure5d_negative_offset.dat};
    \addlegendentry{$f_{\mathrm{echo}}=+0.36\,\mathrm{MHz}$}
    \addplot[scphasordarkred,line width=1.10pt]
      table[x=frequency_mhz,y=normalized_power]
      {figures/data/Figure5d_positive_offset.dat};
    \addlegendentry{$f_{\mathrm{echo}}=-0.37\,\mathrm{MHz}$}
    \addplot[black!58,dotted,line width=0.48pt,forget plot]
      coordinates {(0,0) (0,1.02)};
  \end{axis}

  \node[font=\qtaPlotPanelFont] at
    (\scPhasorPanelCenterOne,\scPhasorPanelLabelY) {(b)};
  \node[font=\qtaPlotPanelFont] at
    (\scPhasorPanelCenterTwo,\scPhasorPanelLabelY) {(c)};
  \node[font=\qtaPlotPanelFont] at
    (\scPhasorPanelCenterThree,\scPhasorPanelLabelY) {(d)};
  \end{pgfinterruptboundingbox}
\end{tikzpicture}

%% file: figures/Figure6a.tex
\pgfmathsetlengthmacro{\figSixACanvasHeight}{6.51cm}
\pgfmathsetlengthmacro{\figSixAAxisLeft}{0.155*\linewidth}
\pgfmathsetlengthmacro{\figSixAAxisBottom}{0.1875*\figSixACanvasHeight}
\pgfmathsetlengthmacro{\figSixAAxisWidth}{0.665*\linewidth}
\pgfmathsetlengthmacro{\figSixAAxisHeight}{0.680*\figSixACanvasHeight}
\begin{tikzpicture}
  \path[use as bounding box] (0,0) rectangle (\linewidth,\figSixACanvasHeight);
  \begin{pgfinterruptboundingbox}
  \definecolor{figsixblue}{RGB}{0,114,189}
  \definecolor{figsixorange}{RGB}{217,83,25}
  \definecolor{figsixdarkorange}{RGB}{162,51,9}
  \begin{axis}[
    name=figsixaleft,
    at={(\figSixAAxisLeft,\figSixAAxisBottom)},
    anchor=south west,
    scale only axis,
    width=\figSixAAxisWidth,
    height=\figSixAAxisHeight,
    xmin=-2,
    xmax=2,
    ymin=-2.05,
    ymax=2.05,
    xlabel={Microwave offset, $f_{\mathrm{MW}}-D_0$ (MHz)},
    ylabel={Echo frequency, $f_{\mathrm{echo}}$ (MHz)},
    tick label style={font=\qtaPlotFont},
    label style={font=\qtaPlotFont},
    axis lines*=left,
    axis line style={black!60},
    tick style={black!60},
    tick align=outside,
    grid=major,
    major grid style={black!9},
    scaled ticks=false,
    xtick={-2,-1.5,-1,-0.5,0,0.5,1,1.5,2},
    ytick={-2,-1,0,1,2},
    every axis plot/.append style={line cap=round},
    clip mode=individual
  ]
    \addplot[
      only marks,
      mark=o,
      mark size=1.95pt,
      draw=figsixblue,
      fill=white,
      line width=0.55pt
    ] table[x=x,y=y]
      {figures/data/Figure6a_frequency.dat};

    \addplot[black,line width=0.74pt,domain=-2:2,samples=2]
      {-x - 0.0147776722132};
    \coordinate (figSixAFitStart) at (axis cs:-2,1.9852223277868);
    \coordinate (figSixAFitEnd) at (axis cs:2,-2.0147776722132);

    \node[
      anchor=south west,
      font=\qtaPlotSmallFont,
      fill=white,
      fill opacity=0.92,
      text opacity=1,
      inner sep=1.5pt
    ] (figSixAFitEquation) at (rel axis cs:0.025,0.045)
      {$f_{\mathrm{echo}}=-(f_{\mathrm{MW}}-D_0)+\Delta f$};
    \draw[
      black!70,
      line width=0.55pt,
      -{Latex[length=1.5mm,width=1.1mm]}
    ] (figSixAFitEquation.north east) --
      ($(figSixAFitStart)!(figSixAFitEquation.north east)!(figSixAFitEnd)$);
  \end{axis}

  \begin{axis}[
    at={(\figSixAAxisLeft,\figSixAAxisBottom)},
    anchor=south west,
    scale only axis,
    width=\figSixAAxisWidth,
    height=\figSixAAxisHeight,
    xmin=-2,
    xmax=2,
    ymin=0,
    ymax=1.8,
    axis x line=none,
    axis y line*=right,
    ylabel={Oscillation amplitude, $A$ (\%)},
    tick label style={font=\qtaPlotFont,text=figsixorange},
    label style={font=\qtaPlotFont,text=figsixorange},
    axis line style={figsixorange},
    tick style={figsixorange},
    tick align=outside,
    scaled ticks=false,
    ytick={0,0.2,0.4,0.6,0.8,1.0,1.2,1.4,1.6,1.8},
    every axis plot/.append style={line cap=round},
    clip mode=individual
  ]
    \addplot[
      only marks,
      mark=square*,
      mark size=1.77pt,
      draw=figsixorange,
      fill=white,
      line width=0.55pt,
      error bars/.cd,
      y dir=both,
      y explicit,
      error bar style={draw=figsixorange,line width=0.50pt},
      error mark=-,
      error mark options={draw=figsixorange,rotate=90,mark size=1.75pt,line
width=0.50pt}
    ] table[
      x=x,
      y=y,
      y error minus=errminus,
      y error plus=errplus
    ] {figures/data/Figure6a_amplitude.dat};

    \addplot[figsixdarkorange,line width=0.80pt]
      table[x=x,y=y]
      {figures/data/Figure6a_amplitude_model.dat};
  \end{axis}
  \end{pgfinterruptboundingbox}
\end{tikzpicture}

%% file: figures/Figure6b.tex
\pgfmathsetlengthmacro{\figSixCanvasHeight}{6.51cm}
\pgfmathsetlengthmacro{\figSixAxisLeft}{0.220*\linewidth}
\pgfmathsetlengthmacro{\figSixAxisBottom}{0.1875*\figSixCanvasHeight}
\pgfmathsetlengthmacro{\figSixAxisWidth}{0.635*\linewidth}
\pgfmathsetlengthmacro{\figSixAxisHeight}{0.680*\figSixCanvasHeight}
\begin{tikzpicture}
  \path[use as bounding box] (0,0) rectangle (\linewidth,\figSixCanvasHeight);
  \begin{pgfinterruptboundingbox}
  \definecolor{tempblue}{RGB}{0,114,189}
  
  \begin{axis}[
    at={(\figSixAxisLeft,\figSixAxisBottom)}, anchor=south west,
    scale only axis, width=\figSixAxisWidth, height=\figSixAxisHeight,
    xmin=19, xmax=39, ymin=-1200, ymax=300,
    xlabel={RTD temperature (deg C)},
    ylabel={Zero-field-splitting shift, $\Delta f$ (kHz)},
    tick label style={font=\qtaPlotFont}, label style={font=\qtaPlotFont},
    axis line style={black!60}, axis lines*=left, tick style={black!60},
    tick align=outside, grid=major, major grid style={black!9}, scaled
ticks=false,
    xtick={20,25,30,35},
    xticklabels={20,25,30,35},
    ytick={-1000,-500,0},
    yticklabels={$-1000$,$-500$,0},
    every axis plot/.append style={line cap=round}, clip mode=individual,
    legend columns=1,
    legend style={draw=none,fill=white,fill opacity=0.92,text opacity=1,
      font=\qtaPlotSmallFont,at={(0.02,0.02)},anchor=south west,
      cells={anchor=west},row sep=-1pt,inner sep=1.5pt}
  ]
    \addplot[black!55,dashed,line width=0.90pt,forget plot]
      table[x=x,y=y] {figures/data/Figure6b_reference.dat};
    \addplot[tempblue,line width=1.05pt,forget plot]
      table[x=x,y=y] {figures/data/Figure6b_fit.dat};
    \addplot[only marks,mark=o,mark size=1.95pt,draw=tempblue,fill=white,
      line width=0.65pt,forget plot,
      error bars/.cd,x dir=both,x explicit,y dir=both,y explicit,
      error bar style={draw=tempblue,line width=0.45pt},
      error mark options={draw=tempblue,rotate=90,mark size=1.1pt,line
width=0.45pt}]
      table[x=x,y=y,x error minus=xneg,x error plus=xpos,
        y error minus=yneg,y error plus=ypos] {figures/data/Figure6b_data.dat};
    \addlegendimage{black!55,dashed,line width=0.90pt}
    \addlegendentry{$-75.0~\mathrm{kHz}/^{\circ}\mathrm{C}$}
    \node[anchor=north east,align=left,font=\qtaPlotSmallFont,
      fill=white,fill opacity=0.90,text opacity=1,inner sep=1.5pt]
      at (rel axis cs:0.98,0.98)
      {$\beta=-71.41\pm1.24~\mathrm{kHz}/^{\circ}\mathrm{C}$};
  \end{axis}
  \end{pgfinterruptboundingbox}
\end{tikzpicture}

%% file: figures/Figure7a.tex
\pgfmathsetlengthmacro{\lddSequenceX}{\linewidth/31.30}
\begin{tikzpicture}[
  x=\lddSequenceX,
  y=0.85cm,
  line cap=round,
  line join=round,
  font=\qtaPlotFont
]
  \definecolor{protocolgreen}{RGB}{181,215,167}
  \definecolor{tracedarkblue}{RGB}{0,79,158}
  \definecolor{pulseblue}{RGB}{139,202,232}
  \definecolor{softblue}{RGB}{238,248,252}
  \path[use as bounding box] (0.00,0.00) rectangle (31.30,2.95);

  \newcommand{\lddpulse}[3]{%
    \path[fill=pulseblue,draw=tracedarkblue,line width=0.45pt]
      plot[domain=-1:1,samples=41,smooth]
        ({#1+#2*\x},{1.03+#3*(cos(90*\x))^2})
      -- ({#1+#2},1.03)
      -- ({#1-#2},1.03)
      -- cycle;
  }
  \newcommand{\timinginterval}[3]{%
    \draw[black!45,line width=0.38pt,
      {Latex[length=1.25mm]}-{Latex[length=1.25mm]}]
      ({#1},1.34) -- ({#2},1.34);
    \node[font=\qtaPlotSmallFont,fill=softblue,inner sep=0.6pt]
      at ({(#1+#2)/2},1.34) {$#3$};
  }

  \node[font=\qtaPlotFont\normalfont\bfseries,text=black!65]
    at (3.55,2.31) {initialize};
  \node[font=\qtaPlotFont\normalfont\bfseries,text=black!65]
    at (15.05,2.31) {{ME-}LDD4B refocusing train};
  \node[font=\qtaPlotFont\normalfont\bfseries,text=black!65]
    at (23.82,2.31) {projection};
  \node[font=\qtaPlotFont\normalfont\bfseries,text=black!65]
    at (28.25,2.31) {readout};

  \fill[softblue] (0.35,0.32) rectangle (30.90,2.08);
  \draw[tracedarkblue!28,line width=0.40pt,rounded corners=2pt]
    (0.35,0.32) rectangle (30.90,2.08);
  \node[font=\qtaPlotFont\normalfont\bfseries,text=tracedarkblue,
    anchor=west,inner sep=0pt] at (0.62,1.36) {$R_{+x}$};
  \draw[black!70,line width=0.55pt] (0.55,1.03) -- (30.67,1.03);

  \fill[protocolgreen] (2.20,1.03) rectangle (5.25,1.85);
  \fill[red!88] (4.18,1.03) rectangle (5.25,1.58);
  \draw[green!40!black,line width=0.30pt]
    (2.20,1.03) rectangle (5.25,1.85);
  \node at (3.19,1.60) {Laser};
  \node[font=\qtaPlotSmallFont,text=white] at (4.72,1.31) {PL};

  \lddpulse{6.15}{0.50}{0.58}
  \lddpulse{8.35}{0.82}{0.58}
  \lddpulse{12.75}{0.82}{-0.58}
  \lddpulse{17.15}{0.82}{-0.58}
  \lddpulse{21.55}{0.82}{0.58}
  \lddpulse{23.75}{0.50}{0.58}

  \pgfmathsetmacro{\lddOuterPulseEdgeOffset}{0.50}
  \timinginterval{6.15+\lddOuterPulseEdgeOffset}{8.35}{\tau}
  \timinginterval{8.35}{12.75}{2\tau}
  \timinginterval{12.75}{17.15}{2\tau}
  \timinginterval{17.15}{21.55}{2\tau}
  \timinginterval{21.55}{23.75-\lddOuterPulseEdgeOffset}{\tau}

  \node[font=\qtaPlotSmallFont,inner sep=0.5pt,text=tracedarkblue]
    at (6.15,0.70) {$\left(\frac{\pi}{2}\right)_{\!x}$};
  \node[font=\qtaPlotSmallFont,inner sep=0.5pt,text=tracedarkblue]
    at (8.35,0.70) {$\pi_x$};
  \node[font=\qtaPlotSmallFont,inner sep=0.5pt,text=tracedarkblue]
    at (12.75,1.82) {$\pi_{-x}$};
  \node[font=\qtaPlotSmallFont,inner sep=0.5pt,text=tracedarkblue]
    at (17.15,1.82) {$\pi_{-x}$};
  \node[font=\qtaPlotSmallFont,inner sep=0.5pt,text=tracedarkblue]
    at (21.55,0.70) {$\pi_x$};
  \node[font=\qtaPlotSmallFont\normalfont\bfseries,inner
sep=0.5pt,text=tracedarkblue]
    at (23.75,0.70) {$\left(\frac{\pi}{2}\right)_{\!+x}$};

  \fill[protocolgreen] (25.10,1.03) rectangle (30.25,1.85);
  \shade[left color=black!42,right color=red!88]
    (25.10,1.03) rectangle (26.17,1.58);
  \draw[green!40!black,line width=0.30pt]
    (25.10,1.03) rectangle (30.25,1.85);
  \node at (28.40,1.60) {Laser};
  \node[font=\qtaPlotSmallFont,text=white] at (25.64,1.31) {PL};

  \draw[line width=0.65pt,-{Latex[length=2.1mm,width=1.5mm]}]
    (0.56,0.13) -- (30.67,0.13);
  \node[anchor=north,inner sep=0pt,font=\qtaPlotFont]
    at (15.62,0.04) {Time};
\end{tikzpicture}

%% file: figures/Figure7bcd.tex
\begingroup
\definecolor{hahnI}{RGB}{0,79,153}
\definecolor{hahnQ}{RGB}{92,166,214}
\definecolor{lddI}{RGB}{184,31,41}
\definecolor{lddQ}{RGB}{237,122,92}
\pgfmathsetlengthmacro{\hahnFullCanvasHeight}{0.6384615*\linewidth}
\pgfmathsetlengthmacro{\hahnVisibleCanvasHeight}{0.5658120*\linewidth}
\pgfmathsetlengthmacro{\hahnLeftX}{0.115*\linewidth}
\pgfmathsetlengthmacro{\hahnLeftWidth}{0.350*\linewidth}
\pgfmathsetlengthmacro{\hahnRightX}{0.615*\linewidth}
\pgfmathsetlengthmacro{\hahnRightSize}{0.355*\linewidth}
\pgfmathsetlengthmacro{\hahnPanelBottom}{0.180*\hahnFullCanvasHeight}
\pgfmathsetlengthmacro{\hahnTraceGap}{0.210*\hahnFullCanvasHeight}
\pgfmathsetlengthmacro{\hahnTraceHeight}{0.5*(\hahnRightSize-\hahnTraceGap)}
\pgfmathsetlengthmacro{\hahnTopBottom}{\hahnPanelBottom+\hahnTraceHeight%
+\hahnTraceGap}
\begin{tikzpicture}
  \path[use as bounding box]
    (0,0) rectangle (\linewidth,\hahnVisibleCanvasHeight);
  \begin{pgfinterruptboundingbox}
    \begin{axis}[
      name=hahntrace,
      at={(\hahnLeftX,\hahnTopBottom)},
      anchor=south west,
      scale only axis,
      width=\hahnLeftWidth,
      height=\hahnTraceHeight,
      xmin=0,
      xmax=40,
      ymin=-2,
      ymax=2,
      xtick={0,10,20,30,40},
      ytick={-2,-1,0,1,2},
      minor x tick num=4,
      minor y tick num=1,
      xlabel={{Evolution time}, $2\tau$ ({$\mu\mathrm{s}$})},
      ylabel={Contrast (\%)},
      title={(b) {ME} {thermal} echo},
      title style={font=\qtaPlotPanelFont,yshift=-1pt},
      tick label style={font=\qtaPlotFont},
      label style={font=\qtaPlotFont},
      axis lines=box,
      axis line style={black!55,line width=0.8pt},
      tick style={black!55},
      tick align=outside,
      tick pos=both,
      grid=major,
      major grid style={black!12},
      scaled ticks=false,
      unbounded coords=discard,
      every axis plot/.append style={line cap=round},
      legend columns=2,
      legend style={
        draw=none,
        fill=white,
        fill opacity=0.88,
        text opacity=1,
        font=\qtaPlotSmallFont,
        at={(0.98,0.98)},
        anchor=north east,
        column sep=0.55em,
        inner sep=1.5pt
      }
    ]
      \addplot[black!45,densely dotted,line width=0.8pt,no marks,forget plot]
        coordinates {(0,0) (40,0)};
      \addplot[hahnI,line width=1.05pt,no marks]
        table[
          x=total_evolution_us,
          y expr=100*\thisrow{contrast_i_movmean3},
          restrict expr to domain={\thisrow{axis_multiplier}}{2:2},
          col sep=comma
        ] {figures/data/Figure7bc.csv};
      \addlegendentry{$C_I$}
      \addplot[hahnQ,dashed,line width=1.05pt,no marks]
        table[
          x=total_evolution_us,
          y expr=100*\thisrow{contrast_q_movmean3},
          restrict expr to domain={\thisrow{axis_multiplier}}{2:2},
          col sep=comma
        ] {figures/data/Figure7bc.csv};
      \addlegendentry{$C_Q$}
      \node[
        anchor=south east,
        font=\qtaPlotSmallFont,
        fill=white,
        fill opacity=0.88,
        text opacity=1,
        inner sep=1.5pt
      ] at (axis description cs:0.98,0.06)
        {{$T_{\mathrm{echo}}=6.5~\mu\mathrm{s}$}};
    \end{axis}

    \begin{axis}[
      name=lddtrace,
      at={(\hahnLeftX,\hahnPanelBottom)},
      anchor=south west,
      scale only axis,
      width=\hahnLeftWidth,
      height=\hahnTraceHeight,
      xmin=0,
      xmax=40,
      ymin=-2,
      ymax=2,
      xtick={0,10,20,30,40},
      ytick={-2,-1,0,1,2},
      minor x tick num=4,
      minor y tick num=1,
      xlabel={{Evolution time}, $8\tau$ ({$\mu\mathrm{s}$})},
      ylabel={Contrast (\%)},
      title={(c) {ME-}LDD4B},
      title style={font=\qtaPlotPanelFont,yshift=-1pt},
      tick label style={font=\qtaPlotFont},
      label style={font=\qtaPlotFont},
      axis lines=box,
      axis line style={black!55,line width=0.8pt},
      tick style={black!55},
      tick align=outside,
      tick pos=both,
      grid=major,
      major grid style={black!12},
      scaled ticks=false,
      unbounded coords=discard,
      every axis plot/.append style={line cap=round},
      legend columns=2,
      legend style={
        draw=none,
        fill=white,
        fill opacity=0.88,
        text opacity=1,
        font=\qtaPlotSmallFont,
        at={(0.98,0.98)},
        anchor=north east,
        column sep=0.55em,
        inner sep=1.5pt
      }
    ]
      \addplot[black!45,densely dotted,line width=0.8pt,no marks,forget plot]
        coordinates {(0,0) (40,0)};
      \addplot[lddI,line width=1.05pt,no marks]
        table[
          x=total_evolution_us,
          y expr=100*\thisrow{contrast_i_movmean3},
          restrict expr to domain={\thisrow{axis_multiplier}}{8:8},
          col sep=comma
        ] {figures/data/Figure7bc.csv};
      \addlegendentry{$C_I$}
      \addplot[lddQ,dashed,line width=1.05pt,no marks]
        table[
          x=total_evolution_us,
          y expr=100*\thisrow{contrast_q_movmean3},
          restrict expr to domain={\thisrow{axis_multiplier}}{8:8},
          col sep=comma
        ] {figures/data/Figure7bc.csv};
      \addlegendentry{$C_Q$}
      \node[
        anchor=south east,
        font=\qtaPlotSmallFont,
        fill=white,
        fill opacity=0.88,
        text opacity=1,
        inner sep=1.5pt
      ] at (axis description cs:0.98,0.06)
        {{$T_{\mathrm{echo}}=14.5~\mu\mathrm{s}$}};
    \end{axis}

    \begin{axis}[
      name=frequencyresponse,
      at={(\hahnRightX,\hahnPanelBottom)},
      anchor=south west,
      scale only axis,
      width=\hahnRightSize,
      height=\hahnRightSize,
      xmin=-0.5,
      xmax=0.4,
      ymin=-0.5,
      ymax=0.5,
      xtick={-0.4,-0.2,0,0.2,0.4},
      ytick={-0.4,-0.2,0,0.2,0.4},
      minor x tick num=1,
      minor y tick num=1,
      xlabel={{Microwave offset}, $f_{\mathrm{MW}}-D_0$ (MHz)},
      ylabel={Echo frequency, $f_{\mathrm{echo}}$ (MHz)},
      title={(d) Echo frequency versus {offset}},
      title style={font=\qtaPlotPanelFont,yshift=-1pt},
      tick label style={font=\qtaPlotFont},
      label style={font=\qtaPlotFont},
      axis lines=box,
      axis line style={black!55,line width=0.8pt},
      tick style={black!55},
      tick align=outside,
      tick pos=both,
      grid=major,
      major grid style={black!12},
      scaled ticks=false,
      unbounded coords=discard,
      every axis plot/.append style={line cap=round},
      legend cell align=left,
      legend style={
        draw=none,
        fill=white,
        fill opacity=0.88,
        text opacity=1,
        font=\qtaPlotSmallFont,
        at={(0.02,0.02)},
        anchor=south west,
        row sep=2pt,
        inner sep=1.5pt
      }
    ]
      \addplot[black!50,densely dotted,line width=0.9pt,no marks,forget plot]
        coordinates {(-0.243193757363319,-0.5) (-0.243193757363319,0.5)};

      \addplot[
        hahnI,
        line width=1.0pt,
        no marks,
        domain=-0.443193757363319:0.356806242636681,
        samples=2
      ] {-x-0.021412514358709504};
      \addlegendentry{\shortstack[l]{{ME} {thermal} echo\\fit, slope = $-1$}}

      \addplot[
        lddI,
        line width=1.0pt,
        no marks,
        domain=-0.443193757363319:0.356806242636681,
        samples=2
      ] {-0.5*x-0.010646090614286154};
      \addlegendentry{\shortstack[l]{{ME-}LDD4B\\fit, slope = $-0.5$}}

      \addplot[
        only marks,
        mark=o,
        mark size=2.10pt,
        draw=hahnI,
        fill=white,
        line width=0.85pt,
        forget plot,
        error bars/.cd,
        y dir=both,
        y explicit,
        error bar style={draw=hahnI,line width=0.85pt},
        error mark=-,
        error mark options={draw=hahnI,rotate=90,mark size=2pt,line
width=0.85pt}
      ] table[
        x=fixed_reference_detuning_MHz,
        y=echo_frequency_MHz,
        y error minus
expr={\thisrow{echo_frequency_MHz}-\thisrow{echo_frequency_ci_low_MHz}},
        y error plus
expr={\thisrow{echo_frequency_ci_high_MHz}-\thisrow{echo_frequency_MHz}},
        restrict expr to domain={\thisrow{axis_multiplier}}{2:2},
        col sep=comma
      ] {figures/data/Figure7d.csv};

      \addplot[
        only marks,
        mark=square,
        mark size=2.10pt,
        draw=lddI,
        fill=white,
        line width=0.85pt,
        forget plot,
        error bars/.cd,
        y dir=both,
        y explicit,
        error bar style={draw=lddI,line width=0.85pt},
        error mark=-,
        error mark options={draw=lddI,rotate=90,mark size=2pt,line
width=0.85pt}
      ] table[
        x=fixed_reference_detuning_MHz,
        y=echo_frequency_MHz,
        y error minus
expr={\thisrow{echo_frequency_MHz}-\thisrow{echo_frequency_ci_low_MHz}},
        y error plus
expr={\thisrow{echo_frequency_ci_high_MHz}-\thisrow{echo_frequency_MHz}},
        restrict expr to domain={\thisrow{axis_multiplier}}{8:8},
        col sep=comma
      ] {figures/data/Figure7d.csv};
    \end{axis}
  \end{pgfinterruptboundingbox}
\end{tikzpicture}
\endgroup

%% file: figures/Figure8.tex
\begingroup
\definecolor{robustone}{RGB}{92,158,201}
\definecolor{robusttwo}{RGB}{66,133,183}
\definecolor{robustthree}{RGB}{44,108,163}
\definecolor{robustfour}{RGB}{26,84,139}
\definecolor{robustfive}{RGB}{13,60,113}
\definecolor{robustsix}{RGB}{5,37,82}

\begin{tikzpicture}
\begin{groupplot}[
    group style={group size=2 by 2,horizontal sep=2.25cm,vertical sep=2.2cm},
    scale only axis,width=0.37\textwidth,height=4.1cm,
    xmin=25,xmax=500,ymin=0,ymax=1.6,
    xtick={100,300,500},ytick={0,0.5,1,1.5},
    xticklabel={\pgfmathprintnumber{\tick}},
    yticklabel={\pgfmathprintnumber{\tick}},
    xlabel={$t_p$ (ns)},
    tick label style={font=\qtaPlotFont},
    label style={font=\qtaPlotFont},
    title style={font=\qtaPlotPanelFont,yshift=-1pt},
    legend style={font=\qtaPlotSmallFont,draw=none,fill=white,
        at={(0.98,0.98)},anchor=north east,inner sep=1pt},
    legend cell align=left,
    axis lines=box,axis line style={black!55,line width=0.8pt},
    tick style={black!55},tick align=outside,tick pos=both,
    grid=major,major grid style={black!12},scaled ticks=false,
    every axis plot/.append style={line width=0.9pt,no marks},
]
\nextgroupplot[
    title={(a) $\Omega_{\mathrm{pk}}$ (MHz)},
    ylabel={$A(t_p)/A_0$},
]
\addplot[robustone,solid] table[x=duration_ns,y=drive_0,col
sep=comma]{figures/data/Figure8acd_model.csv};
\addlegendentry{1}
\addplot[robusttwo,solid] table[x=duration_ns,y=drive_1,col
sep=comma]{figures/data/Figure8acd_model.csv};
\addlegendentry{2}
\addplot[robustthree,solid] table[x=duration_ns,y=drive_2,col
sep=comma]{figures/data/Figure8acd_model.csv};
\addlegendentry{3}
\addplot[robustfour,solid] table[x=duration_ns,y=drive_3,col
sep=comma]{figures/data/Figure8acd_model.csv};
\addlegendentry{4}
\addplot[robustfive,solid] table[x=duration_ns,y=drive_4,col
sep=comma]{figures/data/Figure8acd_model.csv};
\addlegendentry{5}
\addplot[robustsix,solid] table[x=duration_ns,y=drive_5,col
sep=comma]{figures/data/Figure8acd_model.csv};
\addlegendentry{6}
\addplot[black!50,dashed,forget plot] coordinates {(125,0) (125,1.6)};

\nextgroupplot[
    title={(b) Peak timing},
    xmin=125,xmax=1050,ymin=25,ymax=425,
    xtick={250,500,750,1000},ytick={100,200,300,400},
    xlabel={Rabi period $1/\Omega_{\mathrm{pk}}$ (ns)},
    ylabel={\shortstack{Peak endcap duration\\$t_{p,\max}$ (ns)}},
    unbounded coords=discard,filter discard warning=false,
]
\addplot[only marks,robustsix,mark=*,mark size=2.5pt,
    mark options={solid,fill=robustsix},restrict expr to
domain={\coordindex}{0:0}]
    table[x=drive_period_ns,y=sampled_peak_ns,col sep=comma]
    {figures/data/Figure8b_peak_timing.csv};
\addplot[only marks,robustfive,mark=*,mark size=2.5pt,
    mark options={solid,fill=robustfive},restrict expr to
domain={\coordindex}{1:1}]
    table[x=drive_period_ns,y=sampled_peak_ns,col sep=comma]
    {figures/data/Figure8b_peak_timing.csv};
\addplot[only marks,robustfour,mark=*,mark size=2.5pt,
    mark options={solid,fill=robustfour},restrict expr to
domain={\coordindex}{2:2}]
    table[x=drive_period_ns,y=sampled_peak_ns,col sep=comma]
    {figures/data/Figure8b_peak_timing.csv};
\addplot[only marks,robustthree,mark=*,mark size=2.5pt,
    mark options={solid,fill=robustthree},restrict expr to
domain={\coordindex}{3:3}]
    table[x=drive_period_ns,y=sampled_peak_ns,col sep=comma]
    {figures/data/Figure8b_peak_timing.csv};
\addplot[only marks,robusttwo,mark=*,mark size=2.5pt,
    mark options={solid,fill=robusttwo},restrict expr to
domain={\coordindex}{4:4}]
    table[x=drive_period_ns,y=sampled_peak_ns,col sep=comma]
    {figures/data/Figure8b_peak_timing.csv};
\addplot[only marks,robustone,mark=*,mark size=2.5pt,
    mark options={solid,fill=robustone},restrict expr to
domain={\coordindex}{5:5}]
    table[x=drive_period_ns,y=sampled_peak_ns,col sep=comma]
    {figures/data/Figure8b_peak_timing.csv};
\nextgroupplot[
    title={(c) Pair fraction $p_\chi$},
    ylabel={$A(t_p)/A_0$},
]
\addplot[robustone,solid] table[x=duration_ns,y=weight_1,col
sep=comma]{figures/data/Figure8acd_model.csv};
\addlegendentry{0}
\addplot[robusttwo,solid] table[x=duration_ns,y=weight_2,col
sep=comma]{figures/data/Figure8acd_model.csv};
\addlegendentry{0.25}
\addplot[robustfour,solid] table[x=duration_ns,y=weight_3,col
sep=comma]{figures/data/Figure8acd_model.csv};
\addlegendentry{0.5}
\addplot[robustfive,solid] table[x=duration_ns,y=weight_4,col
sep=comma]{figures/data/Figure8acd_model.csv};
\addlegendentry{0.75}
\addplot[robustsix,solid] table[x=duration_ns,y=weight_5,col
sep=comma]{figures/data/Figure8acd_model.csv};
\addlegendentry{1}
\addplot[black!50,dashed,forget plot] coordinates {(125,0) (125,1.6)};

\nextgroupplot[
    title={(d) $E_{\mathrm{iso}}$ (MHz)},yticklabels=\empty,
]
\addplot[robustone,solid] table[x=duration_ns,y=strain_1,col
sep=comma]{figures/data/Figure8acd_model.csv};
\addlegendentry{0}
\addplot[robusttwo,solid] table[x=duration_ns,y=strain_2,col
sep=comma]{figures/data/Figure8acd_model.csv};
\addlegendentry{0.272}
\addplot[robustfour,solid] table[x=duration_ns,y=strain_3,col
sep=comma]{figures/data/Figure8acd_model.csv};
\addlegendentry{0.581}
\addplot[robustfive,solid] table[x=duration_ns,y=strain_4,col
sep=comma]{figures/data/Figure8acd_model.csv};
\addlegendentry{0.891}
\addplot[robustsix,solid] table[x=duration_ns,y=strain_5,col
sep=comma]{figures/data/Figure8acd_model.csv};
\addlegendentry{1.2}
\addplot[black!50,dashed,forget plot] coordinates {(125,0) (125,1.6)};

\end{groupplot}
\end{tikzpicture}
\endgroup

%% file: figures/Figure9.tex
\begingroup
\definecolor{powerlow}{RGB}{92,158,201}
\definecolor{powermid}{RGB}{26,84,139}
\definecolor{powerhigh}{RGB}{5,37,82}
\begin{tikzpicture}
\begin{groupplot}[
    group style={group size=2 by 1,horizontal sep=2.25cm},
    scale only axis,height=6cm,
    xticklabel={\pgfmathprintnumber{\tick}},
    yticklabel={\pgfmathprintnumber{\tick}},
    tick label style={font=\qtaPlotFont},
    label style={font=\qtaPlotFont},
    title style={font=\qtaPlotPanelFont,yshift=-1pt},
    axis lines=box,axis line style={black!55,line width=0.8pt},
    tick style={black!55},tick align=outside,tick pos=both,
    grid=major,major grid style={black!12},scaled ticks=false,
]
\nextgroupplot[
    width=0.45\textwidth,
    title={(a)},
    xmin=75,xmax=320,ymin=0,ymax=1.2,
    xtick={100,150,200,250,300},ytick={0,0.2,0.4,0.6,0.8,1,1.2},
    xlabel={Endcap duration $t_p$ (ns)},
    ylabel={\shortstack{Noise-corrected amplitude\\(percentage points)}},
    legend style={font=\qtaPlotSmallFont,draw=none,fill=white,
        at={(0.97,0.97)},anchor=north east,inner sep=3pt},
    legend cell align=left,
]
\addlegendimage{empty legend}
\addlegendentry{$f_R$ (MHz), $f+2f$ fit}
\addplot+[color=powerlow,solid,line width=0.8pt,mark=*,mark size=1.6pt,
    mark options={solid,fill=powerlow},error bars/.cd,y dir=both,y explicit]
    table[x=endcap_ns,y=corrected_amplitude_pct,y error plus=upper_error_pct,y
error minus=lower_error_pct,col sep=comma]
    {figures/data/Figure9a_low_drive.csv};
\addlegendentry{1.11}
\addplot+[color=powermid,solid,line width=0.8pt,mark=*,mark size=1.6pt,
    mark options={solid,fill=powermid},error bars/.cd,y dir=both,y explicit]
    table[x=endcap_ns,y=corrected_amplitude_pct,y error plus=upper_error_pct,y
error minus=lower_error_pct,col sep=comma]
    {figures/data/Figure9a_middle_drive.csv};
\addlegendentry{1.67}
\addplot+[color=powerhigh,solid,line width=0.8pt,mark=*,mark size=1.6pt,
    mark options={solid,fill=powerhigh},error bars/.cd,y dir=both,y explicit]
    table[x=endcap_ns,y=corrected_amplitude_pct,y error plus=upper_error_pct,y
error minus=lower_error_pct,col sep=comma]
    {figures/data/Figure9a_high_drive.csv};
\addlegendentry{2.84}
\addplot+[color=powerlow,solid,line width=0.8pt,mark=*,mark size=1.6pt,forget
plot,
    mark options={solid,fill=powerlow},error bars/.cd,y dir=both,y explicit]
    table[x=endcap_ns,y=corrected_amplitude_pct,y error plus=upper_error_pct,y
error minus=lower_error_pct,col sep=comma]
    {figures/data/Figure9a_low_drive_extension.csv};
\addplot+[color=powermid,solid,line width=0.8pt,mark=*,mark size=1.6pt,forget
plot,
    mark options={solid,fill=powermid},error bars/.cd,y dir=both,y explicit]
    table[x=endcap_ns,y=corrected_amplitude_pct,y error plus=upper_error_pct,y
error minus=lower_error_pct,col sep=comma]
    {figures/data/Figure9a_middle_drive_extension.csv};
\addplot+[color=powerhigh,solid,line width=0.8pt,mark=*,mark size=1.6pt,forget
plot,
    mark options={solid,fill=powerhigh},error bars/.cd,y dir=both,y explicit]
    table[x=endcap_ns,y=corrected_amplitude_pct,y error plus=upper_error_pct,y
error minus=lower_error_pct,col sep=comma]
    {figures/data/Figure9a_high_drive_extension.csv};
\nextgroupplot[
    width=0.28\textwidth,
    title={(b)},
    xmin=300,xmax=950,ymin=75,ymax=260,
    xtick={300,600,900},ytick={100,150,200,250},
    xlabel={Rabi period $T_R=1/f_R$ (ns)},
    ylabel={\shortstack{Peak endcap duration\\$t_{p,\max}$ (ns)}},
    unbounded coords=discard,filter discard warning=false,
]
\addplot+[only marks,color=powerhigh,mark=*,mark size=2.5pt,
    mark options={solid,fill=powerhigh},restrict expr to
domain={\coordindex}{0:0}]
    table[x=rabi_period_ns,y=sampled_peak_ns,col sep=comma]
    {figures/data/Figure9b_peak_timing.csv};
\addplot+[only marks,color=powermid,mark=*,mark size=2.5pt,
    mark options={solid,fill=powermid},restrict expr to
domain={\coordindex}{1:1}]
    table[x=rabi_period_ns,y=sampled_peak_ns,col sep=comma]
    {figures/data/Figure9b_peak_timing.csv};
\addplot+[only marks,color=powerlow,mark=*,mark size=2.5pt,
    mark options={solid,fill=powerlow},restrict expr to
domain={\coordindex}{2:2}]
    table[x=rabi_period_ns,y=sampled_peak_ns,col sep=comma]
    {figures/data/Figure9b_peak_timing.csv};
\end{groupplot}
\end{tikzpicture}
\endgroup

%% file: main.bbl
\begin{thebibliography}{10}
\expandafter\ifx\csname url\endcsname\relax
  \def\url#1{\burl{#1}}\fi
\expandafter\ifx\csname urlprefix\endcsname\relax\def\urlprefix{URL }\fi
\providecommand{\bibinfo}[2]{#2}
\providecommand{\eprint}[2][]{\url{#2}}
\providecommand{\doi}[1]{\url{https://doi.org/#1}}
\bibcommenthead

\bibitem{Degen2017}
\bibinfo{author}{Degen, C.~L.}, \bibinfo{author}{Reinhard, F.} \&
  \bibinfo{author}{Cappellaro, P.}
\newblock \bibinfo{title}{Quantum sensing}.
\newblock \emph{\bibinfo{journal}{Rev. Mod. Phys.}}
  \textbf{\bibinfo{volume}{89}}, \bibinfo{pages}{035002}
  (\bibinfo{year}{2017}).

\bibitem{Schirhagl2014}
\bibinfo{author}{Schirhagl, R.}, \bibinfo{author}{Chang, K.},
  \bibinfo{author}{Loretz, M.} \& \bibinfo{author}{Degen, C.~L.}
\newblock \bibinfo{title}{Nitrogen-vacancy centers in diamond: nanoscale
  sensors for physics and biology}.
\newblock \emph{\bibinfo{journal}{Annu. Rev. Phys. Chem.}}
  \textbf{\bibinfo{volume}{65}}, \bibinfo{pages}{83--105}
  (\bibinfo{year}{2014}).

\bibitem{Maze2008}
\bibinfo{author}{Maze, J.~R.} \emph{et~al.}
\newblock \bibinfo{title}{Nanoscale magnetic sensing with an individual
  electronic spin in diamond}.
\newblock \emph{\bibinfo{journal}{Nature}} \textbf{\bibinfo{volume}{455}},
  \bibinfo{pages}{644--647} (\bibinfo{year}{2008}).

\bibitem{Balasubramanian2008}
\bibinfo{author}{Balasubramanian, G.} \emph{et~al.}
\newblock \bibinfo{title}{Nanoscale imaging magnetometry with diamond spins
  under ambient conditions}.
\newblock \emph{\bibinfo{journal}{Nature}} \textbf{\bibinfo{volume}{455}},
  \bibinfo{pages}{648--651} (\bibinfo{year}{2008}).

\bibitem{Barry2020}
\bibinfo{author}{Barry, J.~F.} \emph{et~al.}
\newblock \bibinfo{title}{Sensitivity optimization for {NV}-diamond
  magnetometry}.
\newblock \emph{\bibinfo{journal}{Rev. Mod. Phys.}}
  \textbf{\bibinfo{volume}{92}}, \bibinfo{pages}{015004}
  (\bibinfo{year}{2020}).

\bibitem{Kucsko2013}
\bibinfo{author}{Kucsko, G.} \emph{et~al.}
\newblock \bibinfo{title}{Nanometre-scale thermometry in a living cell}.
\newblock \emph{\bibinfo{journal}{Nature}} \textbf{\bibinfo{volume}{500}},
  \bibinfo{pages}{54--58} (\bibinfo{year}{2013}).

\bibitem{Fujiwara2020InVivo}
\bibinfo{author}{Fujiwara, M.} \emph{et~al.}
\newblock \bibinfo{title}{Real-time nanodiamond thermometry probing in vivo
  thermogenic responses}.
\newblock \emph{\bibinfo{journal}{Sci. Adv.}} \textbf{\bibinfo{volume}{6}},
  \bibinfo{pages}{eaba9636} (\bibinfo{year}{2020}).

\bibitem{Hart2026}
\bibinfo{author}{Hart, J.~W.} \emph{et~al.}
\newblock \bibinfo{title}{Microkelvin resolution thermometry at the nanometre
  scale}.
\newblock \bibinfo{howpublished}{arXiv preprint} (\bibinfo{year}{2026}).
\newblock \urlprefix\url{https://arxiv.org/abs/2609.04907}.

\bibitem{Balasubramanian2009}
\bibinfo{author}{Balasubramanian, G.} \emph{et~al.}
\newblock \bibinfo{title}{Ultralong spin coherence time in isotopically
  engineered diamond}.
\newblock \emph{\bibinfo{journal}{Nat. Mater.}} \textbf{\bibinfo{volume}{8}},
  \bibinfo{pages}{383--387} (\bibinfo{year}{2009}).

\bibitem{Lin2021}
\bibinfo{author}{Lin, S.} \emph{et~al.}
\newblock \bibinfo{title}{Temperature-dependent coherence properties of {NV}
  ensemble in diamond up to 600 {K}}.
\newblock \emph{\bibinfo{journal}{Phys. Rev. B}}
  \textbf{\bibinfo{volume}{104}}, \bibinfo{pages}{155430}
  (\bibinfo{year}{2021}).

\bibitem{Wang2021Quadrature}
\bibinfo{author}{Wang, Z.} \emph{et~al.}
\newblock \bibinfo{title}{{AC} sensing using nitrogen-vacancy centers in a
  diamond anvil cell up to 6 {GPa}}.
\newblock \emph{\bibinfo{journal}{Phys. Rev. Applied}}
  \textbf{\bibinfo{volume}{16}}, \bibinfo{pages}{054014}
  (\bibinfo{year}{2021}).

\bibitem{Neumann2013}
\bibinfo{author}{Neumann, P.} \emph{et~al.}
\newblock \bibinfo{title}{High-precision nanoscale temperature sensing using
  single defects in diamond}.
\newblock \emph{\bibinfo{journal}{Nano Lett.}} \textbf{\bibinfo{volume}{13}},
  \bibinfo{pages}{2738--2742} (\bibinfo{year}{2013}).

\bibitem{Toyli2013}
\bibinfo{author}{Toyli, D.~M.}, \bibinfo{author}{de~las Casas, C.~F.},
  \bibinfo{author}{Christle, D.~J.}, \bibinfo{author}{Dobrovitski, V.~V.} \&
  \bibinfo{author}{Awschalom, D.~D.}
\newblock \bibinfo{title}{Fluorescence thermometry enhanced by the quantum
  coherence of single spins in diamond}.
\newblock \emph{\bibinfo{journal}{Proc. Natl Acad. Sci. USA}}
  \textbf{\bibinfo{volume}{110}}, \bibinfo{pages}{8417--8421}
  (\bibinfo{year}{2013}).

\bibitem{Fujiwara2021Thermometry}
\bibinfo{author}{Fujiwara, M.} \& \bibinfo{author}{Shikano, Y.}
\newblock \bibinfo{title}{Diamond quantum thermometry: from foundations to
  applications}.
\newblock \emph{\bibinfo{journal}{Nanotechnology}}
  \textbf{\bibinfo{volume}{32}}, \bibinfo{pages}{482002}
  (\bibinfo{year}{2021}).

\bibitem{Plakhotnik2014}
\bibinfo{author}{Plakhotnik, T.}, \bibinfo{author}{Doherty, M.~W.},
  \bibinfo{author}{Cole, J.~H.}, \bibinfo{author}{Chapman, R.} \&
  \bibinfo{author}{Manson, N.~B.}
\newblock \bibinfo{title}{All-optical thermometry and thermal properties of the
  optically detected spin resonances of the {NV}$^{-}$ center in nanodiamond}.
\newblock \emph{\bibinfo{journal}{Nano Lett.}} \textbf{\bibinfo{volume}{14}},
  \bibinfo{pages}{4989--4996} (\bibinfo{year}{2014}).

\bibitem{Acosta2010}
\bibinfo{author}{Acosta, V.~M.} \emph{et~al.}
\newblock \bibinfo{title}{Temperature dependence of the nitrogen-vacancy
  magnetic resonance in diamond}.
\newblock \emph{\bibinfo{journal}{Phys. Rev. Lett.}}
  \textbf{\bibinfo{volume}{104}}, \bibinfo{pages}{070801}
  (\bibinfo{year}{2010}).

\bibitem{Acosta2011Erratum}
\bibinfo{author}{Acosta, V.~M.} \emph{et~al.}
\newblock \bibinfo{title}{Erratum: temperature dependence of the
  nitrogen-vacancy magnetic resonance in diamond [{Phys. Rev. Lett.} {104},
  070801 (2010)]}.
\newblock \emph{\bibinfo{journal}{Phys. Rev. Lett.}}
  \textbf{\bibinfo{volume}{106}}, \bibinfo{pages}{209901}
  (\bibinfo{year}{2011}).

\bibitem{Hayashi2018}
\bibinfo{author}{Hayashi, K.} \emph{et~al.}
\newblock \bibinfo{title}{Optimization of temperature sensitivity using the
  optically detected magnetic-resonance spectrum of a nitrogen-vacancy center
  ensemble}.
\newblock \emph{\bibinfo{journal}{Phys. Rev. Applied}}
  \textbf{\bibinfo{volume}{10}}, \bibinfo{pages}{034009}
  (\bibinfo{year}{2018}).

\bibitem{Fujiwara2020Thermometry}
\bibinfo{author}{Fujiwara, M.} \emph{et~al.}
\newblock \bibinfo{title}{Real-time estimation of the optically detected
  magnetic resonance shift in diamond quantum thermometry toward biological
  applications}.
\newblock \emph{\bibinfo{journal}{Phys. Rev. Research}}
  \textbf{\bibinfo{volume}{2}}, \bibinfo{pages}{043415} (\bibinfo{year}{2020}).

\bibitem{Wojciechowski2018}
\bibinfo{author}{Wojciechowski, A.~M.} \emph{et~al.}
\newblock \bibinfo{title}{Precision temperature sensing in the presence of
  magnetic field noise and vice-versa using nitrogen-vacancy centers in
  diamond}.
\newblock \emph{\bibinfo{journal}{Appl. Phys. Lett.}}
  \textbf{\bibinfo{volume}{113}}, \bibinfo{pages}{013502}
  (\bibinfo{year}{2018}).

\bibitem{Wang2015Thermometry}
\bibinfo{author}{Wang, J.} \emph{et~al.}
\newblock \bibinfo{title}{High-sensitivity temperature sensing using an
  implanted single nitrogen-vacancy center array in diamond}.
\newblock \emph{\bibinfo{journal}{Phys. Rev. B}} \textbf{\bibinfo{volume}{91}},
  \bibinfo{pages}{155404} (\bibinfo{year}{2015}).

\bibitem{Cerrillo2021}
\bibinfo{author}{Cerrillo, J.}, \bibinfo{author}{Oviedo~Casado, S.} \&
  \bibinfo{author}{Prior, J.}
\newblock \bibinfo{title}{Low field nano-{NMR} via three-level system control}.
\newblock \emph{\bibinfo{journal}{Phys. Rev. Lett.}}
  \textbf{\bibinfo{volume}{126}}, \bibinfo{pages}{220402}
  (\bibinfo{year}{2021}).

\bibitem{Vetter2022}
\bibinfo{author}{Vetter, P.~J.} \emph{et~al.}
\newblock \bibinfo{title}{Zero- and low-field sensing with nitrogen-vacancy
  centers}.
\newblock \emph{\bibinfo{journal}{Phys. Rev. Applied}}
  \textbf{\bibinfo{volume}{17}}, \bibinfo{pages}{044028}
  (\bibinfo{year}{2022}).

\bibitem{Sow2025}
\bibinfo{author}{Sow, M.} \emph{et~al.}
\newblock \bibinfo{title}{Millikelvin intracellular nanothermometry with
  nanodiamonds}.
\newblock \emph{\bibinfo{journal}{Adv. Sci.}} \textbf{\bibinfo{volume}{12}},
  \bibinfo{pages}{e11670} (\bibinfo{year}{2025}).

\bibitem{Childress2025}
\bibinfo{author}{Childress, L.} \emph{et~al.}
\newblock \bibinfo{title}{Bias-field-free operation of nitrogen-vacancy
  ensembles in diamond for accurate vector magnetometry}.
\newblock \emph{\bibinfo{journal}{PRX Quantum}} \textbf{\bibinfo{volume}{6}},
  \bibinfo{pages}{040364} (\bibinfo{year}{2025}).

\bibitem{Doherty2013}
\bibinfo{author}{Doherty, M.~W.} \emph{et~al.}
\newblock \bibinfo{title}{The nitrogen-vacancy colour centre in diamond}.
\newblock \emph{\bibinfo{journal}{Phys. Rep.}} \textbf{\bibinfo{volume}{528}},
  \bibinfo{pages}{1--45} (\bibinfo{year}{2013}).

\bibitem{LopezGarcia2025}
\bibinfo{author}{L{\'o}pez-Garc{\'i}a, A.} \& \bibinfo{author}{Cerrillo, J.}
\newblock \bibinfo{title}{Full qubit control of the double quantum transition
  in {NV} centers for low-field or high-frequency sensing}.
\newblock \emph{\bibinfo{journal}{EPJ Quantum Technol.}}
  \textbf{\bibinfo{volume}{12}}, \bibinfo{pages}{52} (\bibinfo{year}{2025}).

\bibitem{Misonou2020}
\bibinfo{author}{Misonou, D.} \emph{et~al.}
\newblock \bibinfo{title}{Construction and operation of a tabletop system for
  nanoscale magnetometry with single nitrogen-vacancy centers in diamond}.
\newblock \emph{\bibinfo{journal}{AIP Advances}} \textbf{\bibinfo{volume}{10}},
  \bibinfo{pages}{025206} (\bibinfo{year}{2020}).

\bibitem{Ma2019}
\bibinfo{author}{Ma, Z.} \emph{et~al.}
\newblock \bibinfo{title}{Efficient microwave radiation using
  broadened-bandwidth coplanar waveguide resonator on assembly of
  nitrogen-vacancy centers in diamond}.
\newblock \emph{\bibinfo{journal}{Jpn. J. Appl. Phys.}}
  \textbf{\bibinfo{volume}{58}}, \bibinfo{pages}{050919}
  (\bibinfo{year}{2019}).

\bibitem{WMMHR2025}
\bibinfo{author}{{NOAA NCEI Geomagnetic Modeling Team}} \&
  \bibinfo{author}{{British Geological Survey}}.
\newblock \bibinfo{title}{{World Magnetic Model High Resolution 2025}}.
\newblock \bibinfo{howpublished}{NOAA National Centers for Environmental
  Information} (\bibinfo{year}{2024}).
\newblock \urlprefix\url{https://doi.org/10.25921/qb1c-vn52}.

\bibitem{Matsuzaki2016}
\bibinfo{author}{Matsuzaki, Y.} \emph{et~al.}
\newblock \bibinfo{title}{Optically detected magnetic resonance of high-density
  ensemble of {NV}$^{-}$ centers in diamond}.
\newblock \emph{\bibinfo{journal}{J. Phys.: Condens. Matter}}
  \textbf{\bibinfo{volume}{28}}, \bibinfo{pages}{275302}
  (\bibinfo{year}{2016}).

\bibitem{Lang2019}
\bibinfo{author}{Lang, J.~E.} \emph{et~al.}
\newblock \bibinfo{title}{Nonvanishing effect of detuning errors in
  dynamical-decoupling-based quantum sensing experiments}.
\newblock \emph{\bibinfo{journal}{Phys. Rev. A}} \textbf{\bibinfo{volume}{99}},
  \bibinfo{pages}{012110} (\bibinfo{year}{2019}).

\bibitem{Genov2017}
\bibinfo{author}{Genov, G.~T.}, \bibinfo{author}{Schraft, D.},
  \bibinfo{author}{Vitanov, N.~V.} \& \bibinfo{author}{Halfmann, T.}
\newblock \bibinfo{title}{Arbitrarily accurate pulse sequences for robust
  dynamical decoupling}.
\newblock \emph{\bibinfo{journal}{Phys. Rev. Lett.}}
  \textbf{\bibinfo{volume}{118}}, \bibinfo{pages}{133202}
  (\bibinfo{year}{2017}).

\bibitem{Plakhotnik2026}
\bibinfo{author}{Plakhotnik, T.}
\newblock \bibinfo{title}{Nanothermometry in living cells: Physical limits,
  conceptual and material challenges}.
\newblock \emph{\bibinfo{journal}{Advanced Functional Materials}}
  \textbf{\bibinfo{volume}{36}}, \bibinfo{pages}{e28655}
  (\bibinfo{year}{2026}).

\bibitem{Konzelmann2018}
\bibinfo{author}{Konzelmann, P.} \emph{et~al.}
\newblock \bibinfo{title}{Robust and efficient quantum optimal control of spin
  probes in a complex (biological) environment. {Towards} sensing of fast
  temperature fluctuations}.
\newblock \emph{\bibinfo{journal}{New J. Phys.}} \textbf{\bibinfo{volume}{20}},
  \bibinfo{pages}{123013} (\bibinfo{year}{2018}).

\bibitem{Dreau2011}
\bibinfo{author}{Dr{\'e}au, A.} \emph{et~al.}
\newblock \bibinfo{title}{Avoiding power broadening in optically detected
  magnetic resonance of single {NV} defects for enhanced dc magnetic field
  sensitivity}.
\newblock \emph{\bibinfo{journal}{Phys. Rev. B}} \textbf{\bibinfo{volume}{84}},
  \bibinfo{pages}{195204} (\bibinfo{year}{2011}).

\end{thebibliography}
